\documentclass[11pt]{article}
\pdfoutput=1
\usepackage{graphicx}
\graphicspath{{./figures/}}
\usepackage{appendix}
\usepackage[normalem]{ulem}

\usepackage{graphicx}
\usepackage{tabularx}
\usepackage{latexsym,amsmath,amsfonts,amssymb,booktabs}
\usepackage[font=small]{caption}
\usepackage{slashed,upgreek,amscd,cancel,tensor,color}
\usepackage{adjustbox}
\usepackage{soul}
\usepackage{multirow}
\usepackage[numbers,compress,square]{natbib}
\usepackage{epsfig,latexsym}
\usepackage[section]{placeins}
\usepackage{amsmath}
\usepackage[pdfencoding=auto]{hyperref} 
\usepackage{url}
\usepackage{footmisc}
\numberwithin{equation}{section}
\usepackage{doi}
\usepackage{subcaption}
\definecolor{MyBlue}{rgb}{0.15,0.15,0.70}
\definecolor{linkblue}{rgb}{0,0,0.8}
\definecolor{linkgreen}{rgb}{0,0.5,0}

\hypersetup{
	colorlinks=true,
	citecolor=linkgreen,
	linkcolor=linkblue,
	urlcolor=linkblue
}

\input epsf
\newcommand{\kmax}{k_\text{max}}

\usepackage{amssymb}
\usepackage{amsmath}
\usepackage{amsfonts}
\usepackage{upgreek}
\usepackage{latexsym}
\usepackage{stfloats}
\usepackage{afterpage}

\numberwithin{equation}{section}

\newcommand{\hinvMpc}{\,h\, {\rm Mpc}^{-1}\,}

\newcommand{\bea}{\begin{eqnarray}}
	\newcommand{\eea}{\end{eqnarray}}
\newcommand{\be}{\begin{equation}}
	\newcommand{\ee}{\end{equation}}
\newcommand{\fr}[2]{\frac{ #1}{#2}}

\newcommand{\kvec}{\vec{k}}

\newcommand{\eqn}[1]{Eq.~(\ref{#1})}

\newcommand{\fnl}{f_{\rm NL}}

\newcommand{\fnlloc}{f_{\text{NL}}^{\text{loc.}}}
\newcommand{\fnleq}{f_{\text{NL}}^{\text{eq.}}}
\newcommand{\fnlort}{f_{\text{NL}}^{\text{orth.}}}

\def\hinvMpc{h\,{\rm Mpc}^{-1}}
\def\Mpcinvh{h^{-1}{\rm Mpc} }

\definecolor{darkgreen}{rgb}{0.0, 0.5, 0.0}

\definecolor{teal}{rgb}{0.0, 0.5, 0.5}

\definecolor{MattOrange}{rgb}{1.0,0.4,0.2}

\newcommand{\Comment}[1]{{}}

\begin{document}
	\def\thefootnote{\fnsymbol{footnote}}
	\setcounter{page}{1} \baselineskip=15.5pt \thispagestyle{empty}
	
	\vspace*{-25mm}

	\vspace{0.5cm}

	\begin{center}
		
		{\Large \bf {Peeking into the next decade in Large-Scale Structure Cosmology\\ 
		with its Effective Field Theory} } \\
		 [0.7cm]
		{\large Diogo Bragan\c{c}a${}^{1,2}$, Yaniv Donath${}^{3}$, Leonardo Senatore${}^{4}$, and Henry Zheng${}^{1,2}$ \\[0.7cm]}
		
	\end{center}

	\begin{center}
		
		\vspace{.0cm}
		
		\begin{small}
			
			{ \textit{ $^{1}$ Stanford Institute for Theoretical Physics, Physics Department,\\
					Stanford University, Stanford, CA 94306}}
			\vspace{.05in}
			
			{ \textit{ $^{2}$ Kavli Institute for Particle Astrophysics and Cosmology,\\
					SLAC and Stanford University, Menlo Park, CA 94025}}
			\vspace{.05in}
			
			{ \textit{ $^{3}$ Department of Applied Mathematics and Theoretical Physics,\\
					University of Cambridge, Cambridge, CB3 OWA, UK}}
			\vspace{.05in}
			
			{ \textit{ $^{4}$ Institut fur Theoretische Physik, ETH Zurich,
					8093 Zurich, Switzerland}}
			\vspace{.05in}

		\end{small}
	\end{center}

	\vspace{0.5cm}
	
	\begin{abstract}
		After the successful full-shape analyses of BOSS data using the Effective Field Theory of Large-Scale Structure, we investigate what upcoming galaxy surveys might achieve. We introduce a ``perturbativity prior" that ensures that loop terms are as large as theoretically expected, which is effective in the case of a large number of EFT parameters. After validating our technique by comparison with already-performed analyses of BOSS data, we provide Fisher forecasts using the one-loop prediction for power spectrum and bispectrum for two benchmark surveys: DESI and MegaMapper. We find overall great improvements on the cosmological parameters. In particular, we find that MegaMapper (DESI) should obtain at least a 12$\sigma$ ($2\sigma$) evidence for non-vanishing neutrino masses, bound the curvature $\Omega_k$ to 0.0012 (0.012), and primordial inflationary non-Gaussianities as follows: $\fnlloc$ to $\pm 0.26$ (3.3), $\fnleq$ to $\pm16$ (92), $\fnlort$ to $\pm 4.2$~(27). Such measurements would provide much insight on the theory of Inflation. We investigate the limiting factor of shot noise and ignorance of the EFT parameters.
		
	\end{abstract}
	
	\newpage
	
	\tableofcontents
	
	\vspace{.5cm}
	\newpage
	
	\def\thefootnote{\arabic{footnote}}
	\setcounter{footnote}{0}
	
	%
	%
	%
	%

	%
	%
	%
	
	\section{Introduction and Summary}\label{intro}

In the last few years, Large-Scale Structure (LSS) survey data have started to be analyzed using the so-called Effective Field Theory of Large-Scale Structure (EFTofLSS)~\cite{Baumann:2010tm,Carrasco:2012cv}. The approach in which the data are analyzed in the context of this theory is rather simple: all data below a certain maximum wavenumber are used in the Bayesian inference. This technique goes under the name of ``full shape analysis". Such an application to data has allowed a measurement of all cosmological parameters of the $\Lambda$CDM model using just a prior from Big Bang Nucleosynthesis~\cite{DAmico:2019fhj,Ivanov:2019pdj,Colas:2019ret}. The precision and accuracy achieved through this measurement were unexpectedly high, offering a new independent method for determining the Hubble constant with a percent-level precision~\cite{DAmico:2019fhj}, and for measuring $\Omega_m$ with a precision comparable to Planck~\cite{Planck:2018vyg}, among other remarkable achievements. {For example, the EFTofLSS prediction at one-loop order has been used to analyze the BOSS {galaxy} Power Spectrum~\cite{DAmico:2019fhj,Ivanov:2019pdj,Colas:2019ret,Zhang_2022,Carrilho_2023}, and Correlation Function~\cite{Zhang:2021yna,Chen:2021wdi}. 
This was extended to eBOSS in~\cite{Simon:2022csv}.
 The BOSS galaxy-clustering bispectrum monopole was analyzed in~\cite{DAmico:2019fhj,Philcox:2021kcw,Tsedrik_2023} using the EFTofLSS prediction at tree-level, and the one loop level was analyzed in~\cite{damico2022boss,Philcox_2022,Ivanov_2023}. All $\Lambda$CDM cosmological parameters have been measured from these data by only imposing a prior from Big Bang Nucleosynthesis {(BBN)}, reaching quite a remarkable precision. For example, the present amount of matter, $\Omega_m$, and the Hubble constant (see also~\cite{Philcox:2020vvt,DAmico:2020kxu} for subsequent refinements) have error bars that are similar to the ones obtained from the Cosmic Microwave Background (CMB)~\cite{Planck:2018vyg}.
 For clustering and smooth quintessence models, limits on the equation of state $w$ of dark energy of $\lesssim 5\%$ have been set using only late-time measurements~\cite{DAmico:2020kxu,DAmico:2020tty,Simon:2022csv}, similar to the ones from CMB~\cite{Planck:2018vyg}. These measurements establish a new, CMB-independent, method for determining the Hubble constant~\cite{DAmico:2019fhj}, with precision comparable to one from the cosmic ladder~\cite{Riess:2019cxk,Freedman:2019jwv} and CMB. Some models that were proposed to alleviate the tension in the Hubble measurements between the CMB and cosmic ladder (see e.g.~\cite{Verde:2019ivm}) have also been compared to data~\cite{DAmico:2020ods,Ivanov:2020ril,Niedermann:2020qbw,Smith:2020rxx,Simon:2022adh}. {Primordial non-Gaussianities~\cite{DAmico:2022gki,Cabass:2022wjy} and more general extended models beyond the $\Lambda$CDM universe have also been compared to BOSS data in~\cite{Moretti:2023drg,Simon:2022hpr,Piga:2022mge,Chudaykin:2020ghx,Cabass:2022ymb,Semenaite:2022unt,Rubira:2022xhb,Carrilho:2022mon,Simon:2022ftd}.}}

{Concerning the theortical developement, the EFTofLSS was first formulated in Eulerian space~\cite{Baumann:2010tm,Carrasco:2012cv} and later extended to Lagrangian space ~\cite{Porto:2013qua}.
The dark matter power spectrum has been computed at one-, two- and three-loop orders in~\cite{Carrasco:2012cv, Carrasco:2013sva, Carrasco:2013mua, Carroll:2013oxa, Senatore:2014via, Baldauf:2015zga, Foreman:2015lca, Baldauf:2015aha, Baldauf:2015zga, Baldauf:2015aha, Cataneo:2016suz, Lewandowski:2017kes,Konstandin:2019bay}.
These calculations were accompanied by some theoretical developments of the EFTofLSS, such as a careful understanding of renormalization~\cite{Carrasco:2012cv,Pajer:2013jj,Abolhasani:2015mra} (which includes some rather-subtle aspects such as lattice-running~\cite{Carrasco:2012cv} and a better understanding of the velocity field~\cite{Carrasco:2013sva,Mercolli:2013bsa}), of the non-locality in time of the EFTofLSS~\cite{Carrasco:2013sva, Carroll:2013oxa,Senatore:2014eva}, and of several ways for measuring counterterms from simulations~\cite{Carrasco:2012cv,McQuinn:2015tva}.
In addition, an IR-resummation of the long displacement fields is required in order to reproduce the Baryon Acoustic Oscillation (BAO) peak, giving rise to the so-called IR-Resummed EFTofLSS~\cite{Senatore:2014vja,Baldauf:2015xfa,Senatore:2017pbn,Lewandowski:2018ywf,Blas:2016sfa}. An enlightening study was performed in 1+1 dimensions~\cite{McQuinn:2015tva}.
Baryonic effects were also taken into account in~\cite{Lewandowski:2014rca,Braganca:2020nhv}. The dark-matter bispectrum has been computed at one-loop in~\cite{Angulo:2014tfa, Baldauf:2014qfa}, the one-loop trispectrum in~\cite{Bertolini:2016bmt}, and the displacement field in~\cite{Baldauf:2015tla}.
The lensing power spectrum has been computed at two loops in~\cite{Foreman:2015uva}.
Biased tracers, such as halos and galaxies, have been studied in the context of the EFTofLSS in~\cite{ Senatore:2014eva, Mirbabayi:2014zca, Angulo:2015eqa, Fujita:2016dne, Perko:2016puo, Nadler:2017qto,Donath:2020abv} (see also~\cite{McDonald:2009dh}), and the halo and matter power spectrum and bispectrum with all cross correlations in~\cite{Senatore:2014eva, Angulo:2015eqa}. Redshift space distortions have been developed in~\cite{Senatore:2014vja, Lewandowski:2015ziq,Perko:2016puo}. 
The EFTofLSS has also been extended to neutrinos in~\cite{Senatore:2017hyk,deBelsunce:2018xtd}, clustering dark energy in~\cite{Lewandowski:2016yce,Lewandowski:2017kes,Cusin:2017wjg,Bose:2018orj}, and primordial non-Gaussianities in~\cite{Angulo:2015eqa, Assassi:2015jqa, Assassi:2015fma, Bertolini:2015fya, Lewandowski:2015ziq, Bertolini:2016hxg}.
Faster evaluation schemes for the calculation of some of the loop integrals have been developed in~\cite{Simonovic:2017mhp}.
Comparison with high-quality $N$-body simulations to show that the EFTofLSS can accurately recover the cosmological parameters have been performed in~\cite{DAmico:2019fhj,Colas:2019ret,Nishimichi:2020tvu,Chen:2020zjt}.}

So far, the data analysis has mainly focused on the data from BOSS DR12~\cite{BOSS:2016wmc}. While there are certainly more ways in which these data can be analyzed, with many much-improved LSS surveys coming online and being designed, it is natural to ask what kind of measurement the application of the so-far developed EFTofLSS to these data will allow. 
In this paper, we do this by Fisher forecasting the information content of two upcoming surveys, that we take as benchmarks, DESI~\cite{DESI:2016fyo}, and MegaMapper~\cite{Schlegel:2022vrv}, including some forecasts for further analyses on BOSS. 
Our intention is that the results for these two upcoming and planned surveys will give an idea of the capabilities of other further surveys, either already planned or to be planned. 
We primarily focus on the cosmological parameters of the flat $\Lambda$CDM model, while also considering neutrino masses, curvature, and primordial non-Gaussianities ({for a forecast about CDM-isocurvature modes for Euclid and MegaMapper, using just the one-loop power spectrum and tree level bispectrum, see~\cite{Chung:2023syw}}). These are in fact parameters whose detection would allow us to extend or ameliorate the standard model of cosmology and of particle physics. 

For neutrinos, we know from neutrino oscillations that they are massive (see for example~\cite{Kajita:2016cak}), but we do not know the absolute value of their masses. For inflation, curvature should naturally be very small, to the order of the primordial perturbations~$\sim3\cdot 10^{-5}$, though evidence of negative curvature could be extremely interesting, pointing towards the fact that our universe might come out of a bubble nucleation event (see for example~\cite{Bousso:2013uia} and references therein). 
A large positive curvature would essentially rule out eternal inflation, while a large negative curvature would rule out slow-roll eternal inflation~\cite{Kleban:2012ph}. 
Finally, non-Gaussianities could reveal the interaction structure of Inflation, which is actually the most insightful aspect to understand the particle physics origin of this theory. 
Concerning the shape of the non-Gaussianities we will explore, we will analyze the so-called $\fnlloc$ (see e.g.~\cite{Babich:2004gb}), $\fnleq$~\cite{Creminelli:2005hu} and $\fnlort$~\cite{Senatore:2009gt} shapes, which parametrize a large class of non-Gaussianities that can be produced in single field inflation~\cite{Cheung:2007st,Senatore:2009gt}, and, for $\fnlloc$, also in multifield inflation (see e.g.~\cite{Bernardeau:2002jy, Lyth:2002my,Zaldarriaga:2003my,Senatore:2010wk}) (see also recently~\cite{Cabass:2022epm} for a forecast on MegaMapper on these parameters using just the tree-level bispectrum). 
It should be stressed that there exist other shapes of non-Gaussianities that are well motivated (see e.g.~\cite{Flauger:2016idt}), and we leave their exploration to future work.

Let us summarize some important technical aspects of our analysis:
\begin{itemize} 

\item We use the prediction of the EFTofLSS at one loop order for the power spectrum and the bispectrum. 
We provide the Fisher forecasts by utilizing all the multipoles of the line-of-sight angle.

\item The model that we implement is the same as in the analyses of the BOSS data as in~\cite{damico2022boss}. 
In particular, this includes the integration of the one-loop bispectrum integrals as developed in~\cite{Anastasiou:2022udy}. 
The modeling of primordial non-Gaussianities is done as in~\cite{DAmico:2022gki}.

\item We check the accuracy of our predictions against comparison with the posterior obtained by analyzing the BOSS data with the full-likelihood of~\cite{damico2022boss}. We conclude that, approximately, our predictions for the error bars should be roughly accurate to about $30\%$ or $40\%$, once the maximum wavenumber of the analysis has been fixed. 
This error is primarily influenced by our assumption of a diagonal covariance.
 
 \end{itemize}

On top of the overall volume of the survey, we identify two limiting factors that affect the precision of the upcoming measurements. 
One is the {discreteness} of the galaxy field, which induces a shot noise term in the data, and the second is the fact that dozens of EFT parameters, including biases, need to be {fitted} to the observations (these EFT parameters encode the effect at long distances from uncontrolled short distance physics, which includes the relation between galaxy overdensities and matter fields). 
We explore these issues in the following way:

\begin{enumerate}
\item {\bf Shot noise:} In addition to our main analysis, we provide Fisher forecasts with shot noise set to zero, effectively assuming an infinitely dense distribution of galaxies.

\item {\bf EFT parameters:} We investigate the impact of limited knowledge regarding the EFT parameters by conducting Fisher forecasts in the following three ways:
\begin{enumerate}
	\item We set the width of the prior on the EFT parameters to zero, which is equivalent to fixing them. This represents the ultimate reach {in terms of constraining power}.
	
	\item\label{itemgalaxyformation} {\bf Galaxy Formation Prior:} We set the width of the prior on the EFT parameters to~0.3 (rather than about $2$ or 4 as in the normal analyses). 
	This is meant to represent a perhaps realistic prior on the EFT parameters as informed by astrophysical galaxy formation studies. {We note that Halo Occupation Distribution (HOD) informed EFT priors have been studied in~\cite{Zhang:2024thl,Ivanov:2025qie}. For example, in~\cite{Zhang:2024thl}, an EFT prior is constructed from the distribution of best-fits obtained from 320 000 mock galaxy catalogs with 10 000 sets of HOD parameters. Since the HOD-informed priors would introduce only a minor correction to our constraints, we find our simpler, uninformative priors to be sufficiently justified. For the galaxy prior, we adopt a width of approximately 0.3, representing a scenario in which advancements in HOD modelling and galaxy formation simulations are able to reduce the EFT prior width by about an order of magnitude.}
	
	\item\label{itemperturbativityprior} {\bf Perturbativity Prior:} Finally, we introduce a new, theoretically-justified prior on the EFT parameters that we call ``perturbativity prior", which is based on the following reasoning. In the EFTofLSS, it is possible to estimate the correct size of a loop term given the lower order terms by simple scaling formulas. It is self-consistent to impose that the loop terms in the analysis obey this estimate. If the number of parameters to fit to observations is small, this criterium is automatically satisfied once the EFT parameters have been assigned a prior of ${\cal{O}}(1)$, which is the standard way in which priors are set~(see for e.g.~\cite{damico2022boss}). However, when the number of parameters becomes large, say~$n\gg 1$, if each parameter is ${\cal{O}}(1)$, the size of the loop can be a factor of ${\cal{O}}\left(\sqrt{n}\right)$ too large, due to some random accumulation effect. It is easy to convince oneself that the data, which are most effective at the highest wavenumbers where the loop is sizable, might not provide sufficient constraining power to prevent such random accumulation effects from happening. In fact, due to cancellations among the various components of a loop term, it is possible that the loop is small at those high wavenumbers where both data and loop are strong, while the loop could still be too large at intermediate and low wavenumbers. Indeed, at these intermediate and low wavenumbers, the data, being weaker, do not constrain the loop term, having this one become quite smaller in the meantime. We therefore set a Gaussian prior on the overall loop term, by favoring the configurations that satisfy the overall size and scaling as a function of wavenumber of the loop term. 
	While this prior has not yet been tested on simulations or data, it is solidly theoretically justified, and we report results incorporating it.
	\end{enumerate}
\end{enumerate}
		
We find the following results for DESI and MegaMapper for the various cosmological parameters:

\begin{itemize}
\item {$\Omega_k$}: Planck 2018 constrains this parameter to 0.0065~\cite{Planck:2018vyg}. For DESI, we forecast a constraint of about 0.051. MegaMapper instead will reduce this bound to 0.0012, representing an improvement of over 5 times compared to Planck, and just about 1.5 orders of magnitude away from the ultimate limit where it makes sense to measure this parameter, which is the amplitude of the primordial curvature fluctuation. 
We notice however that this bound depends quite strongly on the maximum wavenumber of the analysis. 

\item {$\sum_i m_{\nu_i}$}: Planck 2018 constrains this parameter to {be $<$}0.27 eV~\cite{Planck:2018vyg}. However, the most interesting side of the error for this parameter is the lower one, as it is associated to a detection of non vanishing masses. 
We find that DESI will constraint this parameter to about { $-0.07$ eV from above the reference value (which is what is relevant for detection)} with only the power spectrum (P) and {$-0.05$} eV with the addition of the bispectrum, a significant improvement with respect to Planck. Since our bound depends on the central value of neutrino masses, a more invariant way to cast this bound is that we expect on DESI there will be a guaranteed $2\sigma$-evidence for non vanishing neutrino masses. MegaMapper instead will reduce this bound to 0.008 eV, which should guarantee a $\gtrsim 12\sigma$ detection. After the measurement of the cosmological constant $\Lambda$, this would be the second parameter of the standard model of particle physics that is measured from cosmological observations.

\item {$f_\text{NL}$}: Planck 2018 constraints $\fnlloc$ to $\pm 5$, $\fnleq$ to $\pm47$, $\fnlort$ to $\pm 24$~\cite{Planck:2019kim}. We find that DESI will constraint $\fnlloc$ to $\pm 3.3$, $\fnleq$ to $\pm92$ (or $\pm 114$ without the perturbativity prior), $\fnlort$ to $\pm 27$, which are quite comparable to the limits obtained by Planck. MagaMapper will further reduce these bounds as $\fnlloc$ to $\pm 0.26$, $\fnleq$ to $\pm16$ (or $\pm 18$ without the perturbativity prior), $\fnlort$ to $\pm 4$. 
These are very significant improvements with respect to Planck, that range from factors of almost three for $\fnleq$, six for $\fnlort$, to about a factor of 20 for~$\fnlloc$. 
Such a level of improvement brings with it a clear chance of a discovery of primordial non-Gaussianities, opening the door to a deeper understanding at the particle physics level of the inflationary theory. 
Additionally, the allowed values for non-Gaussianities would begin to be close to that ${\cal{O}}(1)$ which represents the vague but significant threshold beyond which inflation is of the slow-roll kind. 

\end{itemize}

We also study the limiting effect of shot noise and biases. 
We find {that} setting the shot noise to zero for DESI would reduce the error bars of practically all parameters by roughly a factor of two. 
For MegaMapper, the reduction would be of a factor of about three for the $f_{\text{NL}}$'s, and about an order of magnitude for the other cosmological parameters.

Regarding the EFT parameters, assuming perfect prior knowledge of the biases in DESI would lead to varied reductions in the error bars, typically by a factor of two. 
For $\fnleq$, the reduction would be around a factor of five. When considering our galaxy-formation benchmark prior of 0.3 instead of fixed parameters, the improvement is significantly diminished to approximately 30\% for $\fnleq$ and $\fnlort$, with marginal impact on the other parameters.
We find a similar behavior for MegaMapper. 
It appears that the perturbativity prior only captures a fraction of the potential improvement achievable through exact knowledge of the biases.
It would be interesting to see if higher $n$-point functions or higher-order computations can improve on this. 

\vspace{0.6cm}

Overall, this Fisher analysis tells us that even by just using the EFTofLSS at the current level of development, the next decade in LSS surveys could lead to great improvements in our knowledge of cosmological parameters.
	This includes parameters that have not yet been measured, such as neutrino masses, as well as those connected to inflation, such as primordial non-Gaussianities and the curvature of the universe. 
Improvements in the design of surveys to reduce shot noise, or advancements in the measurement of EFT parameters have the potential to further strengthen these already promising results.

{ 

\paragraph{Public Codes:} 
The code to compute the Fisher forecasts is publicly available on GitHub~\footnote{
\url{https://github.com/YDonath/EFTofLSSFisher}}.

}

	\section{Technical aspects of the Fisher Matrix}\label{Fishercont}
	Fisher analyses have become a key tool for forecasting in cosmology. Pioneered in \cite{Tegmark:1997rp}, there have been numerous applications over the past years (for example \cite{Agarwal:2020lov, Yankelevich:2018uaz, Baumann:2017gkg}). We here briefly lay out which methods we will use for our forecasts and what contributes to our estimates.
	
	 At the heart of Fisher forecasts lies the {Cram\'{e}r-Rao} lower bound. It states that the covariance of unbiased estimators for a set of parameters $\theta$ is bounded below by the inverse of the Fisher information matrix $F_{ij}$, defined as the expected value of the Hessian of the log-likelihood
	\bea
	F_{ij}=-\bigg\langle\frac{\partial^2 \log L}{\partial \theta_i \partial \theta_j}\bigg\rangle.
	\eea
	{Under the assumption that the likelihood is Gaussian with mean {$X$} and covariance $C$, we can approximately write the Fisher matrix as \cite{Tegmark:1997rp}{
	\bea\label{Fishersimp}
	F_{ij}=\frac{\partial X^T}{\partial \theta_i}C^{-1}\frac{\partial X}{\partial \theta_j}.
	\eea
	}}
	In order to calculate the Fisher matrix, we need to assume reference parameters $\theta^{\text{ref}}$, on which we evaluate the derivatives. In particular, the reliability of the Fisher forecast depends on this reference cosmology being fairly accurate.
 {An intuitive way to see the reference cosmology dependence and also an alternative derivation for \eqn{Fishersimp} is to start directly with a Gaussian likelihood for an observable $X$ depending on parameters $\theta$, that has a true mean $\tilde{X}$, such that the log-likelihood is given by 
\bea\label{simpexplanation}
-2 \log L = (X(\theta)-\tilde{X})^T C^{-1} (X(\theta)-\tilde{X})+p_1,
\eea
where $p_1$ is a $\theta$-independent constant. We can then assume a reference value $\theta^{\text{ref}}$, and Taylor expand around the reference value to first order, $X(\theta) \simeq X(\theta^{\text{ref}})+\sum_i \frac{\partial X}{\partial \theta_i}(\theta_i-\theta_i^{\text{ref}})$. If $X(\theta^{\text{ref}})$ is accurate we can identify that the $\tilde{X}$ term in \eqn{simpexplanation} cancels with $X(\theta^{\text{ref}})$~(\footnote{Note that the Taylor expansion to first order is sufficient exactly because of this cancelation.}). What remains is the first order term, which we can substitute back into \eqn{simpexplanation}, to get
\bea\label{Likelderiv}
-2 \log L = (\theta-\theta^{\text{ref}})^T F (\theta-\theta^{\text{ref}})+p_1,
\eea
where we get the same formula for the Fisher matrix, $F$, as in \eqn{Fishersimp}. In \eqn{Likelderiv} we can now clearly see that the Fisher matrix is the inverse covariance for the likelihood of the parameter vector $\theta$. As can be seen from the derivation above, the Fisher formalism is sensitive to an accurate reference cosmology in order for $\tilde{X}$ to cancel with $X(\theta^{\text{ref}})$ and also for the Taylor expansion to be an accurate approximation.} Given that we now have precise measurements of all cosmological and EFT parameters \cite{DAmico:2022osl} we have good reason to believe we are making realistic predictions around a realistic reference cosmology. In fact, we will show in Sec.~\ref{valid} that we can reproduce results from previous surveys to great precision. We also checked that slight deviations from this reference cosmology do not greatly alter our results.

	{Note that \eqn{Fishersimp}} is in principle simply an inverse-covariance weighted sum over all available information. Both the mean and the covariance can be modeled using perturbation theory. We will discuss this in Secs.~\ref{powandbis} and \ref{combred}. Then in Sec.~\ref{specatred} we will discuss further ingredients that go into calculating the Fisher matrix, such as fixing the reference bias parameters at different redshifts. We here always refer to the correlators of galaxies in redshift space. 	Therefore, unless explicitly mentioned, we drop the ${}^{h,r}$ index with respect to the notation in \cite{DAmico:2022ukl}, and by $P, B, \delta, ...$ we {always} indicate the quantities for biased tracers in redshift space.
\subsection{Power spectrum and bispectrum}\label{powandbis}
For both the power spectrum and bispectrum there are well-established thin-shell averaged estimators that predict the mean and covariance for \eqn{Fishersimp}. We use the estimators that bin in the momenta, but not in the line of sight~\footnote{Binning in line of sight angles is also possible, see for example \cite{Yankelevich:2018uaz}.}, since we have enough analytical control to integrate over the full line of sight information. 
	
	Furthermore, for both correlators, we use {leading-order} contributions in the covariance, in particular neglecting power spectrum-bispectrum cross-covariance. We will discuss the impact of this approximation in Sec.~\ref{validdiagcov}. This assumption in particular implies that we can write the combined power spectrum and bispectrum Fisher matrix, $F^{P+B}$, as the sum of the individual Fisher matrices
	\bea
	F^{P+B}=F^{P}+F^B,
	\eea 	
	where $F^{P}$ and $F^B$ are the power spectrum and bispectrum Fisher matrices respectively. Next, we will {discuss} these individual contributions. 
	\paragraph{Power spectrum}
	For the power spectrum, the estimator is given by \cite{Tegmark:1997rp,Feldman:1993ky}
	\bea\label{pest}
	\hat{P}(k;\hat{z})=\fr{1}{V_SV_P(k)}\int_{\Delta B_k}d^3q\delta(q;\hat{z})\delta(-q;\hat{z}),
	\eea
	where we used the notation $\Delta B_k=B(0,k+\fr{\Delta k}{2})\backslash B(0,k-\fr{\Delta k}{2})$ and $B(a,r)$ is the ball of radius {$r$} around the point {$a$}. $V_S$ is the survey volume and $V_P(k)=4\pi k^2 \Delta k$ is {a normalization factor given by just the integral on the right-hand side of \eqn{pest} {without the prefactors and }with the galaxy density contrast $\delta\rightarrow 1$, i.e. $V_P(k)= \int_{\Delta B_k}d^3q$}. 
	
	The expected value of the estimator is simply the power spectrum itself and we will evaluate it up to the one-loop order. That is 
	\bea\label{pdef}
	P=P_{\text{Tree}}+P_{1L}.
	\eea
	Here the tree and loop contributions include their respective stochastic and response terms. We use the exact same model as in \cite{DAmico:2022osl,DAmico:2022ukl}. Then, for the covariance, we get
	{\bea\label{covderiv}
	C_{PP}(k,k') &=& \langle\hat{P}(k)\hat{P}(k')\rangle-\langle\hat{P}(k)\rangle\langle\hat{P}(k')\rangle \\ \nonumber
	&=&\fr{1}{V_S^2 V_P^2(k)}\int_{\Delta B_k \times \Delta B_k'}d^3q_1 d^3q_2 \bigg[\langle\delta(q_1;\hat{z})\delta(-q_1;\hat{z})\delta(q_2;\hat{z})\delta(-q_2;\hat{z})\rangle \\ \nonumber 
	&&\qquad-\langle\delta(q_1;\hat{z})\delta(-q_1;\hat{z})\rangle\langle\delta(q_2;\hat{z})\delta(-q_2;\hat{z})\rangle\bigg] \\ \nonumber
	&=&\fr{2}{V_S^2 V_P^2(k)}\int_{\Delta B_k \times \Delta B_k'}d^3q_1 d^3q_2 \langle\delta(q_1;\hat{z})\delta(q_2;\hat{z})\rangle\langle\delta(-q_1;\hat{z})\delta(-q_2;\hat{z})\rangle \\ \nonumber 
	&=&\delta_{k,k'}\fr{4\pi^2}{V_S k^2 \Delta k}P_{\text{Tree}}(k;\hat{z})^2.
	\eea
	Note that, schematically, we used $\langle\delta\delta\delta\delta\rangle=\langle\delta\delta\delta\delta\rangle_c+3\langle\delta\delta\rangle\langle\delta\delta\rangle$, where, by the subindex ${}_c$, we mean the connected correlator. The connected four-point function is the trispectrum, which we neglect here. Of the three disconnected parts one cancels with the $\langle\hat{P}(k)\rangle\langle\hat{P}(k')\rangle$ term in the first line and the other two terms are the same, giving the factor of two in the third line. As mentioned, off-diagonal} contributions to the above covariance are of the order of the trispectrum, and therefore similar in size to the one-loop power spectrum multiplied by a linear power spectrum. To be consistent in the perturbative order, given that we neglect the off-diagonal contributions, we also neglect loop contributions to the diagonal of the covariance. We will see in Sec.~\ref{valid} that this leads to a roughly {$10\%$} effect.
	Finally, by substituting the mean and covariance into \eqn{Fishersimp} we get the power spectrum Fisher matrix
	\bea\label{FullpF}
	F_{ij}^P = \sum_k \frac{k^2 V_S}{4 \pi^2 \Delta k}\int_{-1}^1 \frac{d\mu}{2}\frac{\partial P(k;\hat{z})}{\partial \theta_i}\frac{\partial P(k;\hat{z})}{\partial \theta_j}\frac{1}{\left(P_{\text{Tree}}(k;\hat{z})\right)^2},
	\eea
	with $\mu = \hat{k}\cdot\hat{z}$ the line of sight angle.
	\paragraph{Bispectrum}
	For the bispectrum, we use the estimator \cite{Scoccimarro:1997st,Chan:2016ehg}
{\bea
	\hat{B}(k_1,k_2,k_3;\hat{z})=\fr{1}{V_SV_B(k_1,k_2,k_3)}	\int\limits_{\Delta B_{k_{123}}}d^3q_1d^3q_2d^3q_3\delta_D(q_1+q_2+q_3)\delta(q_1;\hat{z})\delta(q_2;\hat{z})\delta(q_3;\hat{z}),
	\eea}	where we defined $\Delta B_{k_{123}}=\Delta B_{k_1} \times\Delta B_{k_2}\times\Delta B_{k_3}$. Similarly to the power spectrum discussion, the volume $V_B$ is defined by {just} the above integral with $\delta\rightarrow 1$ {, without prefactors} and is given by 
	\bea
	V_B(k_1,k_2,k_3)&=& \int\limits_{\Delta B_{k_{123}}}d^3q_1d^3q_2d^3q_3\delta_D(q_1+q_2+q_3)\\ \nonumber
	&=&8 \pi^2 k_1k_2k_3 \Delta k_1\Delta k_2\Delta k_3 \beta(k_1,k_2,k_3),\eea
	where, without loss of generality, we assume $k_3\geq k_2\geq k_1$ and $\beta(k_1,k_2,k_3)=1$ unless $k_3=k_1+k_2$, in which case it is $\frac{1}{2}$. Again the mean of this estimator is simply the bispectrum itself, which we calculate up to the one-loop order as in \cite{DAmico:2022ukl}:
	\bea\label{bdef}
	B=B_{\text{Tree}}+B_{1L}.
	\eea
	Again the tree and loop contributions include their respective stochastic and response terms. To leading order, the bispectrum-bispectrum covariance is given by~\footnote{{The derivation of the bispectrum covariance follows analogously to \eqn{covderiv}: we expand the six-point function into a sum of products of connected correlators, and neglect terms that are higher order. For more details, see for example \cite{Sefusatti:2006pa}.}}
	\bea\label{bcov}
	C_{BB}(k_1,k_2,k_3,k_1',k_2',k_3')&=&\fr{(2\pi)^6}{V_S V_B(k_1,k_2,k_3)}s_B\prod_{i=1}^3 \left(\delta_{k_i,k_i'}P(k_i;\hat{z})\right),
	\eea
	where $s_B$ is a symmetry factor that is {equal to} 6 for equilateral triangles, 2 for isosceles triangles and 1 otherwise. Again, we do not consider off-diagonal contributions, which would be the connected six-point function, as well as bispectra squared and products of trispectra and power spectra. While these contributions are suppressed relative to the diagonal, they may be sizable in some cases and we therefore check this approximation with respect to {covariances measured in mocks} in Sec.~\ref{valid}. Finally, we bin equally in all $k_i$, so that {if we plug \eqn{bdef} and \eqn{bcov} in to \eqn{Fishersimp}, we get }the bispectrum Fisher matrix at fixed redshift {(see Sec. 4.2 of \cite{Agarwal:2020lov} or Sec. 4.1.3 of \cite{Desjacques:2016bnm} for more details):}
		\bea\label{FullbF}
		F_{ij}^B=\frac{V_S}{(2 \pi)^5}\sum_{(k_1,k_2,k_3)}\fr{1 }{s_B }\int_{-1}^1\int_0^{2\pi}d\mu_1d\phi\fr{\partial B}{\partial \theta^i}\fr{\partial B}{\partial \theta^j}\prod_{i=1}^3\left(\fr{k_i \Delta k}{ P_{\text{Tree}}(k_i;\hat{z})}\right)\begin{cases}
		\fr{1}{2}, \text{if }k_3=k_1+k_2\\
		1, \text{else}\\
	\end{cases},
\eea 
 where $\mu_i(\mu_1,\phi) = \hat{k_i}\cdot\hat{z}$ are the projected momenta, and we omitted writing the arguments of $B$ to avoid clutter. We use the same parametrization of the $\mu_i$ as for example in \cite{ DAmico:2022osl,DAmico:2022ukl}.

	\subsection{Combining redshifts}\label{combred}
	In the previous section, we derived formulas for Fisher matrices for the power spectrum and bispectrum at a single redshift. However, for surveys that cover a range of redshifts, we need to combine the information from different redshift bins to compute the overall Fisher matrix. We will now lay out how we combine these redshifts.
	
	Assuming that all {EFT-}parameters at different redshifts are uncorrelated~\footnote{Note that the {fact that the EFT-}parameters at different redshifts are in principle correlated has been used in \cite{DAmico:2022osl}. But given the large range of redshifts and the mild correction from correlations, we neglect this here. For BOSS, where the redshift binning is finer than for DESI and MegaMapper, and thus the correlation is stronger, this is a 15-20\% effect. Therefore we assume for future surveys the impact will be negligible.}, the full Fisher matrix for a survey with a set of redshift bins can be expressed as the sum of the power spectrum Fisher matrix and the bispectrum Fisher matrix over the redshift bins 
	\bea\label{fullFz}
	F_{\text{survey}} = \sum_{z}{\left(F^P(z)+F^B(z)\right)}.
	\eea
	
	In all forecasts we consider, we split the survey into two sets of redshift bins{, let us call them $\text{bin}_1$ and $\text{bin}_2$}. Let us write $F_{\text{survey}}=F_{\text{survey},1}+F_{\text{survey},2}$ and the sum in \eqn{fullFz} for each of these two Fisher matrices simply runs over the redshifts in that particular set of bins (for BOSS this is, for example, {LOWZ and CMASS, i.e. $\text{bin}_1=\{z\in$ Tab.~\ref{BOSSnumbers} $| z\leq0.45\}$ and $\text{bin}_2=\{z\in$ Tab.~\ref{BOSSnumbers} $| z>0.45\}$}). A common approach is then to define effective redshifts $z_{\text{eff}}$, background number density $n_{b,\text{eff}}$, etc., and simply compute the $F_{\text{survey},i}$ with these {values}. This is a good approximation for the derivatives of the observables, given that the time dependence is largely dominated by the growth factors that have comparably weak time dependence. However, {we find that }this is not a very accurate estimate for the covariance~\footnote{{Apart from the fact that $n_{b,\text{eff}}$ simply gives a very bad estimate for a realistic covariance (i.e. reproducing the measured covariance),} notice also that the power spectrum and bispectrum covariances scale with different powers of $n_b${, and} their measured effective numbers would not be the same.}.{ For example, for the powerspectrum covariances of CMASS and LOWZ, the approximation causes a $5-10\%$ difference.} To emphasize the different redshift-dependent contributions that enter the covariance, let us write the covariance from the previous sections in their full form:
		\bea\label{cov}
	C_{PP}(k,k';z) &=&\delta_{k,k'}\fr{4\pi^2}{V_S(z) k^2 \Delta k}\left((b_1(z)+f(z) \mu^2)^2 P_{11}(k)+\frac{c_1^{St}}{n_b(z)}\right)^2 \\ \nonumber
	C_{BB}(k_i,k_i';z)&=&\fr{32\pi^4 s_B\prod_{i=1}^3 \left(\delta_{k_i,k_i'}\right)}{V_S(z) k_1k_2k_3 \Delta k^3 \beta(k_1,k_2,k_3)} \prod_{i=1}^3 \left((b_1(z)+f (z)\mu_i^2)^2 P_{11}(k_i)+\frac{c_1^{St}}{n_b(z)}\right),\eea
where $P_{11}$ is the linear dark matter power spectrum, $c_1^{St}$ is the tree-level stochastic term and we abbreviated the triangle dependence on the left-hand side $k_i = \{k_1,k_2,k_3\}$. Defining effective numbers as an approximation is not appropriate for the covariance since it is very sensitive to accurate estimates of the survey volume $V_S$, number density $n_b$, and linear bias $b_1$. {Their numerical values} are typically given by survey specifications, and vary greatly with redshift. Specifically, $n_b$ greatly varies with redshift as it depends on the survey target selection and measurements. In contrast, the growth rate $f$ and the linear power spectrum $P_{11}$ have a comparably mild redshift dependence. However, in our analysis, we will nevertheless consider their redshift dependence for completeness.	
 
To summarize, we have weak time dependence in the derivatives and strong redshift dependence in the covariance. Therefore, for {$X\in \{P,B\}$} and $i\in \{1,2\}$ we use the following approximation for the Fisher matrix {of { the two redshift bins}}:~\footnote{Note that we can test the validity of {$\sum_{z\in \text{bin}_r}C_{XX}^{-1}(z)$} as an approximation for the covariance on its own, given that we have measurements for the covariances of BOSS. We in part do this in Sec.~\ref{validdiagcov} and Fig.~\ref{validcov}, where we find very good agreement. \label{sumcovfoot} 
} 
{
\bea\label{sumcov}
(F_{\text{survey},r}^X)_{ij} &=& \sum_{z\in \text{bin}_r}\frac{\partial X^T(z)}{\partial \theta_i}C_{X X}^{-1}(z)\frac{\partial X(z)}{\partial \theta_j} \\ \nonumber
&\simeq&\frac{\partial X^T(z_{\text{eff},r})}{\partial \theta_i}\left(\sum_{z\in \text{bin}_r}C_{XX}^{-1}(z)\right)\frac{\partial X(z_{\text{eff},r})}{\partial \theta_j},
\eea
where $z_{\text{eff},r}$ is the effective redshift for $\text{bin}_r$.} {The vector contractions in \eqn{sumcov}, represent the covariance weighted sum over the available information in the correlator $X$. For the modes, this leads to the sums over $k$ in \eqn{FullpF} and \eqn{FullbF}. We also sum over all redshift space information, which in the continuum limit turns into an integral over the redshift space angles in \eqn{FullpF} and \eqn{FullbF}.} The final full Fisher matrix that we use is $F_{\text{survey}}=F^P_{\text{survey},1}+F^B_{\text{survey},1}+F^P_{\text{survey},2}+F^B_{\text{survey},2}$.		
\subsection{Survey specifications at different redshifts}\label{specatred}
Now that we established how to combine the information from different redshifts, let us discuss what { reference }parameters we {choose} for each of these redshifts. Predictions from the EFTofLSS rely on a number of redshift and survey-dependent parameters. While these factors have been measured to great precision at low redshifts for the BOSS survey, we {need to} discuss how we extrapolate these results to different redshifts and surveys.
	\paragraph{EFT parameters}
	Let us first look at the approximate evolution of all EFT parameters entering the galaxy power spectrum and bispectrum. In particular, we analyze the time dependence of the physical (as opposed to the bare) parameters. There are at least two distinct origins of nuisance parameters in the EFTofLSS. On the one hand, we expand some functions (such as the stress tensor or the galaxy overdensity) in terms of all fields they can depend on, multiplied by parameters. Schematically at linear order, with redshift space distortions implied, this is (see for example \cite{DAmico:2022ukl})
	\bea
	\delta(x,t)=b_1(t) \delta_{\text{dm}}(x,t)+...-c_1^h(t) \frac{k^2}{k_{\text{NL}}^2} \delta_{\text{dm}}(x,t)+...\;,
	\eea	
	where $\delta_{\text{dm}}$ is the dark matter overdensity (note that $\delta$ without indices always denotes the overdensity of galaxies in redshift space).
	We will refer to these parameters as biases (this includes response terms, but not stochastic fields). The BOSS best-fit for the biases, $\vec{b}_{\text{BOSS}}$, has been determined in~\cite{DAmico:2022osl}. The explicit numerical values are in App.~\ref{surv}. 
	When fixing the reference cosmology for surveys at higher redshifts, we rescale the biases according to the estimated linear bias given in the survey specifications \cite{DESI:2016fyo, Ferraro:2019uce}. 
	Specifically for any new survey we set the reference value for the vector of biases $\vec{b}= \{b_1,b_2,...,c_1^h,c_2^h,...\}$ according to
{\bea\label{biasrescale}
	\vec{b}^{\text{ref}}= \frac{b_1^{\text{ref}}}{b_{1,\text{BOSS}}}\vec{b}_{\text{BOSS}}.
	\eea}
	Note that the {$b_1^{\text{ref}}$} in the tables of Sec.~\ref{results}, App.~\ref{surv} and \cite{DESI:2016fyo, Ferraro:2019uce}, account for different redshifts and different galaxy species~\footnote{Of course, for different surveys and thus different galaxy species, the mix of biases will not be perfectly approximated by simply rescaling the BOSS values. We have checked, however, that randomly varying the background values of the EFT-parameters (within a physical range of order one) has a much smaller effect than the precision of the results we present in the later sections.}. For the surveys we consider in this paper, we give the numerical values for {$b_1^{\text{ref}}$} in Tabs.~\ref{tabBOSS}, \ref{fig:DESI_base} and \ref{fig:MMo_base}.

	Contrary to biases, there are parameters coming from correlators of stochastic fields. For example, we have
	\bea
	\langle\delta(k,t)\delta(k',t)\rangle' \supset\langle\epsilon_A(k,t)\epsilon_B(k',t)\rangle' = \frac{1}{n_b}\left(c_{1}^{St}+c_{2}^{St}\frac{k^2}{k_{\text{NL}}^2}+...\right).
	\eea
	We will call these terms stochastic.{ Importantly, since they are Poisson distributed, they are constant in time. Therefore, given that we have a vector of measured values for the stochastic terms from BOSS, we could in principle use these reference values for all redshifts. However, we here make a slight correction relative to the analysis done in \cite{DAmico:2022osl}. There, leading order stochastic terms, for example $c_{1}^{St} $, were varied freely, whereas they should be fixed to one, by definition of ${n_b}$~(\footnote{It is possible to have slight deviations from this condition~\cite{Kokron:2021faa}, which we will study in future work.}). Therefore, for surveys other than BOSS, we fix the leading order stochastic parameters for the power spectrum and bispectrum, i.e. $c_{1}^{St,\text{ref}}=c_{1}^{(222),\text{ref}}=1$, and we also do not take derivatives with respect to these parameters~\footnote{{For BOSS, we will let the leading order stochastic terms vary freely to validate our pipeline against previous data analyses in Sec.~\ref{valid}, i.e. there we take derivatives with respect to these parameters. Instead, when predicting further results in Sec.~\ref{results}, we will keep them fixed for BOSS forecasts as well.} }. For all other terms {in} the vector of stochastic terms $\vec{\epsilon}= \{c_{2}^{St},c_{3}^{St},...\}$ we use }	{\bea
	\vec{\epsilon}^{\ \text{ref}}= \frac{1}{c^{St}_{1,\text{BOSS}}}\vec{\epsilon}_{\text{BOSS}},
	\eea}
	where $\vec{\epsilon}_{\text{BOSS}}$ is the vector of stochastic parameters measured for BOSS~\footnote{In all forecasts for BOSS, we use $\vec{\epsilon}_{\text{BOSS}}$ as reference values.}.
	Explicit numerical values are given in App.~\ref{surv} and for details on the specific parameters see \cite{DAmico:2022ukl}.
	
	As a final remark, we note that we have tested the sensitivity of our results in Sec.~\ref{results} to small shifts in the {reference values of} EFT parameters and found that they do not significantly affect our findings. The parameters that have the greatest impact on our results are those used in the modeling of the covariance, namely $b_1$ and $c_1^{St}$. To ensure accurate predictions {for future surveys, where the best-fit is yet unknown}, we take {$b_1^{\text{ref}}$} directly from survey specifications and {we set} {the reference value of }$c_1^{St}$ to one {(with the exception of BOSS, as mentioned earlier)}. 
		
	\paragraph{Perturbative reach}
	The perturbative reach of the EFTofLSS, parametrized by $\kmax$, can be determined in simulations by setting a threshold for the theory systematic error. This was the approach used for example, in \cite{DAmico:2022osl,DAmico:2019fhj, DAmico:2020tty}. For the BOSS CMASS sample this is~$\kmax = 0.22 \hinvMpc$ at one-loop order and $\kmax = 0.1 \hinvMpc$ at tree level. 
	{In the following, we will lay out how we estimate the $\kmax$ for a different survey at a different redshift, motivated by the method used in \cite{DAmico:2019fhj}. 
	There, roughly, it was imposed that the signal-to-noise of the leading theoretical error should not be sizable in the $k$-bin containing the $\kmax$ (see footnote 23 in \cite{DAmico:2019fhj}). This is a good approximation, assuming that the signal-to-noise of the theoretical error only has a sizeable contribution in the highest $k$-bin. Here, however, we want to limit the full signal-to-noise of the theoretical error over all $k$-bins. The motivation for this is two-fold. First, this approach is binning independent, which is important to be consistent between surveys that have different bin sizes. Second even though the signal-to-noise of the leading theoretical error is well approximated by only considering contributions at large $k$ since it grows very steeply with $k$, it leads to a slight overestimate of the $\kmax$~(\footnote{We note that on small redshift differences for example between CMASS and LOWZ, the two approaches produce the same results. Therefore, the estimates in \cite{DAmico:2019fhj} are accurate.}) since one does not consider the theoretical error at $k<\kmax-\frac{\Delta k}{2}$. We, therefore, consider the theoretical error contribution at all scales. 
	Furthermore, as was used in \cite{DAmico:2022osl}, we will use the same $\kmax$ for the power spectrum and the bispectrum, since we expect the $k$-reach to be the same. To summarize, this means we impose that the signal-to-noise of the theoretical error is the same in all surveys, which then defines the $\kmax$ through { 
\bea\label{findkmaxpre}
\sum_{k = k_{\text{min}}}^{k_{\text{max,1}}}\left(\frac{\sigma_{\text{theory,1}}(k,z_1,...)}{\sigma_{\text{data,1}}(k,z_1,...)}\right)^2 =\sum_{k = k_{\text{min}}}^{k_{\text{max,2}}}\left(\frac{\sigma_{\text{theory,2}}(k,z_2,...)}{\sigma_{\text{data,2}}(k,z_2,...)}\right)^2, 
 \eea}
where {the sum runs over the $k$-bins of the respective survey, the dots represent further dependences, such as the EFT-parameter best-fit, and {{$\sigma_{\text{data/theory,i}}$ are the respective theoretical and data errors that we will discuss next.}}} The leading theoretical error is the next higher loop contribution, and we estimate the data error by perturbatively modeling the covariance. Let us focus on the theoretical error first. To good approximation \cite{Carrasco:2013sva} the $L$-th order loop scaling is given by
{ 
\bea\label{scale}
\sigma_{L-\text{Loop}}(k,z)= P_{11}(k,z) \left(\frac{k}{k_{\text{NL}}(z)}\right)^{(3+n(k,z))L},
\eea	
}
where $n(k,z)$ is the slope of the linear power spectrum around $k$. {For the one-loop analysis, the theoretical error is, therefore, $\sigma_{\text{theory,i}}(k,z_i,...)=\sigma_{2-\text{Loop}}(k,z_i)$.} Note that in \eqn{findkmaxpre}, any constant factor will drop out, so we only care about the scaling. Furthermore given that from the BOSS analysis we have $k_{\text{NL}}^{\text{BOSS}} = 0.7 \hinvMpc$, we can get $k_{\text{NL}}$ at different redshifts by solving
\bea
\int_0^{{k_{\text{NL}}(z)}}dq \;q^2 P_{11}(q,z)=\int_0^{0.7 \hinvMpc}dq \;q^2 P_{11}(q,z=0.57).
\eea
To estimate the data error we use the square root of the covariance estimate from \eqn{cov}{, including the summation over redshifts, mentioned in \eqn{sumcov}.} {We average over the redshiftspace dependence, and do not consider shot noise~\footnote{{Setting shot noise contributions in the covariance to zero} gives more conservative values for $\kmax$. However, in surveys with large shot noise (i.e. we underestimate $\kmax$ more), contributions from higher $k$ are negligible exactly because of this large shot noise.}, that is 
 \bea\label{tilddef}
\sigma_{\text{data},i}(k) &=&\left(\sum_{z \in \text{bin}_i}\int_{-1}^{1}\frac{d\mu}{2}C_{PP,n_b\rightarrow \infty}^{-1}(z,\mu)\right)^{-1/2}\\ \nonumber
&=&\fr{2\pi}{ \sqrt{k^2 \Delta k}}\left(\sum_{z \in \text{bin}_i}\frac{V_S(z)}{P_{11}(k,z)^2}\int_{-1}^{1}\frac{d\mu}{2}\frac{1}{\left(b_1(z)+f(z) \mu^2\right)^4} \right)^{-1/2} \\ \nonumber 
&=:& \frac{\tilde{\sigma}_{\text{data},i}(k)}{\sqrt{\Delta k}}.
\eea
In practice, the sum over {$z \in \text{bin}_i$} runs over the bins mentioned in \eqn{sumcov}. Furthermore, we defined $\tilde{\sigma}_{\text{data},\text{bin}_i}(k)$, since we will take the limit $\Delta k \rightarrow dk$, such that \eqn{findkmaxpre} turns into an integral {with integration measure equal to $dk$.} Then, for}
each survey {bin with effective redshift $z_{\text{eff}}$}, to estimate the {$L$-th loop order $\kmax^L$} , we solve
{ 
\bea\label{findkmax}
&&\int^{\kmax^L} dk \frac{P_{11}(k,z_{\text{eff}})^2}{\tilde{\sigma}^2_{\text{data},i}(k)} \left(\frac{k}{k_{\text{NL}}(z_{\text{eff}})}\right)^{(3+n(k,z_{\text{eff}}))2(L+1)}\\ \nonumber &&\qquad=\int^{ k_{\text{max, CMASS}}^{L}}dk \frac{P_{11}(k,z=0.57)^2}{\tilde{\sigma}^2_{\text{data, CMASS}}(k)} \left(\frac{k}{0.7\hinvMpc }\right)^{(3+n(k,z=0.57))2(L+1)}\ .
 \eea
}

{\subsection{Further ingredients from data analyses}\label{dataeffect}
In principle, we now have all the ingredients to compute Fisher forecasts for a given survey. However, there are a number of aspects related to the data and the data-analysis that we want to consider here, in order to best predict future results. For one, there are priors that we put on cosmological parameters and EFT parameters. For completeness we also briefly discuss the Alcock-Paczynski(AP) effect. }
\paragraph{Priors} We impose priors on the EFT parameters that are very similar to those used in \cite{DAmico:2022osl}. Given that we are assuming a Gaussian likelihood in the Fisher analysis, imposing a Gaussian prior amounts to adding the inverse variance of the prior to the diagonal of the Fisher matrix. The key difference between the priors we have in the Fisher analysis and those in the MCMC is that all of our priors are centered around the best-fit value rather than around zero. However, we have verified that this difference has a negligible impact on the error bars. For the special case of the linear bias $b_1$, we use a log-normal prior of variance 0.8 to ensure its positivity~\footnote{Even though the Fisher formalism only allows for Gaussians, we can put log-normal priors, by analyzing $\log(b_1)$ rather than $b_1$, since the derivative with respect to $\log(b_1)$ can easily be computed. Then imposing a log-normal prior is just a Gaussian on the ``log of the parameter".} For all other EFT parameters, we put a Gaussian prior of width 2, except for response and stochastic terms that are joint between the power spectrum and bispectrum, for which we use a Gaussian prior of width 4. These variance choices are analogous to the ones in \cite{DAmico:2022osl}. 

We anticipate {or hope} that, in the coming years, our understanding of galaxy formation will advance to a level that will allow for stronger priors on the EFT parameters. 
In parts of Secs.~\ref{results} and \ref{pertres} we, therefore, separately use a ``galaxy-formation prior", where we put a Gaussian prior of width $0.3$ on all EFT parameters, except for $b_1$ where we put again a log-normal prior, also with width $0.3$. 
This value of the prior is a benchmark value we deemed reasonably close to what can be potentially achieved.

For the cosmological parameters, we use a Gaussian Big-Bang Nucleosynthesis (BBN) prior on the baryon abundance $\omega_b$ centered around the best-fit value and, with a width of $\sigma_{\text{BBN}}=0.00036$. {Also, we analyze $\log m_{\nu}^{\text{tot.}}:=\log(\sum_i m_{\nu_i}/\text{eV})$ rather than $\sum_i m_{\nu_i}$ which implicitly ensures unbounded positivity, with support $(0,\infty)$, on the neutrino masses, since the logarithm ensures positivity}~\footnote{{In \cite{Colas:2019ret}, a flat prior with, for example, width $[0.06 \text{eV},0.9\text{eV}]$ was used, which would slightly ameliorate the results presented here.}}. In the results Sec.~\ref{results} we transform the predicted error on {$\log m_{\nu}^{\text{tot.}}$} back to {a $68\%$ interval on} $\sum_i m_{\nu_i}$~(\footnote{ \label{neurtrino}Note that the Fisher forecast predicts {$\sum_i m_{\nu_i}\sim \text{Lognormal}\left(\left(\log m_{\nu}^{\text{tot.}}\right)^{\text{ref}},\sigma\left(\log m_{\nu}^{\text{tot.}}\right)^2\right)$}, thus the upper and lower bounds of the $68\%$ confidence interval can be easily computed from the lognormal distribution. However, while the confidence interval bounds for the Gaussian posterior of {$\log m_{\nu}^{\text{tot.}}$} are independent of the reference neutrino mass (to the extent that the Fisher forecast is), the confidence interval for $\sum_i m_{\nu_i}$ derived from the Gaussian of {$\log m_{\nu}^{\text{tot.}}$}, is in fact reference value dependent. To see this, note that the $p$-quantile for a Gaussian is of the form {$\left(\log m_{\nu}^{\text{tot.}}\right)^{\text{ref}} +\sigma\left(\log m_{\nu}^{\text{tot.}}\right)\sqrt{2} \ \text{erf}^{-1}(2p-1)$ } whereas for the lognormal it is of the form {$\sum_i m_{\nu_i}^{\text{ref}} \times \exp\left[\sigma\left( \log m_{\nu}^{\text{tot.}}\right)\sqrt{2} \ \text{erf}^{-1}(2p-1)\right]$}. {Therefore, if we write confidence interval bounds in the Gaussian case, of the form $\left(\left(\log m_{\nu}^{\text{tot.}}\right)^{\text{ref}}\right)^{\sigma_+}_{\sigma_-}$, the error would be $\sigma_{\pm} = \sigma\left(\log m_{\nu}^{\text{tot.}}\right)\sqrt{2} \ \text{erf}^{-1}(2p_{\pm}-1)$, so the $\sigma_{\pm}$ are reference value independent (as long as $\sigma\left(\log m_{\nu}^{\text{tot.}}\right)$ is). In the lognormal case, i.e. errors of the form $\left(\sum_i m_{\nu_i}^{\text{ref}}\right)^{\sigma_+}_{\sigma_-}$, we instead have $\sigma_{\pm} =\sum_i m_{\nu_i}^{\text{ref}} \times \left( \exp\left[\sigma\left(\log m_{\nu}^{\text{tot.}}\right)\sqrt{2} \ \text{erf}^{-1}(2p_{\pm}-1)\right]-1\right)$.} We emphasize, therefore, that the upper and lower bounds of the confidence interval for the lognormal distribution scale linearly with the reference value. Therefore, when in Sec.~\ref{results}, we present confidence intervals $[a,b]$ for {$\sum_i m_{\nu_i}^{\text{ref}} = 0.1 \text{ eV}$}, the confidence interval for {$\sum_i m_{\nu_i}^{\text{ref}} = 0.2 \text{ eV}$} would be {$[2a,2b]$}.}). We do not assume any previous knowledge about the other cosmological parameters. Overall, our choice of priors is almost equivalent to those used in the data analyses \cite{DAmico:2022osl,Colas:2019ret}.

\paragraph{Alcock-Paczynski effect} {Galaxy spectra are measured on celestial coordinates. In order to transform to cartesian coordinates, a reference cosmology needs to be assumed, that might not correspond to the true cosmology. This discrepancy between the reference cosmology and the true cosmology introduces a geometric distortion known as the Alcock-Paczynski (AP) effect \cite{Alcock:1979mp}. In order to account for this effect one has to evaluate the theory model on transformed wave numbers, given by
\bea \label{APtrafo} k = \frac{k^{\rm ref}}{q_\perp} \left[ 1 + (\mu^{\rm ref})^2 \left( \frac{1}{F^2} - 1 \right) \right]^{1/2} \, , 
 \mu = \frac{\mu^{\rm ref}}{F} \left[ 1 + (\mu^{\rm ref})^2 \left( \frac{1}{F^2} - 1 \right) \right]^{-1/2} \, ,
\eea
where 
\bea
 q_{\perp} = \frac{D_A(z) h}{D_A^{\rm ref}(z) h^{\rm ref}} \, , \qquad
 q_{\parallel} = \frac{H^{\rm ref}(z) / h^{\rm ref}}{H(z) /h} \, , \quad F = q_{\parallel} / q_\perp,
\eea
and $D_A$ being the angular diameter distance.

Importantly, this transformation is invertible, and therefore, information preserving. Therefore, the AP effect does not lead to any addition or loss of information, when analyzing all (i.e. the full set of multipoles) information available from a given galaxy statistic at a given order. With the exception of a small part of Sec.~\ref{BOSSres}, in Sec.~\ref{results} we will always analyze the full set of multipoles, where the AP effect is, therefore, irrelevant, {and therefore we do not include it}. 
{The only forecasts where we do not analyze the full set of} multipoles {are }in the beginning of Sec.~\ref{BOSSres}, and Sec.~\ref{valid}. 
The relevance of the AP effect on the {Fisher} forecasts is thus limited to these specific analyses and we, therefore, implement only approximate formulae. 
The general structure of the correlators we analyze are sums of products of rational functions in $k$ and $\mu$, multiplying linear power spectra and loop integrals. For both cases, we use that $\left( \frac{1}{F^2} - 1 \right)$ is very close to zero and we can Taylor expand it. For the rational functions, we use\bea k^n \simeq \left(\frac{k^{\rm ref}}{q_\perp}\right)^n\left(1+\frac{n}{2} (\mu^{\rm ref})^2\left( \frac{1}{F^2} - 1 \right)\right), \quad 
 \mu^n \simeq \left(\frac{\mu^{\rm ref}}{F}\right)^n\left(1-\frac{n}{2} (\mu^{\rm ref})^2\left( \frac{1}{F^2} - 1 \right)\right).
\eea
Instead, when evaluating loop integrals or linear powers spectra, we average the $k$ above over redshift space angles, and we, therefore, evaluate on 
\bea
k_{\text{avg}}= \frac{k^{\rm ref}}{q_\perp}\left(1+\frac{1}{6}\left( \frac{1}{F^2} - 1 \right)\right).
\eea
As we will see in Sec.~\ref{valid}, this approximation is enough to recover large parts of the AP effect.

\section{Pipeline validation against BOSS data analysis}\label{valid}
Ultimately, the constraints derived from the Fisher formalism are an approximation to a more complicated MCMC analysis. For one, an MCMC will in general produce non-Gaussian posteriors (see Fig. 3 in \cite{DAmico:2022gki} for an example). 
In addition, there are several modeling effects that are considered in the data analyses that we do not account for in our Fisher analysis. 
In this section, we quantify how much these unaccounted-for effects contribute to the constraints, and thereby estimate the level of precision we can have confidence in when performing Fisher forecasts.
	
We split this section into two parts. In Sec.~\ref{validfullcov}, we focus on observational effects. To isolate the impact of observational effects on the posterior, we fix the covariance entering the Fisher matrix to the one obtained from the data analysis (i.e. the measured covariance extracted from mocks as in \cite{DAmico:2022osl}). The remaining difference is what we call observational effects, which we do not account for~\footnote{We call them observational effects because the error comes from neglecting window functions and {only using an approximate version of the AP} effect.}. Then in Sec.~\ref{validdiagcov} we validate the modeling of the covariance described in Sec.~\ref{Fishercont}. In particular, we study the extent to which the off-diagonal entries in the covariance impact the Fisher forecast. Throughout this section, we consider the power spectrum multipoles $P_\ell$ for $\ell=0,2$ and the bispectrum monopole $B_0$ at 1-loop. We also focus mainly on constraints of base cosmological parameters ($h$, $\ln(10^{10}A_s)$ and $\Omega_m$). {We expect this} {to be} sufficient to quantify the accuracy of the Fisher forecasts presented in Sec.~\ref{results}. {As mentioned in Sec.~\ref{specatred}, we here vary the leading order stochastic parameters freely, since we compare to an MCMC that does not fix them either. This is in contrast to what we will do in Sec.~\ref{results}.}
	 	 
However, let us discuss here also the validation of our results for primordial non-Gaussianity. While what we discuss in the next
sections is also applicable to non-Gaussianity, we highlight here additional validations specific to non-Gaussianity.
For example, while we use the best-fit of the data analysis from \cite{DAmico:2022osl} as our reference cosmology (see App.~\ref{surv} for details), for $f_{\text{NL}}$ we use $f_{\text{NL}}^{\text{ref}}=0$~(\footnote{We validated that non-zero background values for $f_{\text{NL}}$, such as the ones {allowed by} the BOSS data analysis \cite{DAmico:2022gki} or Planck \cite{Planck:2019kim}, yield very similar constraints.}). The observational effects that we discuss in Sec.~\ref{validfullcov} affect non-Gaussianity constraints minimally. Furthermore, since almost all of the information about non-Gaussianity (with the exception of $f_{\text{NL}}^{\text{loc.}}$) lies in the bispectrum, the exclusion of the power spectrum bispectrum cross-correlation in the covariance is not as sizeable as for other parameters. {In conclusion, we are able to estimate that} our forecast for $f_{\text{NL}}$ is accurate to roughly {$10-25\%$}, {as} we study in more detail in Sec.~\ref{validdiagcov}. 

\subsection{Fisher prediction against full MCMC}\label{validfullcov}
Even when using the same covariance, the same perturbative model, and the same reference cosmology, there are still several effects that lead to a difference between data analysis constraints and Fisher constraints. The most important observational effects are the Alcock-Paczynski(AP) effect and the smoothing effect of the window function convolution. To evaluate the significance of these effects, we perform a Fisher forecast, using the same model, best-fit, priors, covariance, bins, etc. as in \cite{DAmico:2022osl,DAmico:2022ukl}. The sole difference is that in one case we get the posteriors through the Fisher prediction ({and only considering an approximate} AP effect, {and not considering} window functions), and in the other through an MCMC that takes these effects into account~\footnote{As a reference point, we take the {same} chain of the analysis \cite{DAmico:2022osl} (figure 1, $P_\ell+B_0^{1loop}$).}. This in particular means, we avoid most of the discussion from Sec.~\ref{Fishercont}, since we are not estimating the covariance here and also do not change any of the bias parameters nor the $\kmax$. {We highlight, that the AP effect will eventually not affect the results {of the forecasts}, since we consider all multipoles and the transformation in \eqn{APtrafo} is information preserving. We consider it here only because the analysis we use as reference point have been done on a limited number of multipoles.} The results of the MCMC against the Fisher are shown in Fig.~\ref{validMCMC}. 

\label{obs}
 \begin{figure}[!htb]
 \centering
 \includegraphics[width=0.67\textwidth]{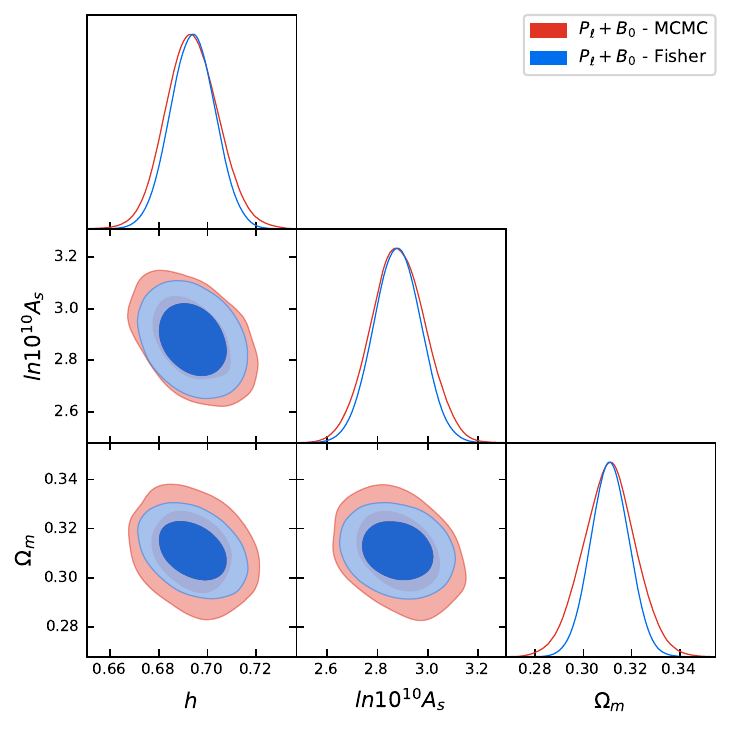}\\
 \hspace{5em}
 {
 \begin{tabular}{|c|c|c|c|} \hline
 $\sigma(\cdot)$ & $h$ & $\ln(10^{10}A_s)$ & $\Omega_m$ \\ \hline
 $P_\ell + B_0 \text{ - MCMC}$ & 0.011 & 0.11 & 0.011 \\ \hline
 $P_\ell + B_0 \text{ - Fisher}$ & 0.0093 & 0.094 & 0.008 \\ \hline
\end{tabular}
}
 \hspace{4em}
 \caption{\footnotesize 
 Triangle plots and errors comparing a Fisher forecast (blue) against the data analysis from \cite{DAmico:2022osl} (red) for base $\Lambda$CDM parameters. For the Fisher forecast the full measured covariance is used including all cross-correlations. We here analyze $\ell=0,2$ power spectrum multipoles and the bispectrum monopole, both at one loop order. We implement the approximate AP effect as discussed in Sec.~\ref{dataeffect}.}
 \label{validMCMC}
 \end{figure}
 
 We observe that the largest discrepancy in the $\sigma$, at {$27\%$}, is found for $\Omega_m${.} However, for the other cosmological parameters, the difference is only around {$15\%$}. We expect a similar level of error, at approximately {$15\%$}, for the other parameters that we analyze in the following sections {and attribute the higher discrepancy on $\Omega_m$ to not fully modelling the AP effect, which most notably depends on $\Omega_m$. However, as argued above this effect will not be relevant for us in later sections}. {We note that if we remove the AP effect and window functions from the BOSS data analysis, the difference to the forecast is less than 5\%~\footnote{Therefore, the non-Gaussian nature of the MCMC-posteriors and the Taylor expansion around the best-fit that the Fisher relies on have minimal impact on the results.}}.
\subsection{Fisher prediction with diagonal covariance}\label{validdiagcov}
For upcoming surveys, the fully measured covariance from mocks is not readily available. Although it is possible to compute these covariances with high accuracy using perturbation theory, the modeling of off-diagonal contributions can be very complex. Therefore, as described in Sec.~\ref{Fishercont}, we do not consider any off-diagonal contributions here~\footnote{However, whenever we analyze multipoles, we do consider multipole cross-correlations, which in multipole space appear as off-diagonal entries. In particular, we consider the $P_0$-$P_2$ cross-correlation.}. Explicitly this means we neglect cross-correlations between the power spectrum and bispectrum and also we neglect cross-correlations between $k$-bins (for the power spectrum and bispectrum respectively)~\footnote{We also neglect any sky correlations since it is a comparably small effect.}. 

By considering the diagonal elements of the measured covariance (actively putting the off-diagonal elements to zero), we can study two effects. First, we can better understand the contribution of off-diagonal elements to the covariance, without yet relying on perturbation theory. Secondly, we can investigate the precision of the analytical diagonal covariance as described in Sec.~\ref{Fishercont} by comparing them to the diagonals of the measured covariance.

The most significant effect of using a diagonal covariance is an overestimation of the bispectrum impact relative to the power spectrum. We find that the off-diagonal elements in the power spectrum covariance have a negligible effect. However, for the bispectrum, the impact is larger. This is expected, as neglecting the power spectrum-bispectrum cross-correlation is equivalent to a scenario where the bispectrum is providing purely independent, new, information from the power spectrum. At least on large enough scales, this is inaccurate as discussed in \cite{Sefusatti:2006pa}. {We check that choosing a higher $k_{\rm min} = 0.1 \hinvMpc$ for both the fully measured covariance as well as the analytical diagonal covariance, parameter constraints obtained now differ only up to 8\%, demonstrating that the diagonal approximation and the omission of the power spectrum-bispectrum cross covariance is only significant on large scales.} In that sense, neglecting the cross-covariance is a double counting of large-scale information. Keeping this in mind, we still want to emphasize that the relative impact of the bispectrum grows rapidly with higher $\kmax$. 

In Fig.~\ref{validcov}, we show the impact of these two isolated effects. First, by reducing the full covariance to just the diagonals, we obtain {about $25\%-30\%$} tighter constraints. Second, if we compare the measured diagonal covariance to the modeled one, the agreement is remarkably good with only {a few percent difference}~\footnote{This strong agreement is {present only when summing} the different redshifts as described in \eqn{sumcov} and footnote \footref{sumcovfoot}. In contrast, the agreement is not as good if we {summarize the survey information to an effective redshift bin}.}. In fact, this could be accounted for by loop order contributions that we did not consider in our analytic modeling of the diagonal covariance which is roughly of the same order. 

 \begin{figure}[t]
 \centering
 \includegraphics[width=0.67\textwidth]{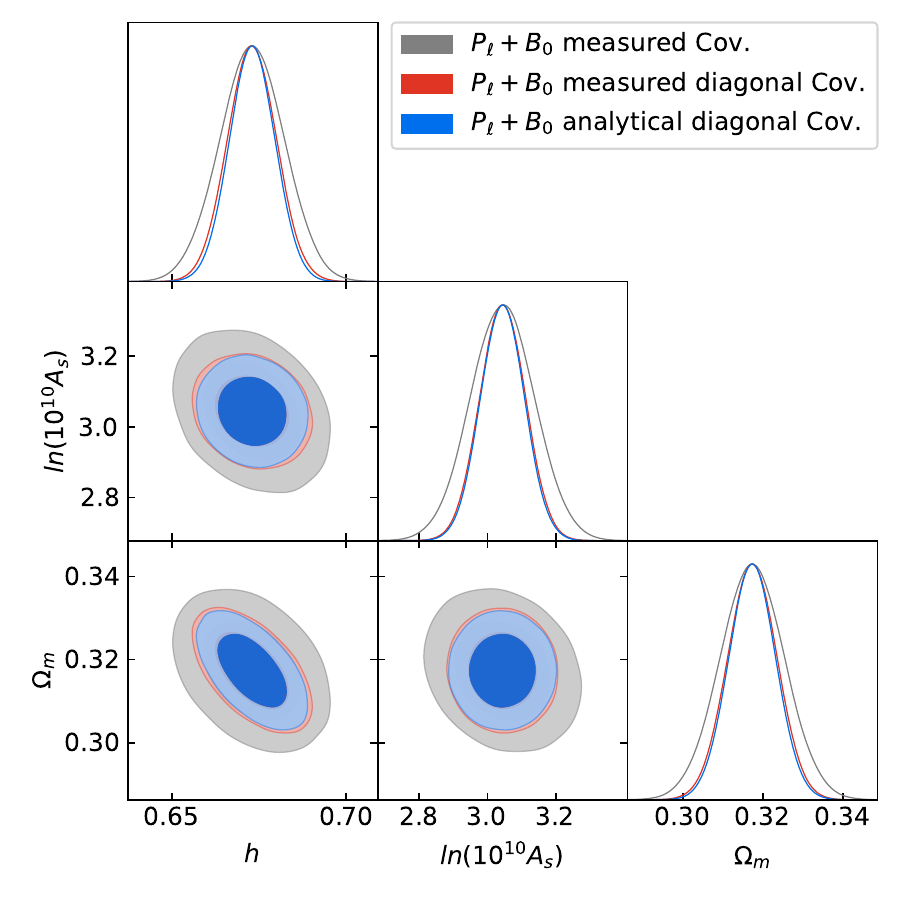}\\
 \hspace{1em}
 { 
 \begin{tabular}{|c|c|c|c|} \hline
 $\sigma(\cdot)$ & $h$ & $\ln(10^{10}A_s)$ & $\Omega_m$ \\ \hline 
 $P_\ell + B_0 \text{ measured Cov.}$ & 0.0093 & 0.094 & 0.008 \\ \hline
 $P_\ell + B_0 \text{ measured diagonal Cov.}$ & 0.0070 & 0.066 & 0.0061 \\ \hline
 $P_\ell + B_0 \text{ analytical diagonal Cov.}$ & 0.0065 & 0.064 & 0.0059 \\ \hline
\end{tabular}
}
 
 \caption{\footnotesize 
 Triangle plots comparing different Fisher forecasts for base $\Lambda$CDM parameters using the $\ell = 0,2$ power spectrum multipoles and the bispectrum monopole, all at one loop order. The plots differ only in their covariances, where we compare the measured covariance (grey), including all cross-covariances, the diagonal of the measured covariance (red), and the analytical prediction for the diagonal covariance (blue). We implement the approximate AP effect as discussed in Sec.~\ref{dataeffect}.} 
 \label{validcov}
 \end{figure}

 By comparing the MCMC data analysis in Fig.~\ref{validMCMC} to the Fisher forecast using the analytical diagonal covariance shown in Fig.~\ref{validcov}, we estimate our confidence in the Fisher results to be about~{$ 40\%$}.
 
 Finally, we perform the same comparison for non-Gaussianity constraints. The MCMC data analysis yields $\left(\sigma(f_{\text{NL}}^{\text{loc.}}), \sigma(f_{\text{NL}}^{\text{eq.}}),\sigma(f_{\text{NL}}^{\text{orth.}})\right)$= (35, 298, 75), while the Fisher forecast with analytical diagonal covariance gives {$\left(\sigma(f_{\text{NL}}^{\text{loc.}}), \sigma(f_{\text{NL}}^{\text{eq.}}),\sigma(f_{\text{NL}}^{\text{orth.}})\right)$= (28, 275, 95).} Thus for non-Gaussianity, we have a much closer agreement at around {$10-25\%$}. 
 
	\section{Results}\label{results}
	{Having validated the Fisher methodology with BOSS data analysis results in Sec.~\ref{valid}, we now use it to predict} the constraining power of DESI and MegaMapper, as well as to provide some additional results for BOSS. As discussed in the previous section, due to observational effects and covariance modeling, we expect the accuracy of cosmological parameter constraints to be roughly {$40\%$}, and the non-Gaussianity parameters to be accurate to {$10-25\%$}. 
	
	While BOSS has now been extensively analyzed, there are still some unexplored questions, which we aim to address here. The full set of cosmological parameters we study in various combinations is {$\{h, \ln(10^{10}A_s), \Omega_m, n_s, \Omega_k, \log m_{\nu}^{\text{tot.}}, f_{\text{NL}}^{\text{loc.}},f_{\text{NL}}^{\text{eq.}}, f_{\text{NL}}^{\text{orth.}}\}$, where we defined $\log m_{\nu}^{\text{tot.}}:=\log(\sum_i m_{\nu_i}/\text{eV})$~\footnote{We approximate the effect of neutrino masses in the analysis solely through the linear power spectrum, and we are not including any further modelling on the level of the bias expansion, or the nonlinear terms.}.} {We refer to {any subset of} the first six parameters in this list (i.e. the list without the non-Gaussianity parameters) as ``base" parameters.} Throughout this section, we use the power spectrum and bispectrum of galaxies in redshift space at 1-loop order. For future surveys like DESI and MegaMapper, we use the full set of multipoles~\footnote{As we will see for the BOSS survey, using the full set of multipoles is roughly equivalent to just using the monopole and quadrupole for both the power spectrum and bispectrum.}, and for BOSS we use either the full set of multipoles or the monopole and quadrupole for the power spectrum and the monopole for the bispectrum~\footnote{When analyzing {a finite number of} multipoles, we implement the approximate AP effect as discussed in Sec.~\ref{dataeffect}. However, when we use the {complete} set of multipoles, in particular for DESI and MegaMapper, the AP effect is irrelevant}. {Furthermore, following the discussion in Sec.~\ref{specatred}, we fix the leading stochastic terms to one, throughout this section. Therefore, the results we will find for BOSS in this section are slightly tighter than what we presented in Sec.~\ref{valid}. }When quoting results for primordial non-Gaussianity, we fix the cosmological parameters, as we will discuss more later this has almost always a negligible effect on our results. {Finally, we note that we use fixed relationships from \cite{DAmico:2022gki,Baldauf:2010vn} between the non-Gaussianity bias parameters $b_i^{f_\text{NL}}$~\footnote{We remind the reader that the parameters $b_i^{f_\text{NL}}$, come from the additional long wavelength field $\tilde{\phi}$, stemming from the non-Gaussianity of the matter field, that now enters the bias expansion. Explicitly, this is~\cite{DAmico:2022gki}
\begin{align}
&[\delta (k ; \hat z ) ]^{\fnl} = b^{\fnl}_1 \fnl \tilde \phi ( k , a_{\rm in}) \\
&\hspace{1in} + \fnl \int \frac{d^3k_1d^3k_2}{(2\pi^3)}\delta_D(\kvec-\kvec_1-\kvec_2) \Bigg( b_1^{\fnl} \left( \frac{\kvec_1 \cdot \kvec_2}{k_2^2} + f \hat z \cdot \kvec \frac{\hat z \cdot \kvec_2}{k_2^2} \right) + b_2^{\fnl} \Bigg) \tilde \phi ( k_1 , a_{\rm in} ) \delta_{\rm{dm}}^{(1)} ( k_2 ) \nonumber \ .
\end{align} } and galaxy bias parameters $b_i$, which we checked to be a negligible approximation with respect to putting an order one prior {centered on $b_i^{f_{\text{NL}, \text{ref}}}$} on $b_i^{f_\text{NL}}$ and then let it vary freely~\footnote{Note that for $f_{\text{NL}}^{\text{ref}}=0$, which is what we use here, in the context of the Fisher forecast, this is not even an approximation, but exact. However, even for non-zero reference values, the change in the error-bar due to this approximation can be quite well understood. If we were only to analyse the power spectrum, we would only constrain the joint parameter $ b_1^{f_\text{NL}}\fnl $ rather than the individual parameters. With the inclusion of the bispectrum this degeneracy is broken due to the presence of $b_2^{f_\text{NL}}\fnl$ {, $b_1 \fnl$, and $ \fnl$ on its own}. The only relevant parameter in the context of this discussion is $\fnlloc$, since for $\fnleq$ and $\fnlort$ almost all information lies in the bispectrum, and we also verified explicitly that the approximation is negligible. In contrast a large part of the constraint on $\fnlloc$ comes from the power spectrum, where however, we can {easily understand} the shift in error bar. Simple error propagation tells us that the error we obtain from the power spectrum {with fixed $b_1^{f_\text{NL}}$ or with the order one prior on $b_1^{f_\text{NL}}$} are related to each other by	\bea
\sigma\left(f_{\text{NL}}^{\text{loc.,full}}\right)&\simeq&\sqrt{\sigma\left(f_{\text{NL}}^{\text{loc.,approx.}}\right)^2+\sigma\left(b_1^{f_\text{NL}}\right)^2\frac{\left(f_{\text{NL}}^{\text{loc.,ref}}\right)^2}{\left(b_1^{f_\text{NL},\text{ref}}\right)^2}},
\eea
where we note that $\sigma\left(b_1^{f_\text{NL}}\right)$ is dominated by the prior width. Therefore these {changes in the error bar} are relevant, if $f_{\text{NL}}^{\text{loc.,approx.}}$ and $b_1^{f_\text{NL}}$ have similar signal to noise $\frac{\sigma\left(f_{\text{NL}}^{\text{loc.,approx.}}\right)}{f_{\text{NL}}^{\text{loc.,ref}}} \simeq \frac{\sigma(b_1^{f_\text{NL}})}{b_1^{f_\text{NL},\text{ref}}}$. For instance with quite large $f_{\text{NL}}^{\text{loc.,ref}}$ both sides of this ratio can roughly be equal to one, as was found in \cite{DAmico:2022gki}. { To be precise, using Planck constraints $\fnlloc= -0.9\pm 5.1$ and a prior $\sigma\left(b_1^{f_\text{NL}}\right)=2$, we find the full change in error bar for $\fnlloc$ with inclusion of both power spectrum and bispectrum at $f_{\text{NL}}^{\text{loc.,ref}}=-0.9$ to be $0.01\%$, $5\%$, $15\%$ for BOSS, DESI and MegaMapper respectively, and even at the Planck 1-$\sigma$ level $f_{\text{NL}}^{\text{loc.,ref}}=-6$, we find changes of $0.5\%$, $47\%$, $54\%$}. {Given in particular that what is important is a detection of non-vanishing $\fnlloc$, and within order one the actual value of $\fnlloc$ is much less important, we conclude that our forecasts are robust even when the $b_i^{f_\text{NL}}$ are free EFT parameters.}}.
 }
	\subsection{BOSS}\label{BOSSres}
	Base cosmological results with the one-loop power spectrum for BOSS have been presented in~\cite{DAmico:2019fhj,Ivanov:2019pdj,Colas:2019ret}, neutrinos have also been analyzed in \cite{Colas:2019ret} and dark energy models in \cite{DAmico:2020tty}. {}{The combination of the power spectrum and bispectrum has led to the measurement of $h$, $\ln(10^{10}A_s)$ and $\Omega_m$ in \cite{DAmico:2022osl}, with the non-Gaussian parameter $f_{\rm{NL}}$ being reported in \cite{DAmico:2022gki} {and at tree level in \cite{Cabass:2022wjy,Cabass:2022ymb}}}. In this section, we present forecasts for the power spectrum and bispectrum with the inclusion of the sum of neutrino masses $\sum_i m_{\nu_i}$, spectral tilt $n_s$ and spatial curvature $\Omega_k$. We also investigate the impact of shot noise and of the EFT parameters and explore the information contained in higher multipoles. The exact numerical values {of the EFT parameters, survey specification and reference cosmology that }we use here and in Sec.~\ref{pertprior} are given in App.~\ref{surv}. Following the binning scheme used in \cite{BOSS:2016wmc}, we divide the sample into two redshift bins. We use the same values of $k_{\text{NL}}$ and $b_1$ for both bins since the redshift difference of the bins is very small. The effective numbers we use are summarized in Tab.~\ref{tabBOSS}.
	\begin{table}[h]
 \centering
\begin{tabular}{|c|c|c|c|c|c|c|c|} \hline
 BOSS: & $z_{\text{eff}}$ & $ n_{b,\text{eff}}[ (\hinvMpc)^3]$ & {$b_1^{\text{ref}}$}&($k^{\text{Tree}}_{\rm max}$, $k^{1L}_{\rm max}$, $k_{\text{NL}}$) $[\hinvMpc]$ &$N^{1L}_{\text{bins}}$&$N^{\text{Tree}}_{\Delta}$&$N^{1L}_{\Delta}$ \\ \hline 
 Bin 1 & 0.32& 2.9$\times 10^{-3}$ & 1.9 & (0.09, 0.20, 0.7) & 18 &9& 62 \\ \hline
 Bin 2 &0.57 & 2.5$\times 10^{-3}$ & 1.9& (0.10, 0.22, 0.7) & 21 &17& 150\\ \hline
\end{tabular}
\caption{\footnotesize BOSS effective survey specifications, calculated according to the formulas in Sec.~\ref{Fishercont} and Tab.~\ref{BOSSnumbers} in App.~\ref{surv}. $n_{b,\text{eff}}$ is the background galaxy number density entering the derivatives (not the covariance), $N_{\text{bins}}$ is the number of $k$-bins we consider for the power spectrum and $N_{\Delta}$ is the number of triangles we consider for the bispectrum.}
 \label{tabBOSS} 
 \end{table}	
 
This section is divided into two parts, based on the type of covariance used. In the first part, we present results using the full measured covariance obtained in \cite{DAmico:2022osl}, which includes all cross-correlations. This {implies that we expect} our results {to be} accurate to about {$15\%$ (and $27\%$ for~$\Omega_m$)} as described in Sec.~\ref{validfullcov}. In the second part, we investigate the impacts of shot noise and higher multipoles using a modeled covariance, as described in Sec.~\ref{validdiagcov}. This allows us to have better analytical control, for example in order to analyze the shot noise influence on the results. 
	\paragraph{Additional results: $n_s$, $\sum_i m_{\nu_i}$ and $\Omega_k$}
	We {present the BOSS forecasts using the} power spectrum monopole and quadrupole, as well as the bispectrum monopole, both at one loop order, {for} parameters that have previously only been analyzed with only the power spectrum (and in some cases with the tree-level bispectrum \cite{DAmico:2019fhj}). The results are summarized in Fig.~\ref{fig:BOSS_ns}.
\begin{figure}[!htb]
 \centering
 \vspace{-1 em}
 \includegraphics[width=0.5\textwidth]{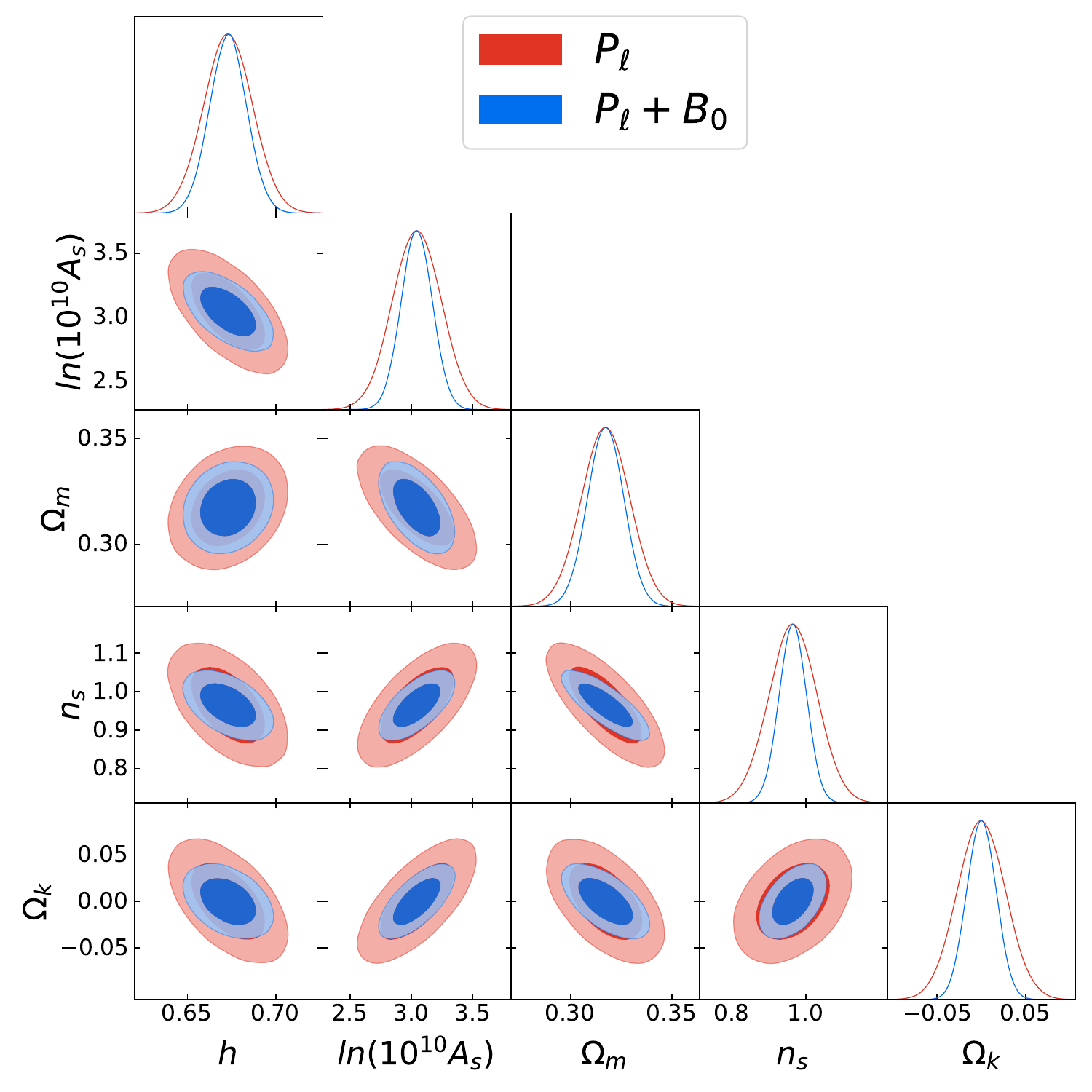} \hspace{-1 em}
 \includegraphics[width=0.49\textwidth]{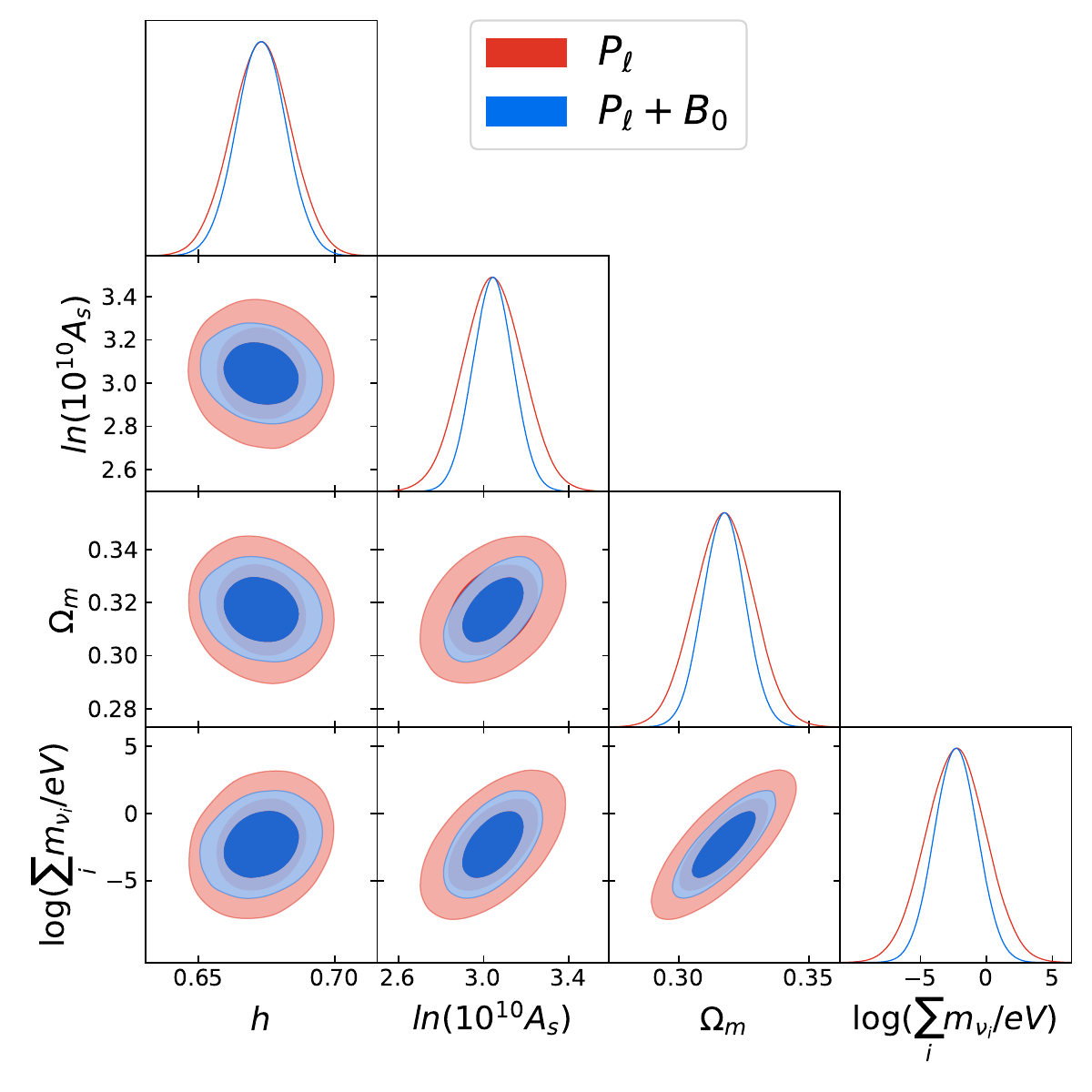}
 \scriptsize
 {
 {\begin{tabular}{|c|c|c|c|c|c|} \hline
 $\sigma(\cdot)$ & $h$ & $\ln(10^{10}A_s)$ & $\Omega_m$ & $n_s$& $\Omega_k$ \\ \hline
 $P_\ell$ & 0.014 & 0.2 & 0.012 & 0.066 & 0.027 \\ \hline
 $P_\ell$+$B_0$ & 0.01 & 0.13 & 0.009 & 0.037 & 0.017\\ \hline
\end{tabular}}
\hspace{0.1cm}
\begin{tabular}{|c|c|c|c|c|c|} \hline
 $\sigma(\cdot)$ & $h$ & $\ln(10^{10}A_s)$ & $\Omega_m$ &$\log m_{\nu}^{\text{tot.}} ({}^{\sigma^+}_{\sigma^-}) $ \\ \hline
 $P_\ell$ & 0.011 & 0.14 & 0.011 & 2.3 (${}^{+0.84}_{-0.089}$)\\ \hline
 $P_\ell$+$B_0$ & 0.0092 & 0.095 & 0.008 & 1.6 ($ {}^{+0.41}_{-0.080}$) \\ \hline
\end{tabular}
 \\
 \vspace{1em}
 \begin{tabular}{|c|c|c|c|} \hline
 $\sigma(\cdot)$ & $f_{\text{NL}}^{\text{loc.}}$&$f_{\text{NL}}^{\text{eq.}}$&$f_{\text{NL}}^{\text{orth.}}$ \\ \hline
 $P_\ell$+$B_0$ &25&266&94 \\ \hline
\end{tabular}
}

 \caption{\footnotesize 
 Triangle plots and errors from Fisher forecasts for BOSS including the spectral tilt and spatial curvature (left), massive neutrinos (right), and primordial non-Gaussianity (bottom). The power spectrum monopole and quadrupole, and the bispectrum monopole were used both at one loop order. {In the table we also report the upper and lower bounds of the $68\%$ confidence interval for the sum of massive neutrinos, i.e $\mathbb{P}\left[\left(\sum_i m_{\nu_i}- \sum_i m_{\nu_i}^{\text{ref}}\right) \in (\sigma^-,\sigma^+)\right] =0.68$.} The covariance used here is the full, measured covariance with all cross-correlations. We implemented the approximate AP effect as discussed in Sec.~\ref{dataeffect}.}
 \label{fig:BOSS_ns}
 \end{figure}	
		
		\paragraph{Impact of shot noise, biases and multipoles}\label{bossadd}
		
		 {For BOSS, we checked that adding the }trispectrum at tree level and the 2-loop power spectrum~\footnote{While currently, we do not have the 2-loop power spectrum for galaxies in redshift space, we can simply run a Fisher analysis on the one loop correlators, with the 2-loop $k_{\rm max}$ reach. This then gives an upper bound estimate for the extra constraining power of the 2-loop correlators. The results on BOSS do not improve even with this optimal estimate, and therefore we believe, a 2-loop analysis will not improve the results on BOSS. \label{2looparg}} do not improve on the measurements~\footnote{{Exploration of the contribution from higher $N$-point functions is currently in progress {\cite{Donath:inprog}}}. }. This is mostly {attributable} to the large shot noise of the survey. However, given the power of the Fisher {formalism}, {we can investigate the effects of certain limits and configurations on parameter constraints.} Of course, we here look at limiting cases, that are unrealistic in reality, but they show where information is lost. In particular, we are interested in the impact of EFT parameters and of the survey shot noise. Throughout this section, we will be using the analytical covariance, which gives us the most control but comes with the caveats mentioned in Sec.~\ref{valid}. We investigate several effects on both base cosmological parameters including $n_s$, and $f_{\text{NL}}$. Just as in \cite{DAmico:2022gki}, unless mentioned otherwise, we fix the cosmological parameters when quoting errors on $f_{\text{NL}}$. 
		 We checked that while $f_{\text{NL}}^{\text{loc.}}$ has a roughly {35\%} error bar reduction due to fixed cosmological parameters, $f_{\text{NL}}^{\text{eq.}}$ and $f_{\text{NL}}^{\text{orth.}}$ are very independent of the other cosmological parameters and their results would {only change by roughly $5-10\%$} if we would not fix the cosmological parameters. 
		
 First, we check the impact of using higher multipoles at one loop {as opposed to} using {only} the power spectrum monopole and quadrupole and the bispectrum monopole. While {there is some improvement with the inclusion of additional multipoles} for the bispectrum, we checked that almost all of this improvement comes from the bispectrum quadrupoles. Still, this improvement is very small, and so we do {not present} the posteriors. The {numerical values} can be found in the table of Fig.~\ref{fig:BOSS_shot}. We can conclude that using the monopole and quadrupole for both the power spectrum and bispectrum, one can extract almost the full redshift space information. {As was shown in \cite{DAmico:2022osl,DAmico:2022gki}, we can already analyze data with the monopole and quadrupole for the power spectrum and bispectrum. Therefore, unless indicated otherwise we analyze using all multipoles.}

 {Next in Fig.~\ref{fig:BOSS_shot}, we show additional constraints in the continuous field limit $n_b \rightarrow \infty$, i.e. having no shot noise.} We roughly halve the error bars for both the base cosmological parameters and $f_{\text{NL}}$. This should serve as motivation to include as many objects into our data sets even if they are faint or somewhat unresolved. 
 This will also become important in Sec.~\ref{MMores}.
 
 Lastly, the EFTofLSS, like any EFT, will {need} a larger number of parameters when going to higher perturbative orders. This is in principle not a problem as long as they are independent enough from the parameters of interest. {It is interesting to investigate {how better knowledge of these parameters would impact the results. We put the ``{galaxy-formation} prior" mentioned in Sec.~\ref{dataeffect}, where we put stronger priors on all EFT parameters motivated by {hopefully-realistic} future {knowledge} on galaxy formation. We also take }the limit in which these parameters are fixed, in other words representing the scenario in which all ``nuisance" parameters are known and measured exactly with no error. This, in a sense, is the theoretical upper bound for the EFTofLSS at a given order.} It is interesting to note from the results presented in the table in Fig.~\ref{fig:BOSS_shot} that the biases have varying impacts on different cosmological parameters. Specifically, the biases overwhelmingly affect the primordial parameters, $A_s, n_s$ and $f_{\text{NL}}$. As we will see in Sec.~\ref{pertprior}, $f_{\text{NL}}$-constraints are more sensitive to the EFT parameters than $A_s$ and $n_s$. 
 This is not so surprising, considering that the functional form of EFT counterterms resemble the functional form induced by primordial non-Gaussianities.

 \begin{figure}[!htb]
 \centering
 \includegraphics[width=0.5\textwidth]{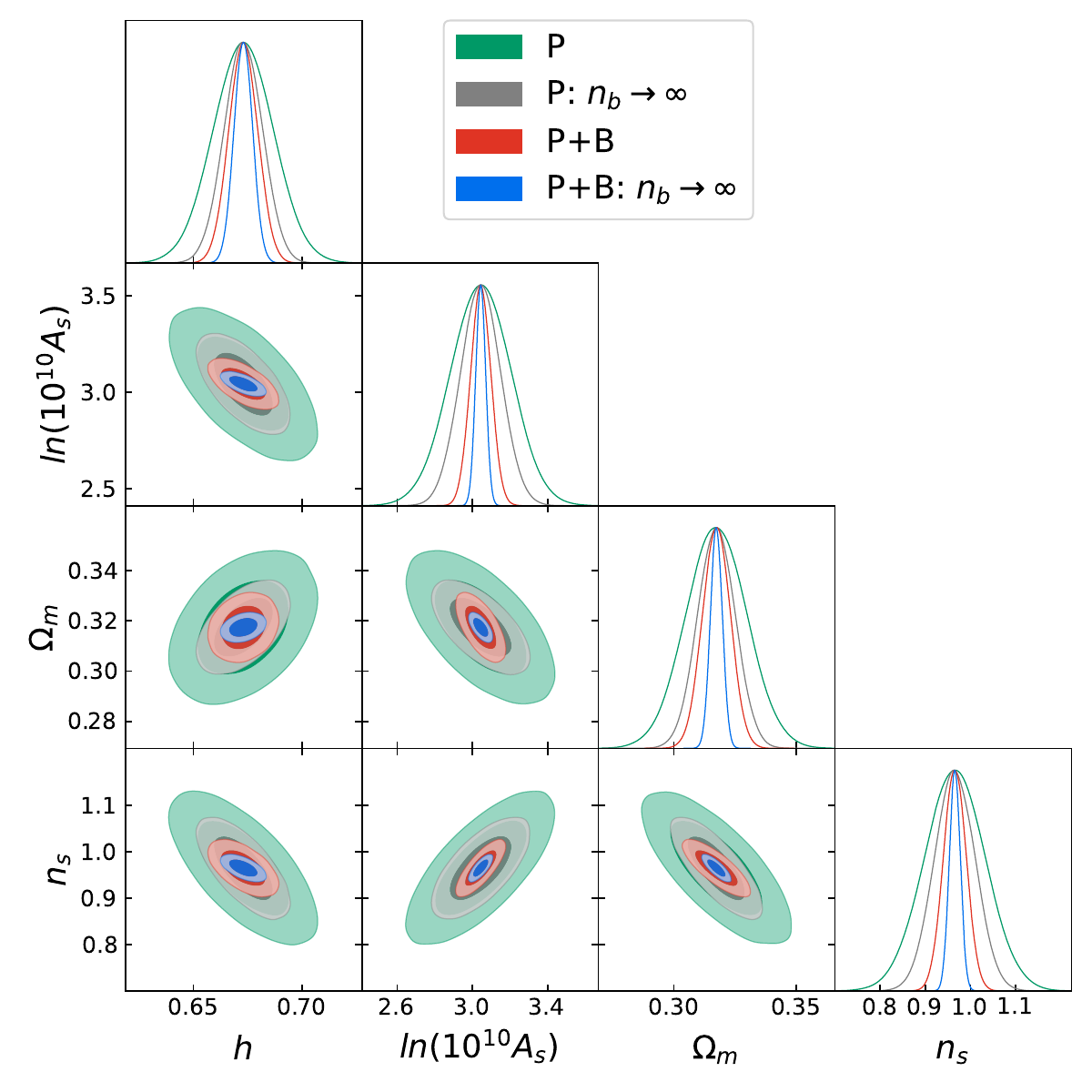} \hspace{-1em}
 \includegraphics[width=0.5\textwidth]{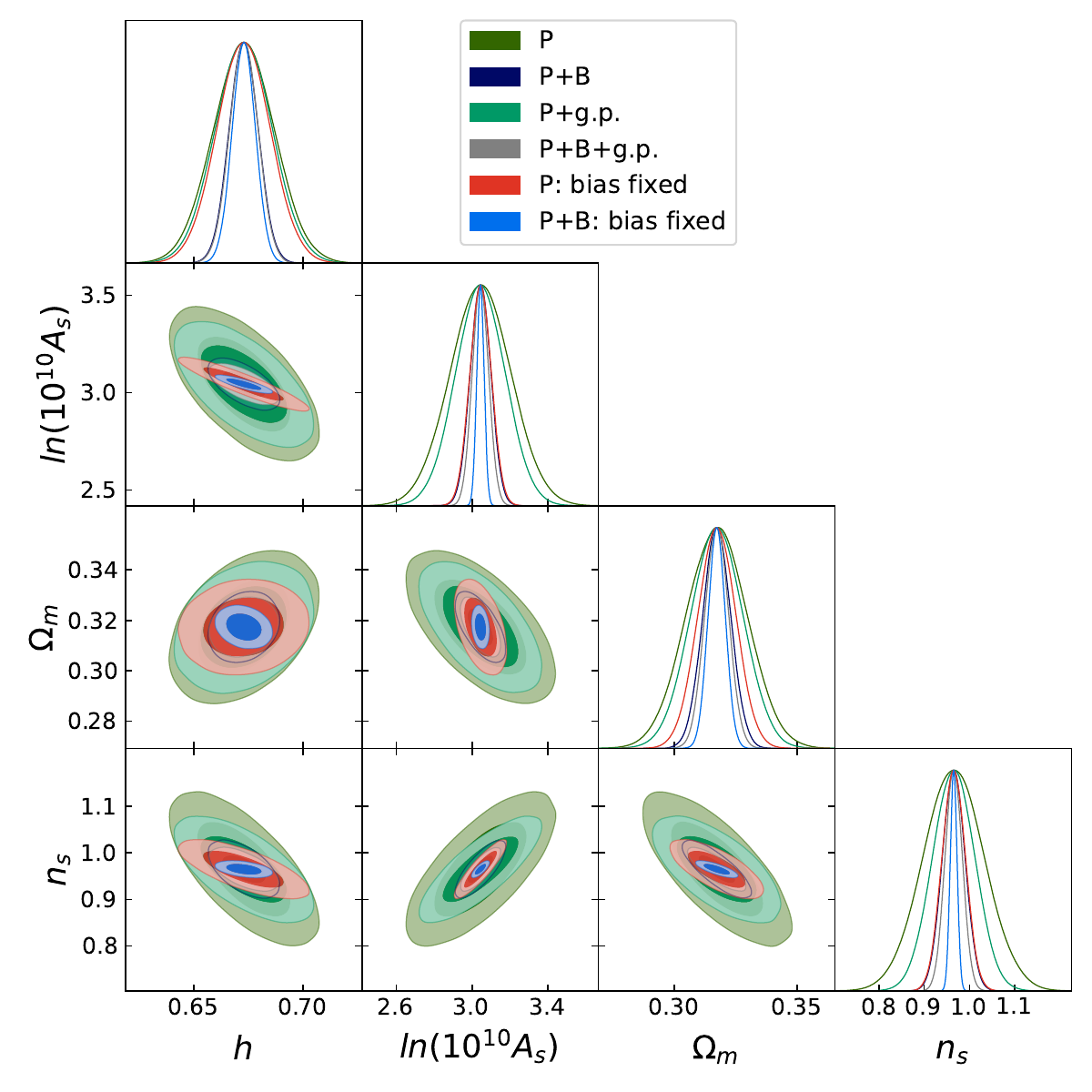}
 \scriptsize
 {
\begin{tabular}{|c|c|c|c|c|c|c|c|} \hline
 $\sigma(\cdot)$ & $h$ & $\ln(10^{10}A_s)$ & $\Omega_m$& $n_s$ &$f_{\text{NL}}^{\text{loc.}}$&$f_{\text{NL}}^{\text{eq.}}$&$f_{\text{NL}}^{\text{orth.}}$ \\ \hline
$P_\ell$ & 0.014 & 0.16 & 0.012 & 0.067& -&-&- \\
 $P$ & 0.014 & 0.16 & 0.012 & 0.067& -&-&- \\
 $P+ g.p.: $ & 0.013 & 0.13 & 0.011 & 0.047 & -&-&- \\ 
 $P: $ bias fixed & 0.012&0.06& 0.008& 0.026 & -&-&- \\ 
 $P: n_b \rightarrow \infty$ & 0.009 & 0.11 & 0.008 & 0.045 & -&-&- \\\hline
 $P_\ell$+$B_0$& 0.0067 & 0.092 & 0.0060 & 0.029& 25&266&94 \\
 $P$+$B$ & 0.0067 & 0.055 & 0.0057 & 0.025& 23&253&67 \\
 $P+B+ g.p.: $ & 0.0065&0.043& 0.0049& 0.019 & 19&187&51 \\ 
 $P+B:$ bias fixed & 0.0054 & 0.019 & 0.0035 & 0.007 & 4.8&45&26 \\ 
 $P+B: n_b \rightarrow \infty$ & 0.0043 & 0.026 & 0.0024 & 0.012& 11&147&41 \\\hline
 \end{tabular}
 }

\caption{\footnotesize 
 Triangle plots and errors from several different Fisher forecasts for BOSS using the analytical covariance. We compare base results to results obtained without shot noise (left) and with biases fixed {or with a ``galaxy-formation prior" (g.p.)} (right). {{In the table, we also show the impact of including higher multipoles on the power spectrum and bispectrum and also see the impact on $f_{\text{NL}}$.}} For the constraints on $f_{\text{NL}}$, we fix the other cosmological parameters. }

 \label{fig:BOSS_shot}
 \end{figure}

\subsection{DESI}\label{DESIres}
	
We now turn to predict the performance of upcoming surveys, starting with the imminent DESI survey. 
We base our results on the Emission Line Galaxies (ELGs) sample, which is the largest of the DESI surveys \cite{DESI:2016fyo}. 
We note that while we are able to derive the value for the linear bias through specifications given in \cite{DESI:2016fyo}, we do not have values for the other EFT parameters. 
We therefore shift all biases according to the method described in Sec.~\ref{specatred}, i.e. we shift them all according to the change in the linear bias with respect to the BOSS best-fit. 
The final numerical values we use for the DESI forecast are given in Tab.~\ref{fig:DESI_base}. 
For all the future surveys we use $k_{\text{min}}=0.001\hinvMpc $ for the power spectrum and $k_{\text{min}}=0.02 \hinvMpc $ for the bispectrum. 
We bin with $\Delta k = 0.005\hinvMpc $ for the power spectrum and $\Delta k = 0.02 \hinvMpc $ for the bispectrum. 
For binning consistency, we have checked that using a smaller binning does not affect our results. {To reduce binning effects, ideally one would always average the observables over a $k$-bin. However, given the numerical complexity of doing such a procedure for every bin, especially for the bispectrum, we instead evaluate on a effective number $k_{\text{eff}}$ to approximate this averaging. This is analogous to the method used for example in \cite{DAmico:2022osl}, with the difference that we also evaluate on $k_{\text{eff}}$ the tree level contribution, rather than averaging it. The effect is minimal.} 
Throughout this section, we use the analytical covariance from \eqn{cov} and \eqn{sumcov} with all the ingredients discussed in Sec.~\ref{specatred}. 
We remind that the validity of the covariance was discussed in Sec.~\ref{valid}.
		
\begin{table}[h]
 \centering
\begin{tabular}{|c|c|c|c|c|c|c|c|} \hline
 DESI: & $z_{\text{eff}}$ & $ n_{b,\text{eff}}[ (\hinvMpc)^3]$ &{$b_1^{\text{ref}}$}&($k^{\text{Tree}}_{\rm max}$, $k^{1L}_{\rm max}$, $k_{\text{NL}}$) $[\hinvMpc]$ &$N^{1L}_{\text{bins}}$&$N^{\text{Tree}}_{\Delta}$& $N^{1L}_{\Delta}$ \\ \hline 
 Bin 1 & 0.84& 8.0$\times 10^{-4}$ & 1.3 & (0.08, 0.18, 0.9) & 37 &17& 115 \\ \hline
 Bin 2 &1.23 & 3.2$\times 10^{-4}$ & 1.5& (0.09, 0.23, 1.3) & 45 &17& 191\\ \hline
\end{tabular}
 \caption{\footnotesize 
 DESI effective survey specifications, calculated according to the formulas in Sec.~\ref{Fishercont} and Tab.~\ref{DESInumbers} in App.~\ref{surv}. $n_{b,\text{eff}}$ is the background galaxy number density entering the derivatives (not the covariance), $N_{\text{bins}}$ is the number of $k$-bins we consider for the power spectrum and $N_{\Delta}$ is the number of triangles we consider for the bispectrum. }
 \label{fig:DESI_base}
 \end{table}	
 \paragraph{Results}
 We present results, including the spectral tilt, spatial curvature, neutrino masses, and {non}-Gaussianity in Fig.~\ref{fig:desi}, using all multipoles. 
 The results for $f_{\text{NL}}$ were obtained with fixed cosmological parameters. 
 Analyzing $f_{\text{NL}}$ in combination with cosmological parameters changes the $f_{\text{NL}}$ constraints by less than {8\%}. 
 For neutrino masses, with the caveats discussed in footnote~\footref{neurtrino}, it seems likely that DESI is already able to detect massive neutrinos at the 2$\sigma$ level. 
 
 \begin{figure}[!htb]
 \centering
 \includegraphics[width=0.49\textwidth]{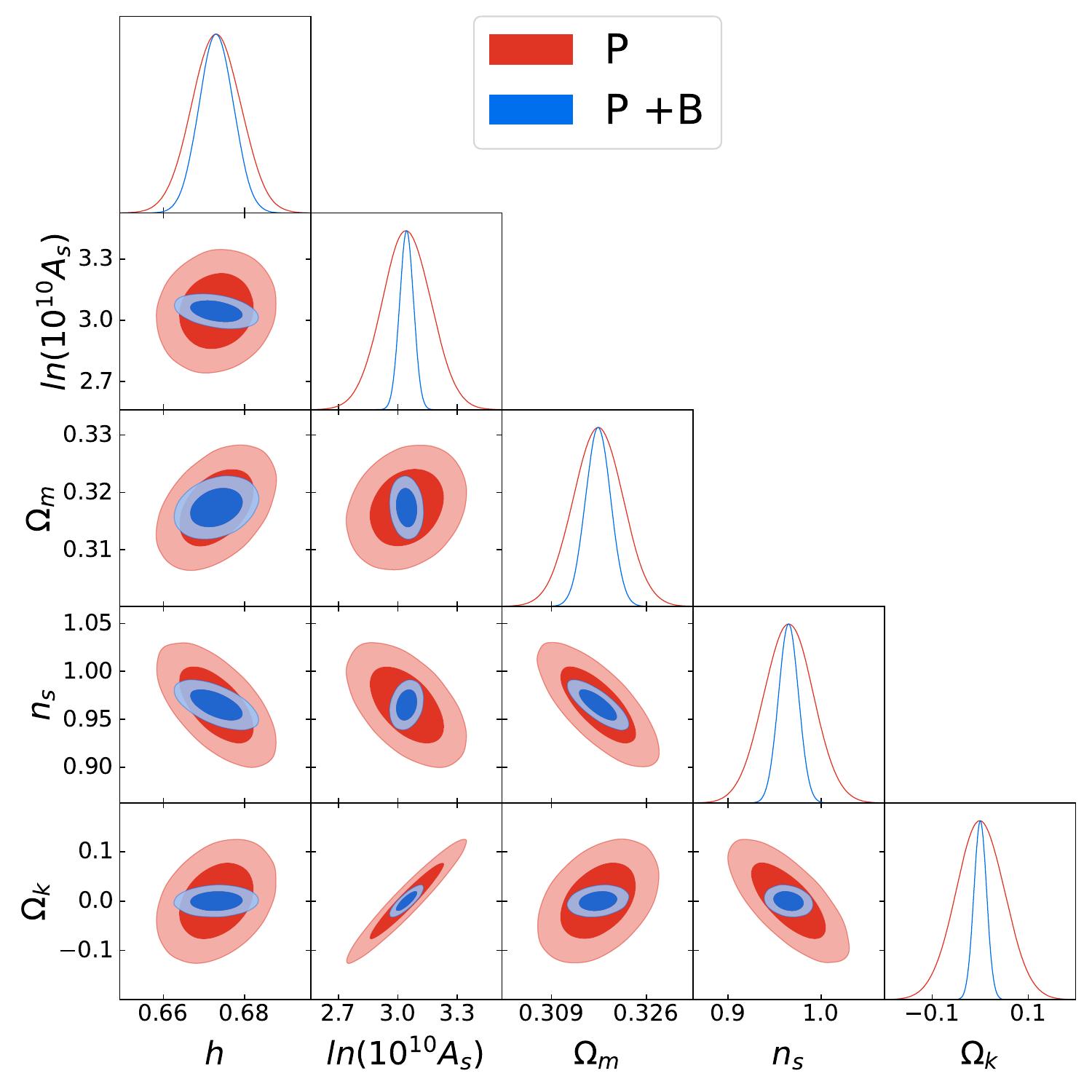} \hspace{-1 em}
 \includegraphics[width=0.5\textwidth]{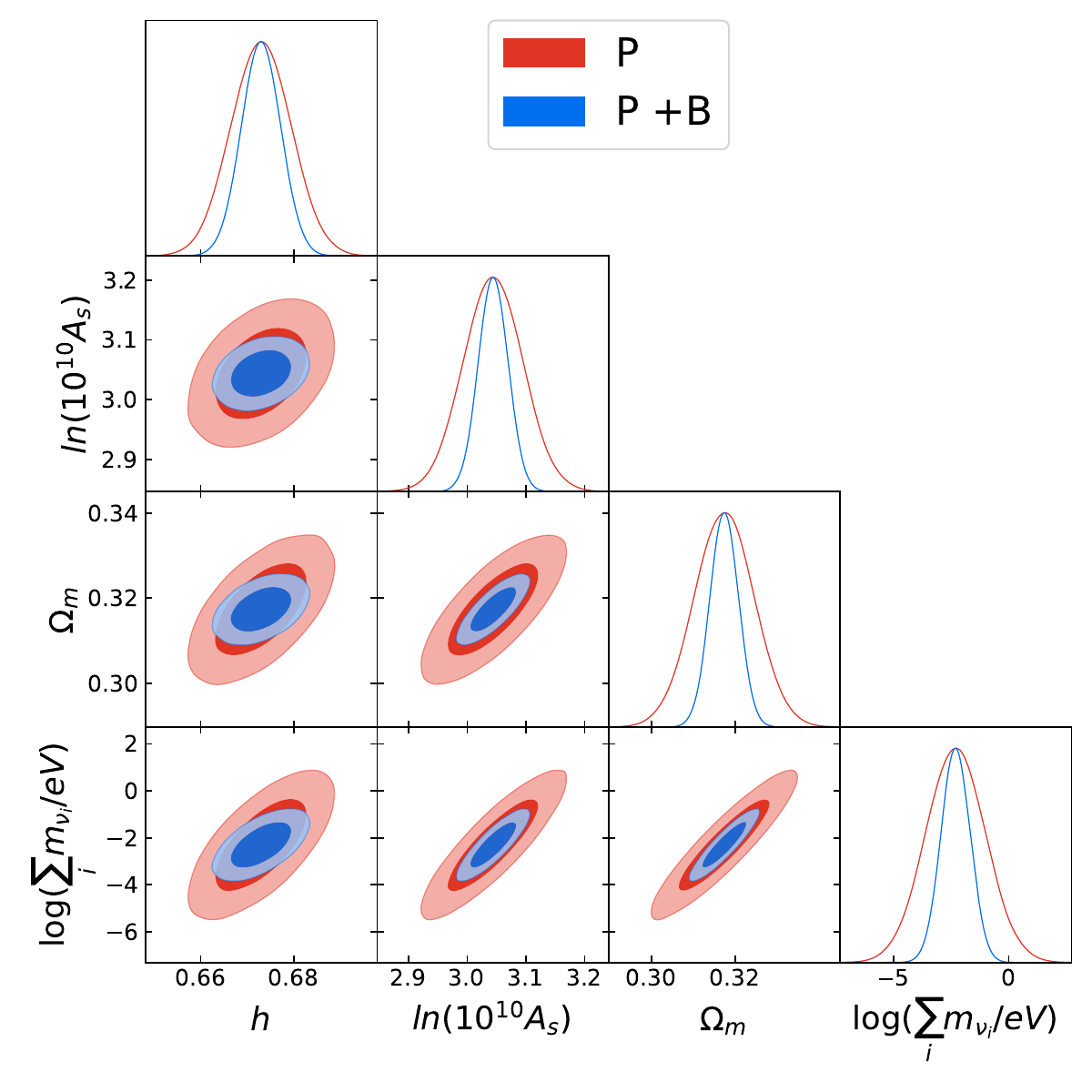}
 \scriptsize
 {{\begin{tabular}{|c|c|c|c|c|c|} \hline
 $\sigma(\cdot)$ & $h$ & $\ln(10^{10}A_s)$ & $\Omega_m$ & $n_s$& $\Omega_k$ \\ \hline
 $P$ & 0.0061 & 0.12 & 0.0045 & 0.027 & 0.051 \\ \hline
 $P$+$B$ & 0.0042 & 0.035 & 0.0023 & 0.011 & 0.013\\ \hline
\end{tabular}}
\hspace{0.1cm}
\begin{tabular}{|c|c|c|c|c|} \hline
 $\sigma(\cdot)$ & $h$ & $\ln(10^{10}A_s)$ & $\Omega_m$ &$\log m_{\nu}^{\text{tot.}} ({}^{\sigma^+}_{\sigma^-}) $ \\ \hline
 $P$ & 0.0065 & 0.051 & 0.0072 & 1.3 (${}^{+0.26}_{-0.072}$) \\ \hline
 $P$+$B$ & 0.0042 & 0.025 & 0.0034 & 0.63 (${}^{+0.087}_{-0.047}$)\\ \hline
\end{tabular} \\

 \vspace{1 em}
 \begin{tabular}{|c|c|c|c|} \hline
 $\sigma(\cdot)$ & $f_{\text{NL}}^{\text{loc.}}$&$f_{\text{NL}}^{\text{eq.}}$&$f_{\text{NL}}^{\text{orth.}}$ \\ \hline
 $P$+$B$ & 3.5 & 114 & 30 \\ \hline
\end{tabular}
}
 \caption{\footnotesize 
 Triangle plots and errors from Fisher forecasts for DESI including the spectral tilt and spatial curvature (left) and massive neutrinos (right) and Non-Gaussianity (bottom). {In the table we also report the upper and lower bounds of the $68\%$ confidence interval for the sum of massive neutrinos, i.e $\mathbb{P}\left[\left(\sum_i m_{\nu_i}- \sum_i m_{\nu_i}^{\text{ref}}\right) \in (\sigma^-,\sigma^+)\right] =0.68$.} We use all power spectrum and bispectrum multipoles at one loop order for the above results and use the analytical covariance without cross-correlations.}
 \label{fig:desi}
 \end{figure}	
 
 \paragraph{Impact of shot noise and biases}\label{sec:DESIadd}
 Similar to Sec.~\ref{bossadd}, it is interesting to investigate constraints with the ``galaxy-formation prior" (g.p.) putting stronger priors on EFT parameters, and look at the theoretical limits of fixed biases and zero shot noise for DESI. 
 As shown in Fig.~\ref{fig:DESI_shot}, the g.p. mostly affects $f_{\text{NL}}^{\text{eq.}}$ and $f_{\text{NL}}^{\text{orth.}}$. However, in both the zero shot noise~\footnote{We note that due to still large shot noise, the 2-loop analysis for DESI does not much improve the results, which we verified with the same method as mentioned in footnote \footref{2looparg}.} and fixed bias limits, we observe improvements of roughly a factor of 2-3 in $\ln(10^{10}A_{s}), \Omega_{m}$ and $n_{s}$, while $h$ improves less significantly in either of these limits. 
 These results are consistent with those obtained for BOSS in Sec.~\ref{bossadd} since, as we noted, the biases have larger degeneracies with $\ln(10^{10}A_{s})$, $n_{s}$, and non-Gaussianities. We find that the effect of shot noise accounts for approximately {50\%} of the constraints on $f_{\text{NL}}$. 
 Interestingly, fixing the bias parameters has a striking effect on non-Gaussianities, particularly for $f_{\text{NL}}^{\text{eq.}}$, for which we would obtain a six-fold reduction
in the error bars. {In combination with the results from the g.p., this strongly motivates the need for tighter priors and therefore better measurements of biases when performing the analysis of DESI in the near future. {In order to {further} improve on this aspect,} we present in Sec.~\ref{pertprior} the $f_{\text{NL}}$ constraints forecasted with the EFT-motivated perturbativity prior.

 \begin{figure}[h]
 \centering
 \includegraphics[width=0.5\textwidth]{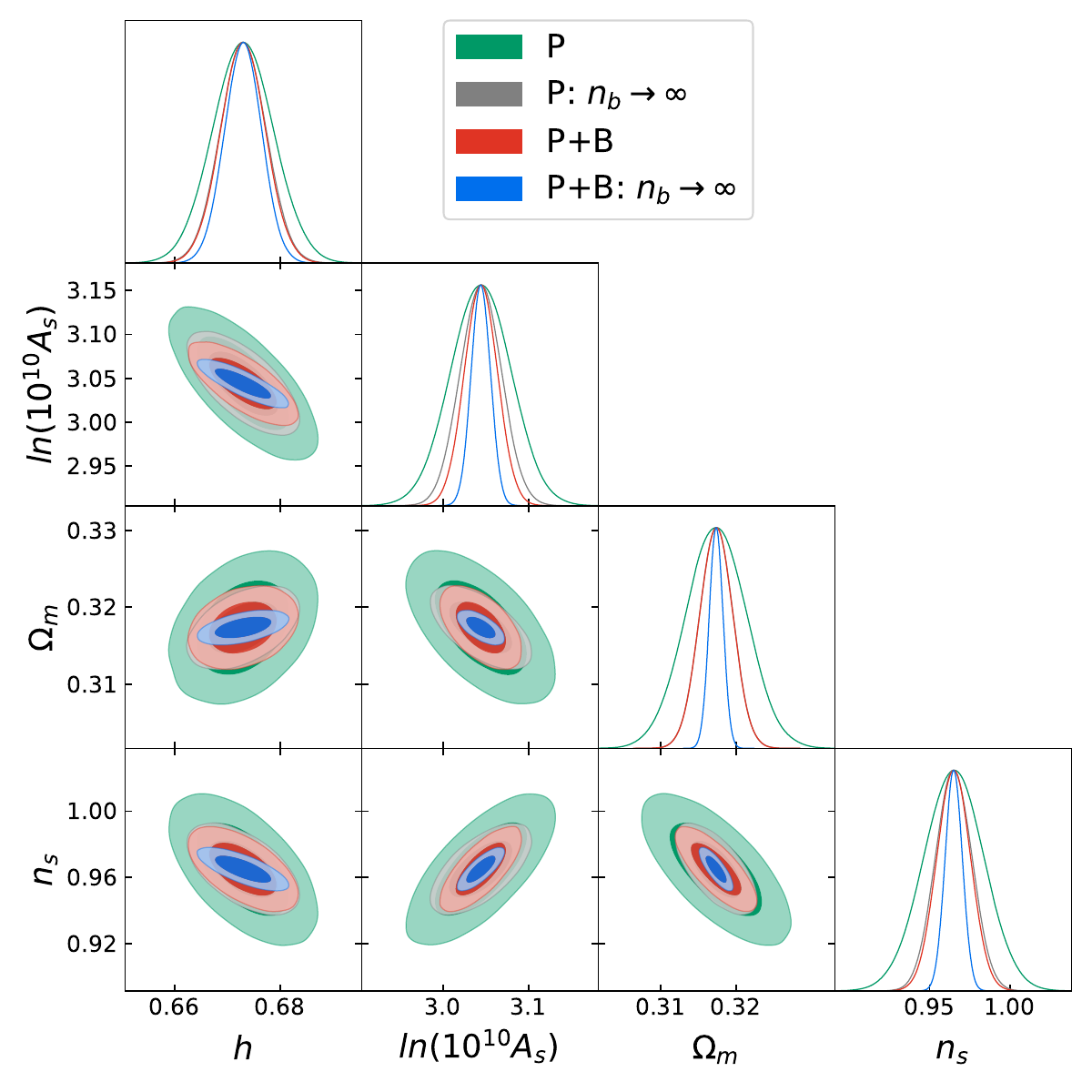} \hspace{-1em}
 \includegraphics[width=0.5\textwidth]{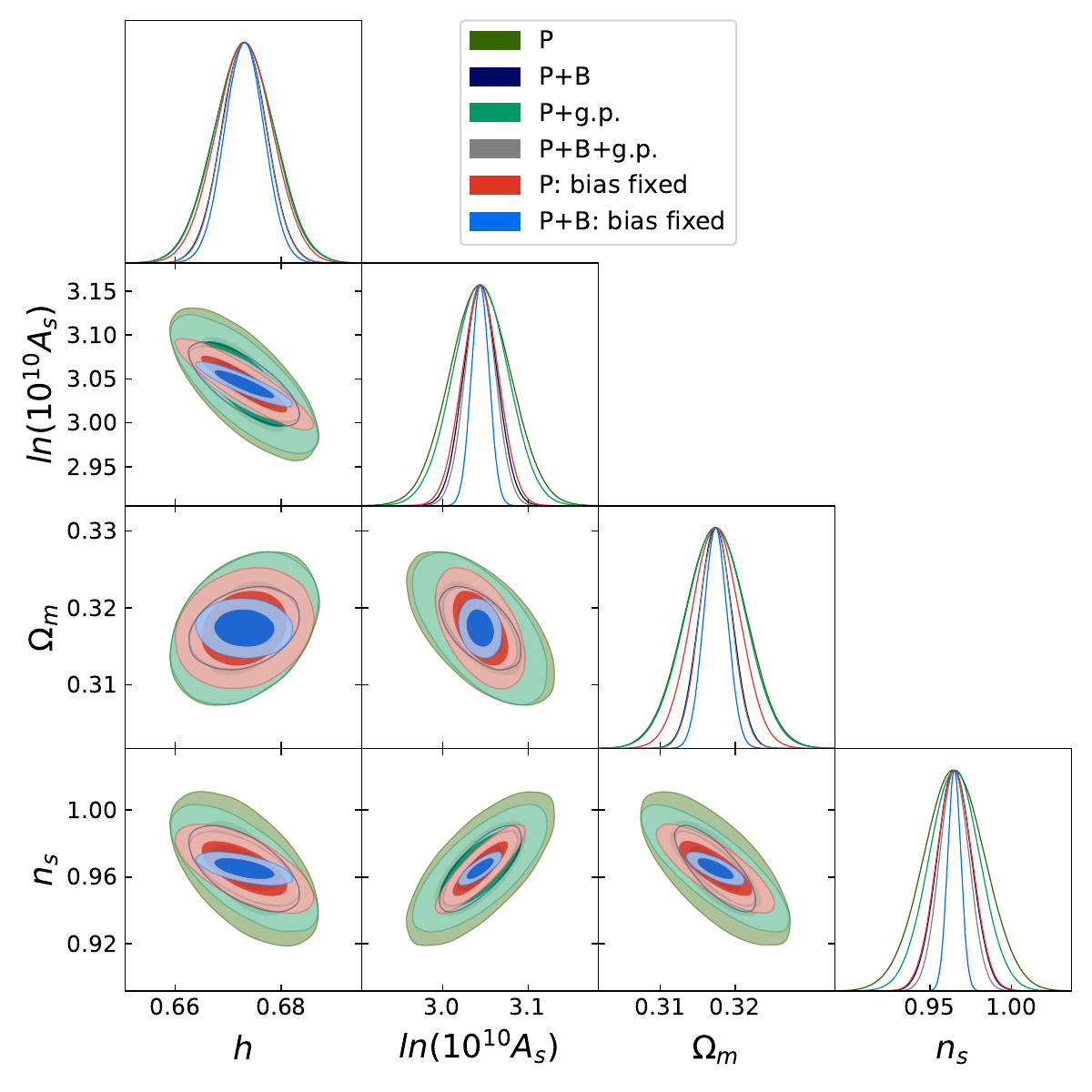}

 \scriptsize
 {
\begin{tabular}{|c|c|c|c|c|c|c|c|} \hline
 $\sigma(\cdot)$ & $h$ & $\ln(10^{10}A_s)$ & $\Omega_m$& $n_s$ &$f_{\text{NL}}^{\text{loc.}}$&$f_{\text{NL}}^{\text{eq.}}$&$f_{\text{NL}}^{\text{orth.}}$ \\ \hline
 $P$ & 0.0057 & 0.036 & 0.0041 & 0.019 & -&-&- \\
 $P+ g.p.: $ & 0.0056 & 0.032 & 0.0040 & 0.016 & -&-&- \\ 
 $P: $ bias fixed & 0.0053& 0.021 & 0.0032 & 0.011 & -&-&- \\ 
 $P: n_b \rightarrow \infty$ & 0.0044 & 0.024 & 0.0022 & 0.011 & -&-&- \\ \hline
 $P$+$B$ & 0.0042 & 0.020 & 0.0022 & 0.010 & 3.5 & 114 & 30 \\
 $P+B+ g.p.: $ & 0.0042 & 0.018 & 0.0022 & 0.009 & 3.4&83&23 \\ 
 $P+B:$ bias fixed &0.0037& 0.010 & 0.0016 & 0.004 & 2.0&21&11 \\ 
 $P+B: n_b \rightarrow \infty$ & 0.0035 & 0.011 & 0.0009 & 0.005& 1.7&67&17 \\ \hline
 \end{tabular}
 }
\caption{\footnotesize 
 Triangle plots and errors from several different Fisher forecasts for DESI. We compare base results to results obtained without shot noise (left) and with biases fixed {or with a ``galaxy-formation prior" (g.p.)} (right). In the table, we also show the impact of including higher multipoles on the power spectrum and bispectrum and also see the impact on $f_{\text{NL}}$. For the constraints on $f_{\text{NL}}$, we fix the other cosmological parameters. }
\label{fig:DESI_shot}
 \end{figure}

	\subsection{MegaMapper}\label{MMores}
	For MegaMapper, we base our Fisher forecasts on the two scenarios mentioned in \cite{Schlegel:2022vrv} (there called ``idealized" and ``fiducial"), which we call the optimistic (MMo) and pessimistic (MMp) scenarios. {These two scenarios are in turn based on the specifications presented in Tab. 1 (MMo) and Tab. 2 (MMp) of \cite{Ferraro:2019uce}. It is important to note that these specifications are preliminary and may differ from the final specifications. We caution that our results are based on these preliminary specifications and may need to be revised as more information becomes available.}
	We find that the {constraints predicted by the }two scenarios differ by {$30-40\%$}. Given the similarity of the results in these two situations, we present here the results in the optimistic scenario, leaving the pessimistic scenario in App.~\ref{MMpbase}. Thus, the {numerical values} that we will use in this section were derived from Tab. 1 of~\cite{Ferraro:2019uce} and methods from Sec.~\ref{Fishercont}. They are given in Tab.~\ref{fig:MMo_base}. 
	
\begin{table}[h]
 \centering
 {
\begin{tabular}{|c|c|c|c|c|c|c|c|} \hline
 MMo: & $z_{\text{eff}}$ & $ n_{b,\text{eff}}[ (\hinvMpc)^3]$ & {$b_1^{\text{ref}}$}&($k^{\text{Tree}}_{\rm max}$, $k^{1L}_{\rm max}$, $k_{\text{NL}}$) $[\hinvMpc]$ &$N^{1L}_{\text{bins}}$&$N^{\text{Tree}}_{\Delta}$& $N^{1L}_{\Delta}$ \\ \hline 
 Bin 1 & 2.4& 1.8$\times 10^{-3}$ & 3.1 & (0.14, 0.36, 3.2) & 73 &62& 696 \\ \hline
 Bin 2 &4.3 & 1.1$\times 10^{-4}$ & 6.3& (0.28, 0.76, 10.1) & 153 &294& 5491\\ \hline
\end{tabular}
}
 \caption{\footnotesize 
 MegaMapper effective survey specifications, calculated according to the formulas in Sec.~\ref{Fishercont} and Tab.~\ref{MMonumbers} in App.~\ref{surv}. $n_{b,\text{eff}}$ is the background galaxy number density entering the derivatives (not the covariance), $N_{\text{bins}}$ is the number of $k$-bins we consider for the power spectrum at 1-loop and $N_{\Delta}$ is the number of triangles we consider for the bispectrum at 1-loop.} \label{fig:MMo_base}
 \end{table}

	As in the DESI forecast, we shift the rest of the biases parameters according to the method described in Sec.~\ref{specatred}. Furthermore, we again use $k_{\text{min}}=0.001\hinvMpc $ for the power spectrum and $k_{\text{min}}=0.02 \hinvMpc $ for the bispectrum, as well as $\Delta k =0.005\hinvMpc $ for the power spectrum and $\Delta k =0.02 \hinvMpc $ for the bispectrum. { Again, to reduce binning effects, we evaluate on $k_{\text{eff}}$. }The results for $f_{\text{NL}}$ were again obtained with fixed cosmological parameters. {Analyzing $f_{\text{NL}}$ in combination with cosmological parameters changes the $f_{\text{NL}}$ constraints by less than 3\%. {Finally, just like for the DESI forecasts, we use the analytical covariance from \eqn{cov} and \eqn{sumcov}, following the discussion in Sec.~\ref{specatred} and its precision discussed in Sec.~\ref{valid}.}

	\paragraph{{Results}}\label{sec:mmdisc}
	
	{We present base results for MegaMapper in a similar format to the previous sections in Fig.~\ref{fig:mm}. We see that the bispectrum contains significant constraining power.} As mentioned in Sec.~\ref{validdiagcov}, we expect that the constraints presented here will be an overestimate as we are neglecting cross-correlations. Nevertheless, the impact of the bispectrum at higher $k_{\rm max}$ becomes relatively more important, and therefore continues to be a very important tool for future data analyses. 
	
 {In particular, shown in in Fig.~\ref{fig:mm}, the inclusion of the bispectrum allows for very tight constraints on neutrino masses. Even with the caveats discussed in footnote~\footref{neurtrino}, neutrino mass detection with MegaMapper seems very likely.} 
	
 \begin{figure}[!htb]
 \centering
 \includegraphics[width=0.49\textwidth]{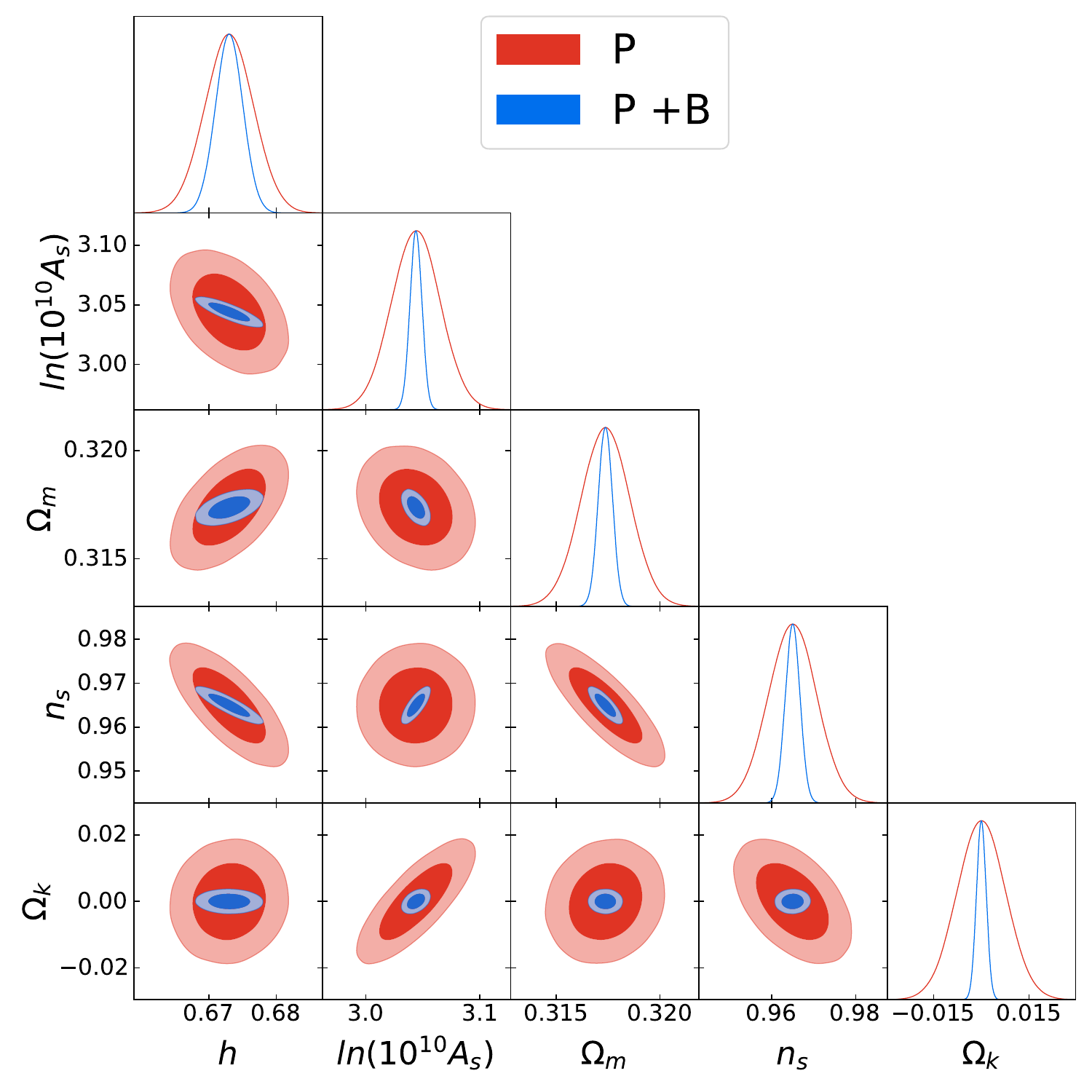} \hspace{-1 em}
 \includegraphics[width=0.5\textwidth]{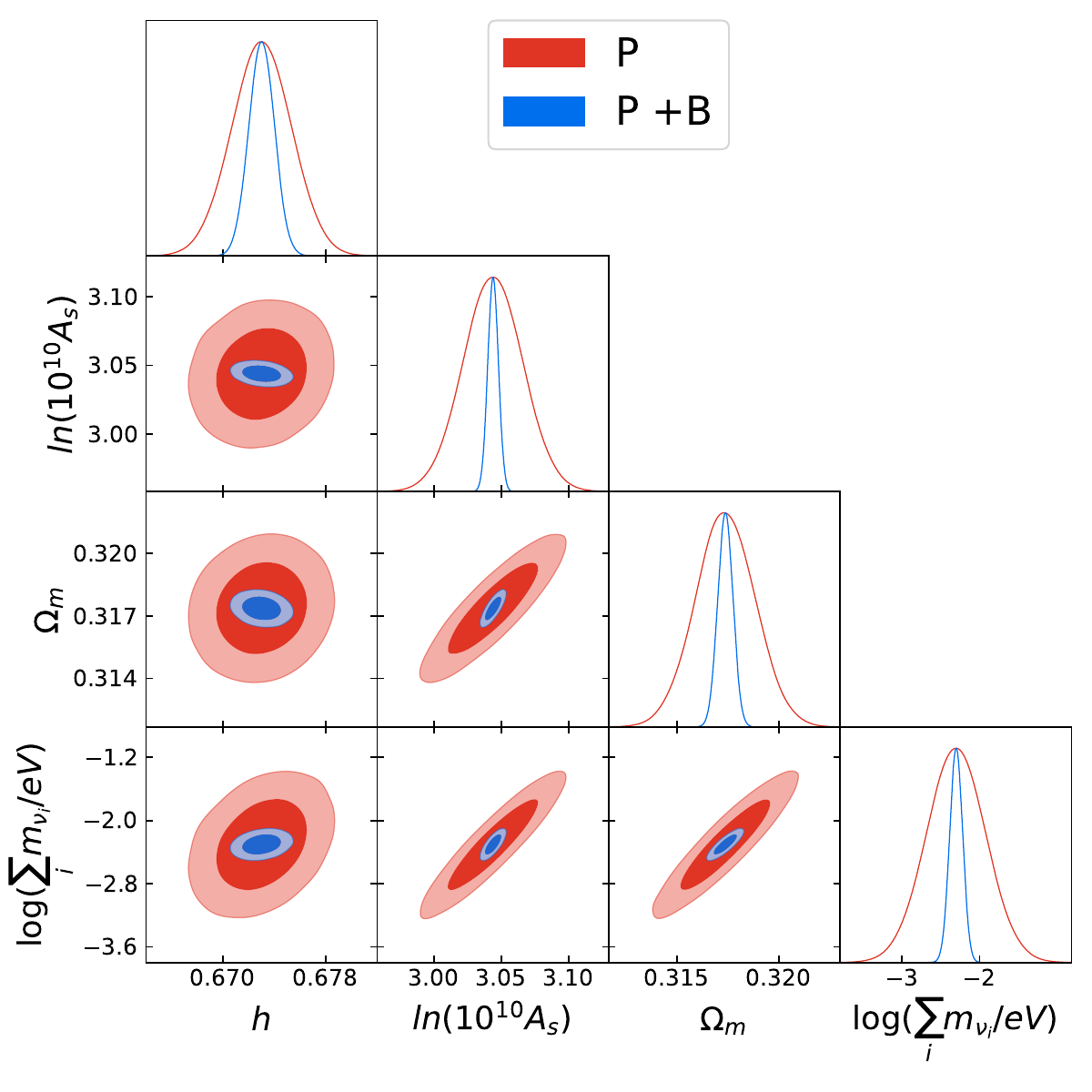}
 \scriptsize
 
 {
{\begin{tabular}{|c|c|c|c|c|c|} \hline
 $\sigma(\cdot)$ & $h$ & $\ln(10^{10}A_s)$ & $\Omega_m$ & $n_s$& $\Omega_k$ \\ \hline
 $P$ & 0.0036 & 0.021 & 0.0012 & 0.006 & 0.0076 \\ \hline
 $P+B$ & 0.0021 & 0.0052 & 0.0003 & 0.002 & 0.0015 \\ \hline
\end{tabular}}
{\begin{tabular}{|c|c|c|c|c|} \hline
 $\sigma(\cdot)$ & $h$ & $\ln(10^{10}A_s)$ & $\Omega_m$ &$\log m_{\nu}^{\text{tot.}} ({}^{\sigma^+}_{\sigma^-}) $ \\ \hline
 $P$ & 0.0023 & 0.022 & 0.0015 & 0.38 (${}^{+0.046}_{-0.031}$) \\ \hline
 $P$+$B$ & 0.0010 & 0.0039 & 0.0004 & 0.083 (${}^{ +0.009}_{-0.008}$) \\ \hline
\end{tabular}}
 \\
 \vspace{1em}
 {
 \begin{tabular}{|c|c|c|c|} \hline
 $\sigma(\cdot)$ & $f_{\text{NL}}^{\text{loc.}}$&$f_{\text{NL}}^{\text{eq.}}$&$f_{\text{NL}}^{\text{orth.}}$ \\ \hline
 $P$+$B$ & 0.27 & 17.7 & 4.6 \\ \hline
\end{tabular}}
}
 \caption{\footnotesize 
 Triangle plots and errors from Fisher forecasts for MegaMapper including the spectral tilt and spatial curvature (left) and massive neutrinos (right) and non-Gaussianity (bottom). 
 {In the table we also report the upper and lower bounds of the $68\%$ confidence interval for the sum of massive neutrinos, i.e $\mathbb{P}\left[\left(\sum_i m_{\nu_i}- \sum_i m_{\nu_i}^{\text{ref}}\right) \in (\sigma^-,\sigma^+)\right] =0.68$. } We use all power spectrum and bispectrum multipoles for the above results and use the analytical covariance without cross-correlations.}
 \label{fig:mm}
 \end{figure}

 \paragraph{Impact of shot noise and biases}
 Given the long timeline until results will be available for MegaMapper, and target selection is yet to happen, we will discuss some aspects that might improve results {as was discussed for DESI in Sec.~\ref{sec:DESIadd}}. In particular, while the perturbative reach is far greater at higher redshifts, as can be seen from Tab.~\ref{fig:MMo_spec}, the shot noise, especially for the higher redshift bin, is extremely large~\footnote{{This also means that the 2-loop analysis for MegaMapper just marginally improves on this results at $<20\%$ error bar reduction, which we verified with the same method as mentioned in footnote \footref{2looparg}.} }. {We, therefore, present the limiting case of zero shot noise to better understand the possible gain achievable by reducing the currently estimated shot noise. {Equally motivated by the long timeline of MegaMapper, we present results with stronger bias priors, anticipating the better understanding of galaxy formation until the data release.} Along with the zero shot noise {and ``galaxy-formation prior" }results, we also present the impact of fixing biases in Fig.~\ref{fig:MMo_spec}.}

 \begin{figure}[!htb]
 \centering
 \includegraphics[width=0.5\textwidth]{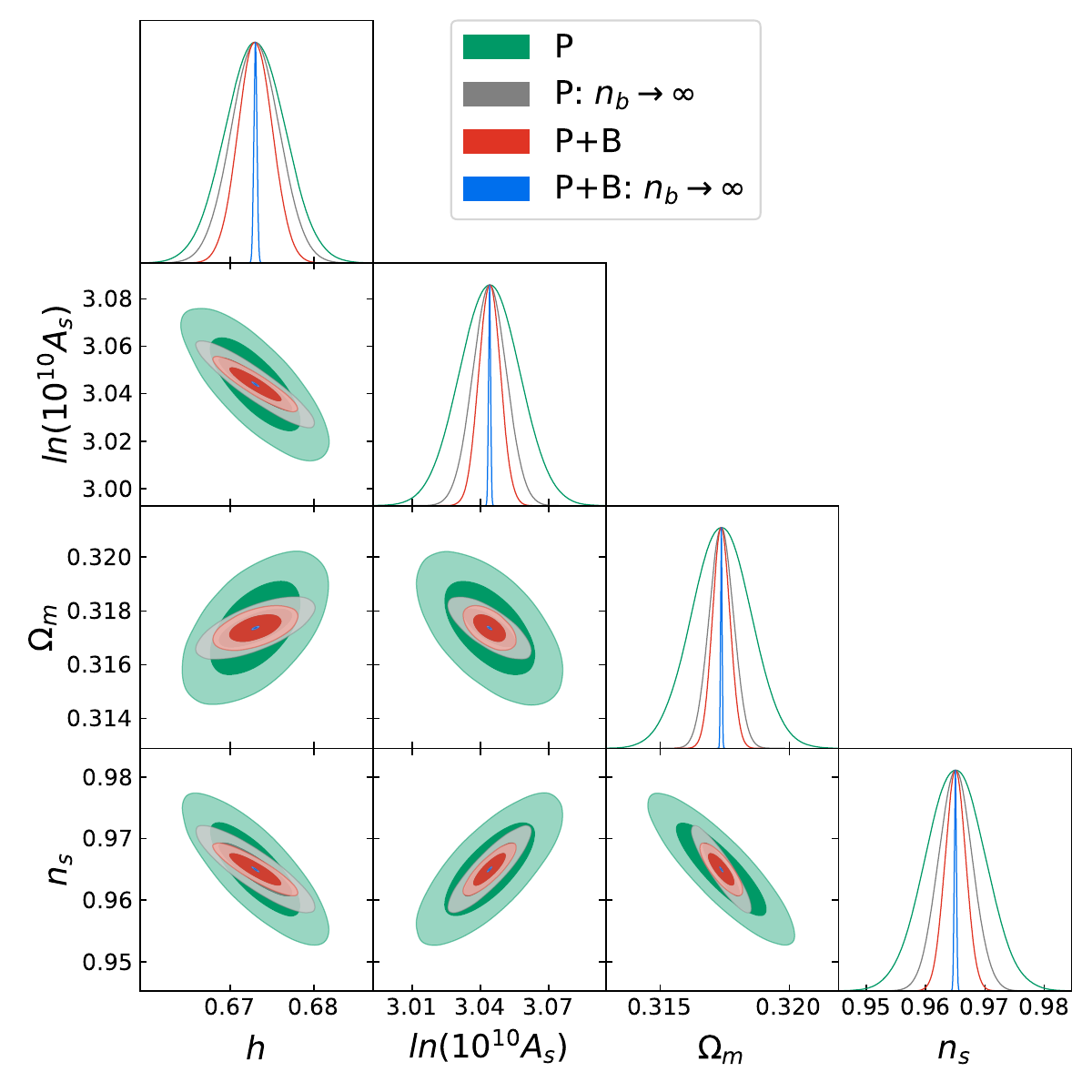} \hspace{-1em}
 \includegraphics[width=0.5\textwidth]{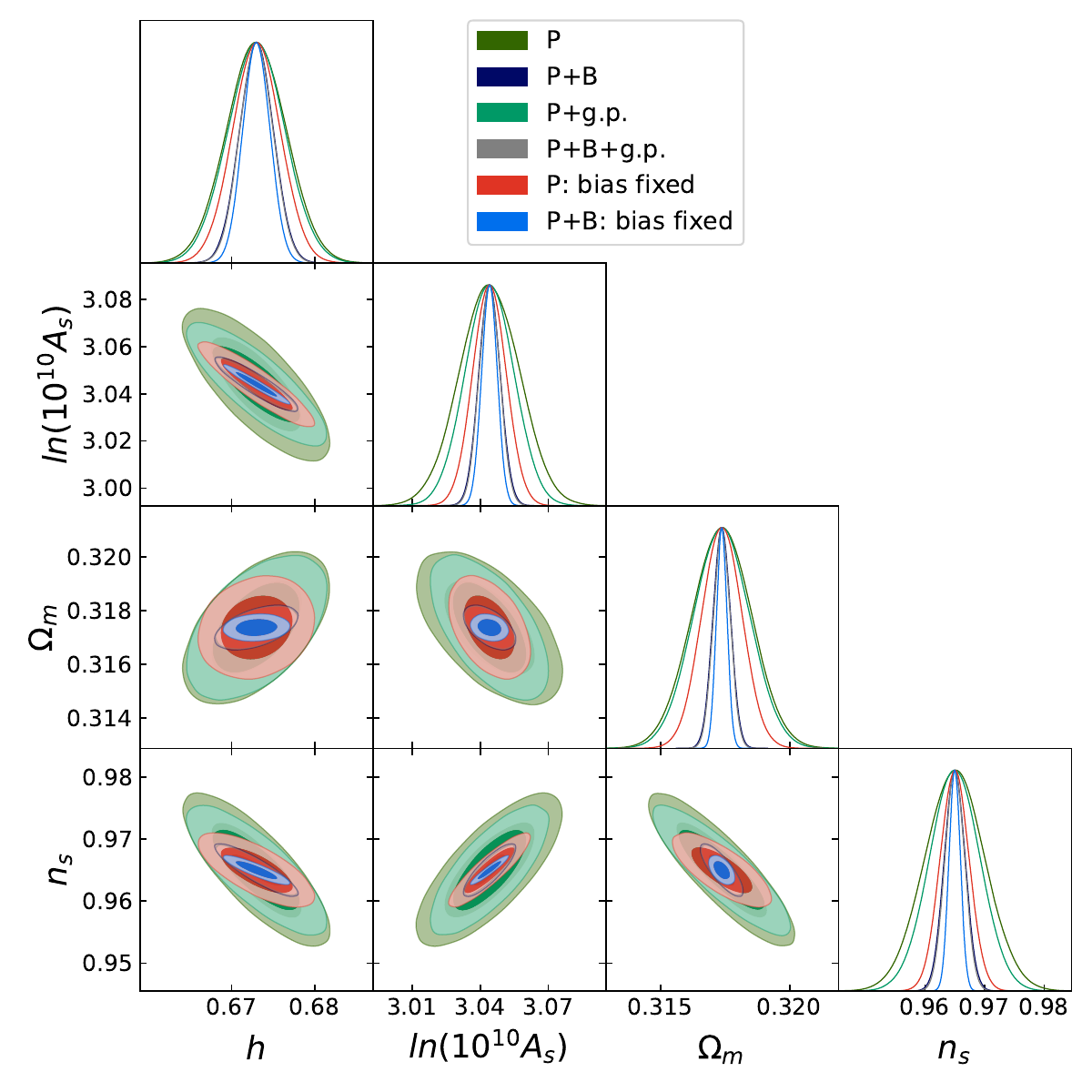}

 \scriptsize
 {
{\begin{tabular}{|c|c|c|c|c|c|c|c|} \hline
 $\sigma(\cdot)$ & $h$ & $\ln(10^{10}A_s)$ & $\Omega_m$& $n_s$ &$f_{\text{NL}}^{\text{loc.}}$&$f_{\text{NL}}^{\text{eq.}}$&$f_{\text{NL}}^{\text{orth.}}$ \\ \hline
 $P$ & 0.0036 & 0.013 & 0.0012 & 0.005 & -&-&- \\
 $P+ g.p.: $ & 0.0034 & 0.011 & 0.0011 & 0.0043 & -&-&- \\ 
 $P: $ bias fixed & 0.0029&0.0074& 0.00079& 0.0024 & -&-&- \\ 
 $P: n_b \rightarrow \infty$ & 0.0029 & 0.0074 & 0.00047 & 0.0029 & -&-&- \\ \hline
 $P$+$B$ & 0.0021 & 0.0047 & 0.00034 & 0.0017 & 0.27 & 18 & 4.6 \\
 $P+B+ g.p.: $ & 0.0020 & 0.0045 & 0.00033 & 0.016 & 0.26&13&3.6 \\ 
 $P+B:$ bias fixed &0.0016& 0.0034 & 0.00021 & 0.0010 & 0.17&3.6&1.7 \\ 
 $P+B: n_b \rightarrow \infty$ & 0.00019 & 0.00045 & 0.000029 & 0.00018 & 0.11&5.4&1.5 \\ \hline
 \end{tabular}}
 }
\caption{\footnotesize 
 Triangle plots and errors from several different Fisher forecasts for MegaMapper. We compare base results to results obtained without shot noise (left) and with biases fixed {or with a ``galaxy-formation prior" (g.p.)} (right). In the table, we also show the impact of including higher multipoles on the power spectrum and bispectrum and also see the impact on $f_{\text{NL}}$. For the constraints on $f_{\text{NL}}$, we fix the other cosmological parameters.}
\label{fig:MMo_spec}
 \end{figure}	
 
 We see that {stronger bias priors mostly have an effect on $f_{\text{NL}}^{\text{eq.}}$ and $f_{\text{NL}}^{\text{orth.}}$. Going further and} fixing the biases we would again, roughly, reduce the error bar by a factor 2, with again the exception of $f_{\text{NL}}^{\text{eq.}}$ where the dependence is much stronger. This again motivates the perturbativity prior we discuss in Sec.~\ref{pertprior}. {This is very similar to the case of BOSS and DESI shown in Secs.~\ref{bossadd} and \ref{sec:DESIadd}. Thus, the relative gain of {putting the ``galaxy-formation prior" or }fixing the biases is very similar among the three surveys we consider.}
 
 {However, shot noise affects the three surveys very differently. In particular, for MegaMapper, shot noise is quite significant for some cosmological parameters. Especially for the base parameters, we can see from the table in Fig.~\ref{fig:MMo_spec}, that reduction of shot noise for MegaMapper can lead to a $\sim$10-fold error bar reduction. Instead for non-Gaussianity parameters, while shot noise still seems to be in an important factor, it is comparably less significant. In particular, the effect of setting shot noise to zero is similar to fixing the biases when analyzing $f_{\text{NL}}$.}

\section{Further constraining $f_{\text{NL}}$ with a perturbativity prior}\label{pertprior}

As we have seen in Sec.~\ref{results}, in particular Figs.~\ref{fig:BOSS_shot}, \ref{fig:DESI_shot} and \ref{fig:MMo_spec}, fixing the biases leads to stronger constraints on the primordial parameters $\ln(10^{10} A_s)$ and $n_s$ and to vast improvements on some $f_{\text{NL}}$ parameters. We will see in this section that some non-Gaussianity parameters are greatly affected by EFT parameter constraints. In particular, small improvements on the constraints on the bias parameters can lead to {significant} improvements on $f_{\text{NL}}^{\text{eq.}}$ and $f_{\text{NL}}^{\text{orth.}}$. We are thus motivated to place stronger (and physically justifiable) priors on the nuisance parameters in order to further constrain single-field inflation. 

As mentioned in Sec.~\ref{specatred}, we put independent priors on the EFT parameters, restricting their individual size. This is motivated by the fact that the EFTofLSS predicts these parameters to be of order one. However, given that the MCMC explores the full parameter space in a random walk, the final loop contribution can be $\sqrt{n}$ larger than the truth, where $n$ is the number of EFT parameters. We aim to address the issue that such parameter configurations are unphysical yet can still fit the data well. This happens because, at {intermediate and low $k$'s, where each term is not too small, even a too-large loop is comparably small with respect to the data error that scales like $k^{-3/2}$. Therefore, only at large $k$ where the data error is sufficiently small, would parameter configurations exhibiting $\sqrt{n}$ enhancements to the loop be ruled out. However, there exist parameter configurations exhibiting $\sqrt{n}$ larger contributions at lower $k$'s that cancel out at large $k$, making the loop appear to have the correct size at those scales. Therefore, a loop contribution that is too large at {low} $k$ can still fit the data well, but would go unnoticed, even though it would clearly be unphysical. {Through scaling relations, this then translates to an overestimate of the expected higher loop contribution.} {This argument is shown {for the estimated 2-loop contribution} in Fig.~\ref{fig:argument}}. 
 \begin{figure}[!htb]
 \centering
 \includegraphics[width=0.8\textwidth]{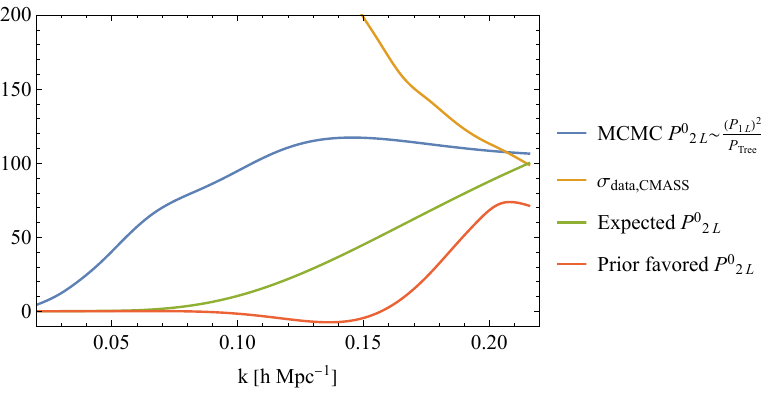}
 \caption{\footnotesize {Plot showing various {two-loop monopole power spectra, $P_{2L}^0$ } against the {CMASS} data error, $\sigma_{ \text{data,CMASS}}$~(orange). {As an example of a typical MCMC{, the BOSS CMASS $P_{2L}^0$ was estimated using the relation $P_{2L} \sim \frac{P^{2}_{1L}}{P^{\rm Tree}}$, and} is shown in blue}}.
The expected $P^0_{2L}$ size (green) has been computed using $P_{2L} = \frac{(P^{C}_{1L})^2}{P^{\rm Tree}}$
from \eqn{finalcorr}, and an example of a curve that stays within this
prior, representing a $P^0_{2L}$ that would
be allowed by the perturbativity prior, is shown in red, purely as an
example to be contrasted with the blue curve.}
\label{fig:argument}
 \end{figure}	

We, therefore, propose an additional prior, which we call ``perturbativity prior", on the size of the loop contributions, {aimed at being effective} in the intermediate and low $k$ regions where the $\sqrt{n}$ enhancements are not sufficiently restricted by the data analysis. To formulate this ``perturbativity prior'', we use the fact that, as we sample the different bias configurations, the two-loop contribution can have a maximal size, and therefore a maximal signal-to-noise. 
This maximal signal-to-noise was what defined the $\kmax$ in \eqn{findkmaxpre}. 
In this section, we show that by using appropriate scaling relations between the two-loop and one-loop contributions, we can translate this threshold for the two-loop contribution into a prior on the size of the one-loop contribution.

\subsection{Contribution to the Fisher matrix}
We impose a perturbativity prior for the power spectrum and bispectrum respectively, and the procedure is the same in both cases. 
We therefore keep the derivation generic, for the loop of some observable, $X_{1L}$, where $X \in \{P,B\}$. In a later step, we will derive an estimate for the correct size of the loop, denoted by $X_{1L}^C$. As mentioned in the previous section, this estimate will come from a threshold for the signal-to-noise of two-loop contributions, through which we can infer properties about the correct one-loop contributions. The quantity we want to constrain is $X_{1L}$, whereas $X_{1L}^C$ we assume to be estimated before the data analysis. We then impose that on average, $X_{1L}$ is close to $X_{1L}^C$, therefore, we impose a Gaussian prior
\bea\label{masterprior}
\frac{1}{N_{X}}\sum_{k_i}\int_{-1}^1\int_0^{2\pi}\frac{d\mu_i}{2}\frac{d\phi}{2\pi}\frac{X_{1L}(k_i;\hat{z})}{X_{1L}^C(k_i;\hat{z})}\sim\mathcal{N}(1,1),
\eea
where $k_i\in\{k, (k_1,k_2,k_3)\}$, $\mu_i\in\{\mu, \mu_1\}$ and $N_{X}\in\{N_{\text{bins}},N_{\Delta}\}$ for the power spectrum and bispectrum respectively. {We here implement the real space part of the perturbativity prior}~\footnote{We note that the real space perturbativity prior is on its own only restricting the size of the real space correlators. However, given that the size of the full redshift space observables is highly dependent on the real space EFT-parameter values, there is little room left for the full redshift space contribution to be large, if the real space contribution is restricted enough. We therefore expect that the full redshift space prior is highly correlated with the real space one, and, therefore, only do the real space version here. {Adding the redshift space part is straightforward.}}. For the remainder of this section, we, therefore, always refer to real space quantities, indicated by dropping the $\hat{z}$ argument. The real space prior we then impose is given by 
\bea\label{masterprior2}
\frac{1}{N_{X}}\sum_{k_i}\frac{X_{1L}(k_i)}{X_{1L}^C(k_i)}\sim\mathcal{N}(1,1).
\eea
One way to write the prior above is to impose the same, independent, prior for each bin. That is, we impose{ 
\bea\label{eqbin}
X_{1L}(k_i)\sim\mathcal{N}\left( X_{1L}^C(k_i), N_X \left(X_{1L}^C(k_i)\right)^2\right),
\eea}
which implies \eqn{masterprior2}.

In order to use this prior for the Fisher matrix, we assume once again that our reference cosmology is accurate and we assume $X_{1L}^{\text{ref}}\simeq X_{1L}^C$~(\footnote{Given that the best-fit we currently have was not obtained with the use of the perturbativity prior, this approximation is not guaranteed to be justified. However, as we will see in Fig.~\ref{fig:priorMaster}, the perturbativity prior only mildly affects the errors of EFT parameters. Therefore, we assume that the best-fit values are also only slightly modified, still making the reference values we use here, a good approximation. {This issue will disappear once a data analysis with this additional prior is performed.}}). Then, for fixed $k_i$, similarly to \eqn{simpexplanation}, we can Taylor expand to get~\footnote{{ We would not need to do this Taylor expansion in an actual data analysis.}} 
\bea\label{priorfin}
-2 \log \text{Prior} &=& \frac{1}{N_X \left(X_{1L}^C\right)^2}\left(X_{1L}(\theta)-X_{1L}^C\right)^2 +r_1\\ \nonumber
&\simeq&(\theta-\theta^{\text{ref}})^T F^{X,\text{pert.}} (\theta-\theta^{\text{ref}})+r_1,
\eea
where {$F^{X,\text{pert.}}_{ij}=\frac{1}{N_X \left(X_{1L}^C\right)^2}\frac{\partial X_{1L}}{\partial \theta_i} \frac{\partial X_{1L}}{\partial \theta_j}$} and $r_1$ is a parameter-independent constant. Therefore, {summing over all bins, }the perturbativity prior that we finally implement in the Fisher forecast is given by
{\bea\label{pcontri}
	F_{ij}^{\text{\text{pert.}}}&=&\frac{1}{N_{\text{bins}}}\sum_k \frac{1}{P_{1L}^C(k)^2}\frac{\partial P_{1L}(k)}{\partial \theta_i}\frac{\partial P_{1L}(k)}{\partial \theta_j}\\ \nonumber
	&&\quad+\frac{1}{N_{\Delta}}\sum_{k_1,k_2,k_3} \frac{1}{B_{1L}^C(k_1,k_2,k_3)^2}\frac{\partial B_{1L}(k_1,k_2,k_3)}{\partial \theta_i}\frac{\partial B_{1L}(k_1,k_2,k_3)}{\partial \theta_j}.
	\eea
{In order to be able to implement this prior, we need to} derive the estimates for $P_{1L}^C$ and $B_{1L}^C$. We derive the threshold for the size of the 1-loop contributions {from limiting the 2-loop signal-to-noise ratio}. This is similar to how we determined the $\kmax$ in Sec.~\ref{specatred}. There, we demanded that the signal-to-noise of the 2-loop {contribution for} any new survey does not exceed its signal-to-noise of BOSS CMASS, where we know it is negligible. In particular, the maximal signal-to-noise that has previously been chosen in the data analysis to determine the $\kmax$ was $\frac{1}{9}$, which is also what we use here. Explicitly,
	\bea\label{rawest}
	&&\int_{0}^{\kmax}\left(\frac{P_{2L}(k)}{\tilde{\sigma}_{P,\text{data}}(k)}\right)^2 dk \simeq\frac{1}{9}, \\ \nonumber
	&&\int_{\nu_B}\left(\frac{B_{2L}(k_1,k_2,k_3)}{\tilde{\sigma}_{B,\text{data}}(k_1,k_2,k_3)}\right)^2 dk_1dk_2dk_3 \simeq\frac{1}{9},
	\eea
	where $\tilde{\sigma}_{X,\text{data}}$ here is defined as in \eqn{tilddef}, using that in the continuum limit we get $\Delta k\rightarrow dk $, and $\nu_B$ is the set of all triangles with maximal wavenumber {smaller than or equal to} $\kmax$. We then get the estimates for the correct {one-}loop contributions, through the {approximate} size relations between two-loop and one-loop, $P_{2L}\sim \frac{{P^{C}_{1L}}^{2}}{P^{NS}_{\text{Tree}}}$ and $B_{2L}\sim \frac{{B^{C}_{1L}}P^{C}_{1L}}{P^{NS}_{\text{Tree}}}$~(\footnote{Note that here the numerators have the usual stochastic contributions, but $P^{NS}_{\text{Tree}}$ in the denominators has no shot noise.}). Finally, in order to perform the integrals in \eqn{rawest}, we assume scaling functions $S^X(k_i)$, defined by $\frac{S^X(k_i) }{S^X(\kmax) }= \frac{X^C_{1L}(k_i)}{X^C_{1L}(\kmax)}$~(\footnote{{One could have normalized this scaling factor to make it unitless, {\it i.e.} $S^X(k_i) \to \frac{S^X(k_i) }{S^X(\kmax) }$ without changing anything in the final formulas}.}), that should approximate the $k$ dependencies of $X^C_{1L}(k_i)$. Plugging in the size relations and scaling approximation into \eqn{rawest}, we can solve for the $X^C_{1L}(k_i)$ to get
\bea\label{finalcorr}
&&P_{1L}^C(k)=S^P(k)\left(9 \int_0^{\kmax}\left(\frac{S^P(k')^2}{P^{NS}_{\text{Tree}}(k')\tilde{\sigma}_{P,\text{data}}(k')}\right)^2 dk'\right)^{-1/4}, \\ \nonumber
&&B_{1L}^C=S^B(k_1,k_2,k_3)\left(9 \int_{\nu_B}\left(\frac{S^B(k_1',k_2',k_3')}{\tilde{\sigma}_{B,\text{data}}(k_1',k_2',k_3')}\frac{1}{3}\left(\frac{P_{1L}^C(k_1')}{P^{NS}_{\text{Tree}}(k_1')}+2p.\right)\right)^2 dk_1'dk_2'dk_3'\right)^{-1/2},
\eea
where we dropped the $k$-dependence for $B_{1L}^C$ to avoid clutter and in the bottom line we symmetrize the $\frac{P_{1L}^C}{P^{NS}_{\text{Tree}}}$.
Note that $P_{1L}^C$ and $B_{1L}^C$ do not depend on the overall size of the scaling estimates $S^X$, {as we are normalizing it at $k_{\rm max}$}. 

There are well established estimates for the behaviour of the power spectrum loop \cite{Carrasco:2013sva} and bispectrum loop \cite{Steele:2021lnz} in a scaling universe. For biased tracers we adapt this to
\bea
S^P_{\text{int}}(k)&=&P_{\text{Tree}}(k)\left(\frac{k}{k_{\text{NL}}}\right)^{3+n(k)}, \\ \nonumber
S^B_{\text{int}}(k_1,k_2,k_3)&=&B_{\text{Tree}}(k_1,k_2,k_3)\left(\left(\frac{k_1}{k_{\text{NL}}}\right)^{3+n(k_1)}+ 2 p.\right),
\eea
which is very similar to the one used in \eqn{scale}, but with $P_{\text{Tree}}$ and $B_{\text{Tree}}$ being the (real space) biased tracers tree-level power spectrum and bispectrum. In order to have both the right IR and UV behaviour of the loop contributions we also include a loop counter term to the estimate. {For the scaling of the counterterms,} we use 
\bea
S^P_c(k)&=&-2 b_1 \beta P_{11}(k)\left(\frac{k}{k_{\text{NL}}}\right)^2, \\ \nonumber
S^B_c(k_1,k_2,k_3)&=&-2 b_1^2 \beta \left(P_{11}(k_1)P_{11}(k_2)\left(\frac{k_3}{k_{\text{NL}}}\right)^{2}+ 2 p.\right),
\eea
where we use the reference value $\beta_{\text{BOSS}}=1$ and rescale with \eqn{biasrescale} for other surveys~\footnote{The {representative} counter terms here correspond to the terms multiplied by $c_1^h$ and $c_3^h$ in \cite{DAmico:2022ukl}.}. Finally, given that $S^P$ and $S^B$ should be upper bounds on the the scaling of the loops, we want to avoid cancelations and ensure positivity. Therefore, the final scalings we implement are
\bea
S^P(k)&=&\max(|S^P_{\text{int}}(k)|,|S^P_{c}(k)|), \\ \nonumber
S^B(k_1,k_2,k_3)&=&\max(|S^B_{\text{int}}(k_1,k_2,k_3)|,|S^B_{c}(k_1,k_2,k_3)|).
\eea
} {A so derived and estimated $P_{2L} = \frac{(P^{C}_{1L})^2}{P^{\rm Tree}}$, with $P^{C}_{1L}$ from \eqn{finalcorr}, is being plotted in
Fig.~\ref{fig:argument}.}
\subsection{Results}\label{pertres}
 {While the perturbativity prior further constrains both cosmological parameters and bias parameters, the largest effect comes from further constraining particular bias parameters. We present this improvement in Fig.~\ref{fig:priorMaster}. Indeed,
there we can see that small improvements on particular EFT parameters~\footnote{We remind the reader that $c_2$ and $c_4$ are the second-order biases that enter the tree-level bispectrum, alongside the linear bias $b_1$. For more details see \cite{DAmico:2022osl,DAmico:2022ukl}.} lead to vast improvements on $f_{\text{NL}}^{\text{eq.}}$. In the tables of Fig.~\ref{fig:priorMaster} we also show the effect on the other types of non-Gaussianity we analyze. Each analysis was performed with fixed cosmological parameters and each type of non-Gaussianity was analyzed separately. For completeness, and to stress the importance of the bispectrum loop, we also compare it with bispectrum tree-level constraints. The survey specifications used for the tree-level analysis are also in Tabs.~\ref{tabBOSS}, \ref{fig:DESI_base}, and \ref{fig:MMo_base}. In the following, we will present results for each survey based on the plots in Fig.~\ref{fig:priorMaster}.
	 
Finally, as we also discussed in Sec.~\ref{results}, base cosmological parameters are less affected by constraints on bias parameters. However, we find that the inclusion of the perturbativity prior can have a relevant effect on them, which we discuss in App.~\ref{ppbase}. {In particular, we find that DESI will constrain curvature to 0.012 and MegaMapper to 0.0012.}
	 
\begin{figure}[htb]
 \centering
 \hspace{-0.3cm}
 \begin{minipage}{0.48\textwidth}
 \centering
 \caption*{BOSS}
 \includegraphics[width=\textwidth]{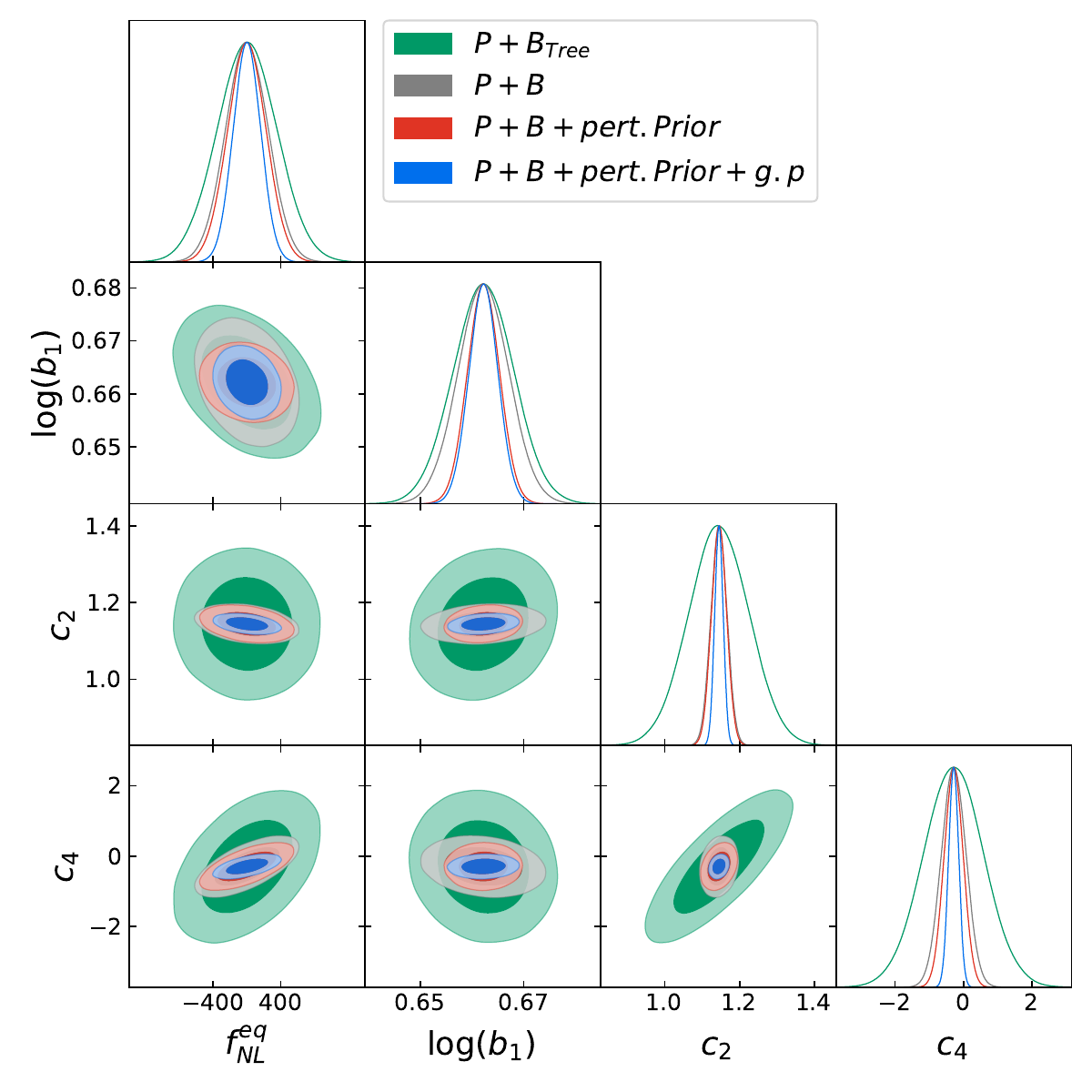}
 \end{minipage}
 \hfill
 \begin{minipage}{0.48\textwidth}
 \centering
 \caption*{DESI}
 \includegraphics[width=\textwidth]{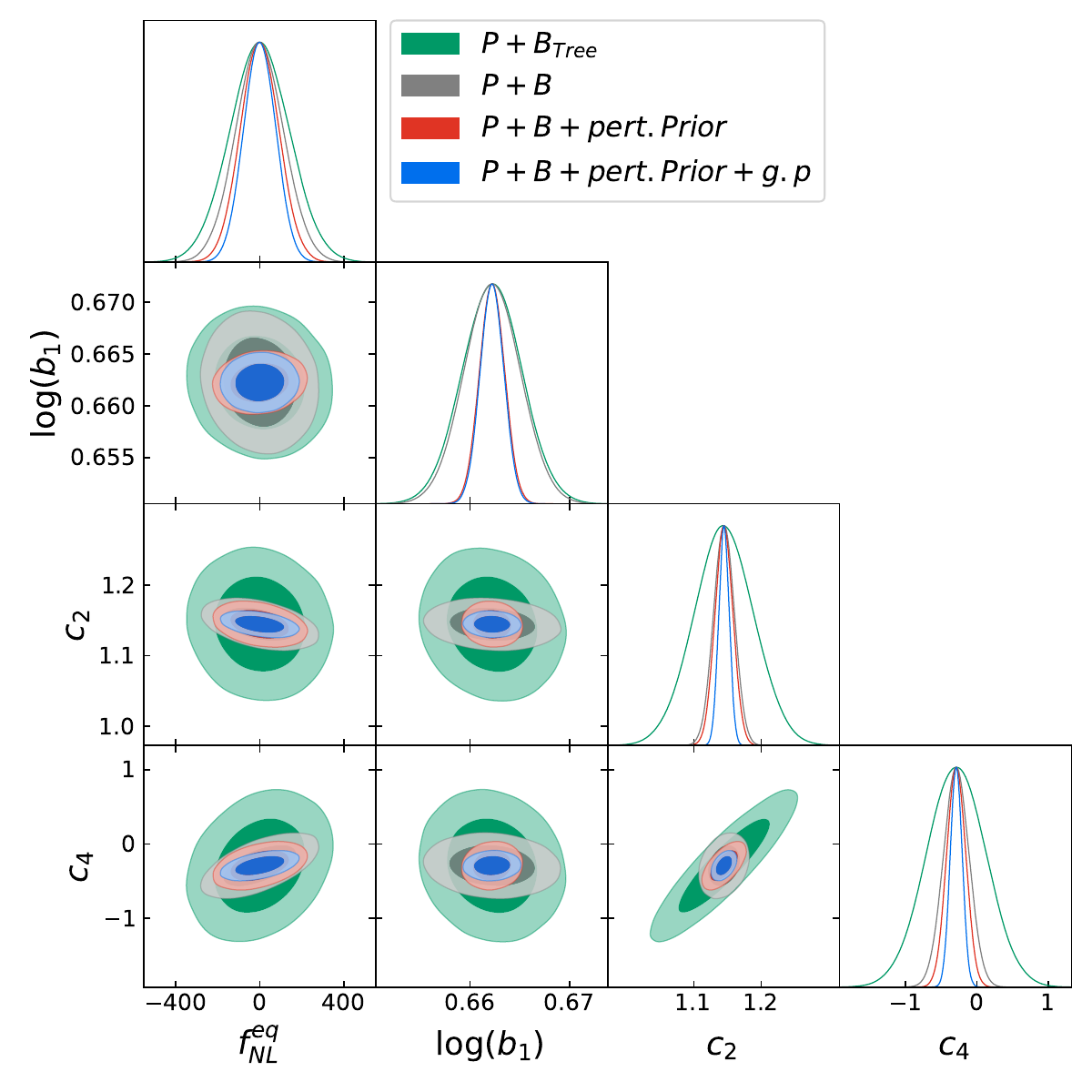}
 \end{minipage}
 \vspace{0.5cm}
 \begin{minipage}{0.47\textwidth}
 \centering
 
 \caption*{MegaMapper}
 \hspace{-0.625cm}
 \includegraphics[width=\textwidth]{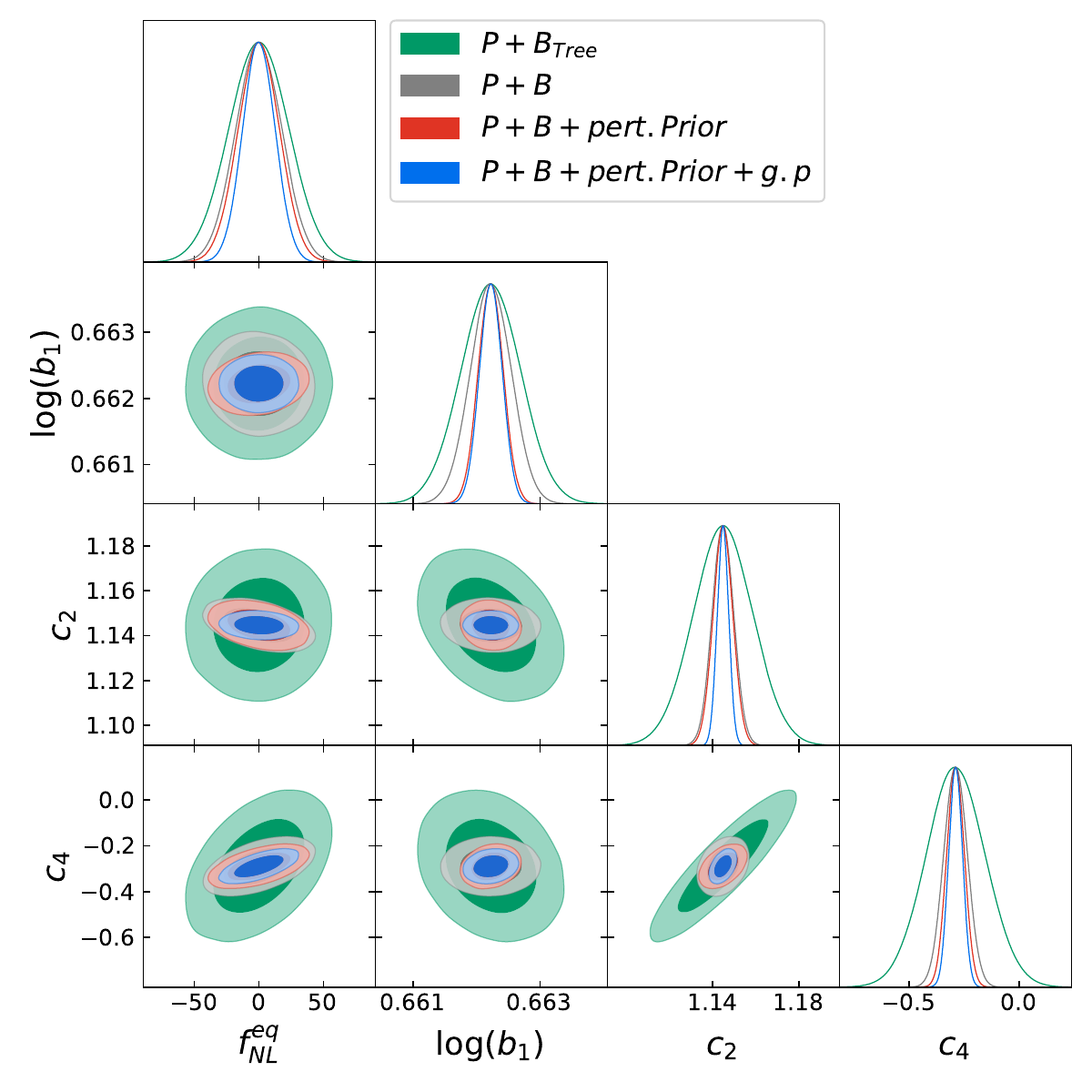}
 \end{minipage}
 \hfill
 \hspace{-2em}
 \begin{minipage}{0.53\textwidth}
 \centering
{
{\fontsize{7.7}{10}\selectfont
{
\begin{tabular}{|c|c|c|c|c|c|c|}
\hline
BOSS: $\sigma(\cdot)$ & $f_{\text{NL}}^{\text{loc.}}$ & $f_{\text{NL}}^{\text{eq.}}$ & $f_{\text{NL}}^{\text{orth.}}$ & $\log(b_1)$ & $c_2$ & $c_4$ \\ 
\hline
$P$+$B_{\text{Tree}}$ & 37 & 357&142 & 0.006 & 0.081 & 0.88 \\ 
\hline
$P$+$B$ & 23 & 253&67 & 0.005 & 0.021 & 0.36 \\
\hline
$P$+$B$+{p.p.} & 17 & 228 &62& 0.003 & 0.020 & 0.28 \\
\hline
$P$+$B$+{p.p.+g.p. }& 15& 164&49 & 0.003 & 0.011 & 0.15 \\
\hline
\end{tabular}}
}
 \vspace{0.5cm}
 
 {\fontsize{7.7}{10}\selectfont
 {
\begin{tabular}{|c|c|c|c|c|c|c|}
\hline
DESI: $\sigma(\cdot)$ & $f_{\text{NL}}^{\text{loc.}}$ & $f_{\text{NL}}^{\text{eq.}}$ & $f_{\text{NL}}^{\text{orth.}}$ & $\log(b_1)$ & $c_2$ & $c_4$ \\ 
\hline
$P$+$B_{\text{Tree}}$ & 3.61 & 142 & 71.5 & 0.003 & 0.04 & 0.4 \\ 
\hline
$P$+$B$ & 3.46 & 114 & 30.2 & 0.003 & 0.02 & 0.2 \\
\hline
$P$+$B${+p.p.} & 3.27 & 91.5 & 27.0 & 0.001 & 0.01 & 0.1 \\
\hline
$P$+$B${+p.p.+g.p.} & 3.19& 77.0 &21.8& 0.001 & 0.008 & 0.08 \\
\hline
\end{tabular}}
}
 \vspace{0.5cm}

{\fontsize{7.7}{10}\selectfont
{\begin{tabular}{|c|c|c|c|c|c|c|}
\hline
MMo: $\sigma(\cdot)$ & $f_{\text{NL}}^{\text{loc.}}$ & $f_{\text{NL}}^{\text{eq.}}$ & $f_{\text{NL}}^{\text{orth.}}$ & $\log(b_1)$ & $c_2$ & $c_4$ \\ 
\hline
$P$+$B_{\text{Tree}}$ & 0.29 & 23.4 & 8.7 & 0.0005 & 0.01 & 0.14 \\ 
\hline
$P$+$B$ & 0.27 & 17.7 & 4.6 & 0.0003 & 0.005 & 0.05 \\
\hline
$P+B$+{p.p.} & 0.26 & 16.0& 4.2 & 0.0002 & 0.005 & 0.04 \\
\hline
$P$+$B$+{p.p.+g.p.} & 0.26 & 12.6 & 3.4& 0.0002 & 0.003 & 0.03 \\
\hline
\end{tabular}}
}}
 \end{minipage}
 \hspace{0.15cm}
 
 \caption{Triangle plots and errors from Fisher forecasts for BOSS (top left), DESI (top right), and MegaMapper (bottom left), for the equilateral type of non-Gaussianity, and leading bias parameters. We also show errors on other non-Gaussianity parameters in the tables. Each analysis was done with cosmological parameters fixed and each non-Gaussianity parameter was analyzed separately. We always include the power spectrum at one loop order with the addition of either the tree-level bispectrum the loop bispectrum or the loop bispectrum with a perturbativity prior (p.p.) also in combination with the ``galaxy-formation prior" (g.p.). We use all power spectrum and bispectrum multipoles in each case and use the analytical covariance without cross-correlations.}
 \label{fig:priorMaster}
\end{figure}

\paragraph{BOSS}
{As shown in the table of Fig.~\ref{fig:priorMaster}, the one-loop bispectrum already significantly improves the constraints on non-Gaussianities by $\sim$ {$30-50\%$ }with respect to the tree-level analysis. {{Additionally} including the perturbativity prior to the one-loop bispectrum yields a further {$\sim$ 24\% reduction in $\sigma$ for $f_{\text{NL}}^{\text{loc.}}$, and $\sim$ 10\% reduction in $\sigma$ for $f_{\text{NL}}^{\text{eq.}}$ and $f_{\text{NL}}^{\text{orth.}}$.} {The ``galaxy-formation prior" would further reduce the error by {$\sim$ {$20-30\%$} for $f_{\text{NL}}^{\text{eq.}}$ and $f_{\text{NL}}^{\text{orth.}}$, and 14\% for $f_{\text{NL}}^{\text{loc.}}$}.} The addition of the loop breaks the degeneracy between $c_2$ and $c_4$~(\footnote{The breaking of this degeneracy due to the loop bispectrum is already present in \cite{DAmico:2022osl}.}), greatly improving constraints on both parameters, which translates to stronger constraints on $f_{\text{NL}}^{\text{eq.}}$. {Furthermore,} the inclusion of the perturbativity prior further breaks this degeneracy and improves upon the constraints on $b_1$, which leads to additional improvements on the $f_{\text{NL}}$ parameters.}
 \paragraph{DESI}\label{sec:DESI_prior} 
As it was seen in BOSS, the inclusion of the loop bispectrum and perturbativity prior breaks the degeneracy between $c_2$ and $c_4$ and tightens the constraints on EFT parameters for DESI as well. However, the resulting effect on $f_{\text{NL}}$ parameters is different. Unlike BOSS, the inclusion of the one-loop bispectrum and perturbativity prior to DESI does not uniformly tighten all $f_{\text{NL}}$ errors. For one, most of the information to constrain $f_{\text{NL}}^{\text{loc.}}$ is contained in the power spectrum and the tree-level bispectrum. {In contrast}, the one-loop bispectrum does improve the constraint of $f_{\text{NL}}^{\text{eq.}}$ and $f_{\text{NL}}^{\text{orth.}}$ by {19\% and 58\%} respectively. The perturbativity prior reduces these errors further by {20\% and 11\% }and the future galaxy formation prior by another {16\% and 19\%}. 

\paragraph{MegaMapper}

MegaMapper results are more similar to DESI than BOSS. $f_{\text{NL}}^{\text{loc.}}$ is mostly constrained through the power spectrum and does not improve much with {the addition of }the {bispectrum }loop or the perturbativity prior. Additionally, as was seen for DESI, the inclusion of
the bispectrum loop leads to a more significant improvement in the constraint on $f_{\text{NL}}^{\text{orth.}}$ compared to $f_{\text{NL}}^{\text{eq.}}$, with {47\% and 24\%} improvements respectively. To place these results into context, note that the tree level results we obtain are {in agreement with} those obtained in \cite{Cabass:2022epm}~\footnote{The results we present here would correspond to their MegaMapper - B results, with fixed cosmological parameters and free bias parameters. The disagreement with our results is {$<10\%$}.}. Note, however, the tree-level $\kmax$ we estimate here is a bit lower, thus we predict slightly less optimistic constraints for the tree-level bispectrum, which makes the addition of the loop more important. 
{The loop again breaks the degeneracy between $c_2$ and $c_4$, and the perturbativity prior enhances the constraints on $c_2$ in particular,} thereby improving the constraints on
$f_{\text{NL}}^{\text{eq.}}$ and $f_{\text{NL}}^{\text{orth.}}$ by $9\%$} and the ``galaxy-formation prior" further reduces the errors by {$21\%$ and $18\%$} respectively.

	\section*{Acknowledgements}
	 We are very grateful for conversations with Guido D'Amico, Eva Silverstein and Pierre Zhang and for providing their results from the BOSS MCMC. Y.D. would like to thank Enrico Pajer, Blake Sherwin, Paul Shellard, James Fergusson, and Daniel Glazer for useful discussions. Y.D. acknowledges support from the STFC. LS is supported by the SNSF grant $200021\_213120$.

	\appendix
	
	\section{Survey details and best-fits}\label{surv}
	Shown below are the exact parameter values that we use in the analyzes for each survey. Regarding cosmological parameters, for all surveys, we use the following best-fit as the reference cosmology~\footnote{All cosmological parameters with the exception of $f_{\text NL}$ and $\sum_i m_{\nu_i}$ are fixed to Planck preferred values \cite{Planck:2018vyg}.} for the cosmological parameters (noted with a ${}^{\text{ref}}$).
	\bea
	&&h^{\text{ref}}=0.673 \, , \qquad
	 \ln(10^{10}A_s)^{\text{ref}}=3.044 \, , \qquad
	 \Omega_m^{\text{ref}}=0.317 \, ,\qquad
	 n_s^{\text{ref}}=0.965 \, , \\ \nonumber 
	 &&\Omega_k^{\text{ref}}=0 \, , \qquad
	 \omega_b^{\text{ref}}=0.022 \, , \qquad	
	{\sum_i m_{\nu_i}^{\text{ref}}=0.1}\,\text{eV} \, ,\qquad 
	 f_{\text{NL}}^{\text{loc.},\text{ref}}=f_{\text{NL}}^{\text{eq.},\text{ref}}=f_{\text{NL}}^{\text{orth.},\text{ref}}=0 \, .
	\eea 
	
	Furthermore, in Tab.~\ref{EFTparams} we give the EFT parameter best-fit that we use, coming from \cite{DAmico:2022osl}. Note, this is what we denote as $\vec{b}_{\text{BOSS}}$ ($\texttt{b1}$ - $ \texttt{Bc14}$) and $\vec{\epsilon}_{\text{BOSS}}$ ($ \texttt{Bd1}$ - $ \texttt{Be12}$) in Sec.~\ref{specatred}. Furthermore, we use the notation as in \texttt{PyBird} \cite{DAmico:2020kxu}. For a conversion to the notation of \cite{DAmico:2022osl}, see App. D.4 of \cite{DAmico:2022osl}. The best-fit values of the 41 EFT parameters are given in Tab.~\ref{EFTparams}.
\begin{table}[h]
\centering
\begin{tabular}{@{}ll|ll|ll|ll@{}}
\toprule
Parameter & Value & Parameter & Value & Parameter & Value & Parameter & Value \\
\midrule
$\texttt{b1}$ & $1.94$ & $\texttt{c2}$ & $1.14$ & $\texttt{b3}$ & $-0.37$ & $\texttt{c4}$ & $-0.29$ \\
$\texttt{b4}$ & $0.13$ & $\texttt{b6}$ & $-0.35$ & $\texttt{b7}$ & $0.22$ & $\texttt{b8}$ & $-0.30$ \\
$\texttt{b9}$ & $0.015$ & $\texttt{b10}$ & $0.043$ & $\texttt{b11}$ & $0.036$ & $\texttt{Bc1}$ & $5.46$ \\
$\texttt{Bc2}$ & $-1.54$ & $\texttt{Bc3}$ & $1.31$ & $\texttt{Bc4}$ & $-0.48$ & $\texttt{Bc5}$ & $0.11$ \\
$\texttt{Bc6}$ & $0.87$ & $\texttt{Bc7}$ & $-0.46$ & $\texttt{Bc8}$ & $0.44$ & $\texttt{Bc9}$ & $-0.42$ \\
$\texttt{Bc10}$ & $-0.65$ & $\texttt{Bc11}$ & $-0.088$ & $\texttt{Bc12}$ & $-0.37$ & $\texttt{Bc13}$ & $-0.16$ \\
$\texttt{Bc14}$ & $-0.20$ & $\texttt{Bd1}$ & $5.4$ & $\texttt{Bd2}$ & $-0.72$ & $\texttt{Bd3}$ & $-0.43$ \\
$\texttt{ce2}$ & $0.55$ & $\texttt{Be1}$ & $1.69$ & $\texttt{Be2}$ & $0.91$ & $\texttt{Be3}$ & $0.074$ \\
$\texttt{Be4}$ & $-0.14$ & $\texttt{Be5}$ & $6.07$ & $\texttt{Be6}$ & $-0.093$ & $\texttt{Be7}$ & $-0.97$ \\
$\texttt{Be8}$ & $0.26$ & $\texttt{Be9}$ & $0.26$ & $\texttt{Be10}$ & $-0.15$ & $\texttt{Be11}$ & $0.43$ \\
$\texttt{Be12}$ & $-0.43$ & & & & & & \\
\bottomrule
\end{tabular}
 \caption{\footnotesize 
Best-fit EFT parameters.} \label{EFTparams}
 \end{table}

Next, we show the survey specifications that were used in each survey. With the methods from Sec.~\ref{Fishercont}, this builds the basis for the numerical values that we use in the forecasts.
\paragraph{BOSS}
	For the BOSS survey, we use the survey specifications as presented in \cite{BOSS:2016wmc,Reid:2015gra,BOSS:2016psr,Font-Ribera:2013rwa}. We display them in the same way they enter our formulas in Tab.~\ref{BOSSnumbers}. 
\begin{table}[h]
 \centering
 \begin{tabular}{|c|c|c|c|c|c|c|c|} \hline
 $z$ & $N_g$ & $V [ (\Mpcinvh)^3]$& $n_b [ (\hinvMpc)^3]$ & $b_1$ \\ \hline 
 0.05 & 7370 & $2.55\times 10^7$ & $2.9 \times 10^{-4}$& 1.48 \\
 0.15 & 47560 & $1.64\times 10^8$ & $2.9 \times 10^{-4}$ & 1.56 \\
 0.25 & 120600 & $4.02\times 10^8$ & $3.0 \times 10^{-4}$ & 1.65 \\
 0.35 & 214016 & $7.04\times 10^8$ & $3.0 \times 10^{-4}$& 1.73 \\
 0.45 & 287040 & $1.04\times 10^9$ & $2.8 \times 10^{-4}$& 1.83 \\
 0.55 & 445740 & $1.38\times 10^9$ & $3.2 \times 10^{-4}$ & 1.92 \\
 0.65 & 206400 & $1.72\times 10^9$ & $1.2 \times 10^{-4}$ & 2.02 \\
 0.75 & 20400 & $2.04\times 10^9$ & $1.0 \times 10^{-5}$ & 2.12\\ \hline
\end{tabular}
 \caption{\footnotesize 
 Survey details for each redshift bin for BOSS. We show the number of mapped galaxies $N_g$, the volume of the redshift bin $V$ as well as the number density $n_b$ and the linear bias $b_1$.} \label{BOSSnumbers}
 \end{table}
\paragraph{DESI}
	As mentioned in the main text, for DESI we focus on the largest sample which is the set of Emission Line Galaxies. The numerical values we use are calculated from table 2.3 in \cite{DESI:2016fyo}. We present the specifications in Tab.~\ref{DESInumbers}. 

	\begin{table}[h]
 \centering

	 \begin{tabular}{|c|c|c|c|c|c|c|c|} \hline
 $z$ & $N_g$ & $V [ (\Mpcinvh)^3]$& $n_b [ (\hinvMpc)^3]$ & $b_1$ \\ \hline 
 0.65 & 432600 & $2.63\times 10^9$ &$1.64 \times 10^{-4}$ & 1.18 \\
 0.75 & $3.18\times 10^6$ & $3.15\times 10^9$ &$1.01 \times 10^{-3}$ & 1.23 \\
 0.85 & $2.70\times 10^6$ & $3.65\times 10^9$ &$7.38 \times 10^{-4}$ & 1.29 \\
 0.95 & $2.93\times 10^6$ & $4.1\times 10^9$ & $7.15 \times 10^{-4}$ & 1.35 \\
 1.05 & $2.02\times 10^6$ & $4.52\times 10^9$ & $4.46 \times 10^{-4}$& 1.41 \\
 1.15 & $1.89\times 10^6$ & $4.89\times 10^9$ & $3.87 \times 10^{-4}$ & 1.47 \\
 1.25 & $1.87\times 10^6$ & $5.22\times 10^9$ &$3.59 \times 10^{-4}$ & 1.53 \\
 1.35 & 732200 & $5.5\times 10^9$ & $1.33 \times 10^{-4}$ & 1.59 \\
 1.45 & 652400 & $5.75\times 10^9$ & $1.13 \times 10^{-4}$ & 1.65 \\
 1.55 & 460600 & $5.97\times 10^9$ & $7.71 \times 10^{-5}$ & 1.72 \\
 1.65 & 176400 & $6.15\times 10^9$ &$2.87 \times 10^{-5}$ & 1.78 \\ \hline
 \end{tabular}
 \caption{\footnotesize 
 Survey details for each redshift bin for DESI. We show the number of mapped galaxies $N_g$, the volume of the redshift bin $V$ as well as the number density $n_b$ and the linear bias $b_1$.} \label{DESInumbers}
 \end{table}

	\paragraph{MegaMapper} Finally, for MegaMapper, the specifications are still to be finalized given the early stage of the experiment compared to BOSS or DESI. We take the numerical values from \cite{Ferraro:2019uce,Schlegel:2022vrv}, where as mentioned in the main text there is an ``idealized'', and a ``fiducial'' scenario. The specifications for the optimistic (or ``idealized") scenario are in Tab.~\ref{MMonumbers}. They are based on Tab. 1 of \cite{Ferraro:2019uce}. For the ``fiducial" or what we call ``pessimistic" scenario, we refer to table 2 in \cite{Ferraro:2019uce}.
	\begin{table}[h]
 \centering

	 \begin{tabular}{|c|c|c|c|c|c|c|c|} \hline
 $z$ & $N_g$ & $V [ (\Mpcinvh)^3]$& $n_b [ (\hinvMpc)^3]$ & $b_1$ \\ \hline 
 2 & $6.75\times 10^7$ & $2.70\times 10^{10}$ &$2.5\times 10^{-3}$ & 2.5 \\
 2.5 & $3.32\times 10^7$ & $2.76\times 10^{10}$ & $1.2\times 10^{-3}$& 3.3 \\
 3 & $1.63\times 10^7$ & $2.72\times 10^{10}$ &$6\times 10^{-4}$ & 4.1 \\
 3.5 & $7.88\times 10^6$ & $2.63\times 10^{10}$ & $3\times 10^{-4}$& 4.9 \\
 4 & $3.76\times 10^6$ & $2.51\times 10^{10}$ & $1.5\times 10^{-4}$ & 5.8 \\
 4.5 & $1.90\times 10^6$ & $2.38\times 10^{10}$ & $8\times 10^{-5}$ & 6.6 \\
 5 & 901730 & $2.25\times 10^{10}$ & $4\times 10^{-5}$ & 7.4 \\ \hline
\end{tabular}
 \caption{\footnotesize 
 Survey details for each redshift bin for MegaMapper (optimistic). We show the number of mapped galaxies $N_g$, the volume of the redshift bin $V$ as well as the number density $n_b$ and the linear bias $b_1$.} \label{MMonumbers}
 \end{table}

	\section{Further analyses}
	Two further analyses, which to us do not carry the same significance as those presented in the main sections, are presented here for completeness. In Sec.~\ref{MMpbase} we present results for the ``pessimistic" scenario for MegaMapper as opposed to the ``optimistic" scenario presented in Sec.~\ref{MMores}. In Sec.~\ref{ppbase} we also present the impact of the perturbativity prior, as discussed in Sec.~\ref{pertprior}, on base cosmological parameters.
 	\subsection{MegaMapper ``pessimistic" results} \label{MMpbase}
	The survey specifications for the ``pessimistic” scenario given in Tab.~\ref{fig:MMp_base} are determined with the methods described in Sec.~\ref{Fishercont}. 	
		\begin{table}[h]
 \centering
\begin{tabular}{|c|c|c|c|c|c|c|c|} \hline
 MMp: & $z_{\text{eff}}$ & $ n_{b,\text{eff}}[ (\hinvMpc)^3]$ & $b_1$&($k^{\text{Tree}}_{\rm max}$, $k^{1L}_{\rm max}$, $k_{\text{NL}}$) $[\hinvMpc]$ &$N^{1L}_{\text{bins}}$&$N^{\text{Tree}}_{\Delta}$& $N^{1L}_{\Delta}$ \\ \hline 
 Bin 1 & 2.1& 8.8$\times 10^{-4}$ & 2.7 & (0.12, 0.31, 2.2) & 62 &43& 428 \\ \hline
 Bin 2 &4.3 & 8.4$\times 10^{-5}$ & 4.0& (0.21, 0.57, 8.2) & 114 &150& 2331\\ \hline
\end{tabular}
 \caption{\footnotesize 
 MegaMapper ``pessimistic" effective survey specifications, calculated according to the formulas in Sec.~\ref{Fishercont}. $n_{b,\text{eff}}$ is the background galaxy number density entering the derivatives (not the covariance), $N_{\text{bins}}$ is the number of $k$-bins we consider for the power spectrum at 1-loop and $N_{\Delta}$ is the number of triangles we consider for the bispectrum at 1-loop.} \label{fig:MMp_base}
 \end{table}
 
 To avoid redundancy with the discussion of the optimistic scenario, we here simply focus on base results for the pessimistic MegaMapper scenario. The discussion on fixing biases, shot noise and the inclusion of the perturbativity prior is comparable to the optimistic case. The only difference are the absolute values, while the relative gains are similar. We present results in Fig.~\ref{fig:mmp}, where we present the same base parameters as in the main section. Comparing to the figure and tables in Fig.~\ref{fig:mm} for almost all parameters we see only a {$30 - 40\%$} difference compared to the optimistic case. The non-Gaussianity scenarios all differ by roughly {$40-45\%$}, independent on whether we use the tree-level bispectrum, the loop, or the perturbativity prior.

 \begin{figure}[!htb]
 \centering
 \includegraphics[width=0.49\textwidth]{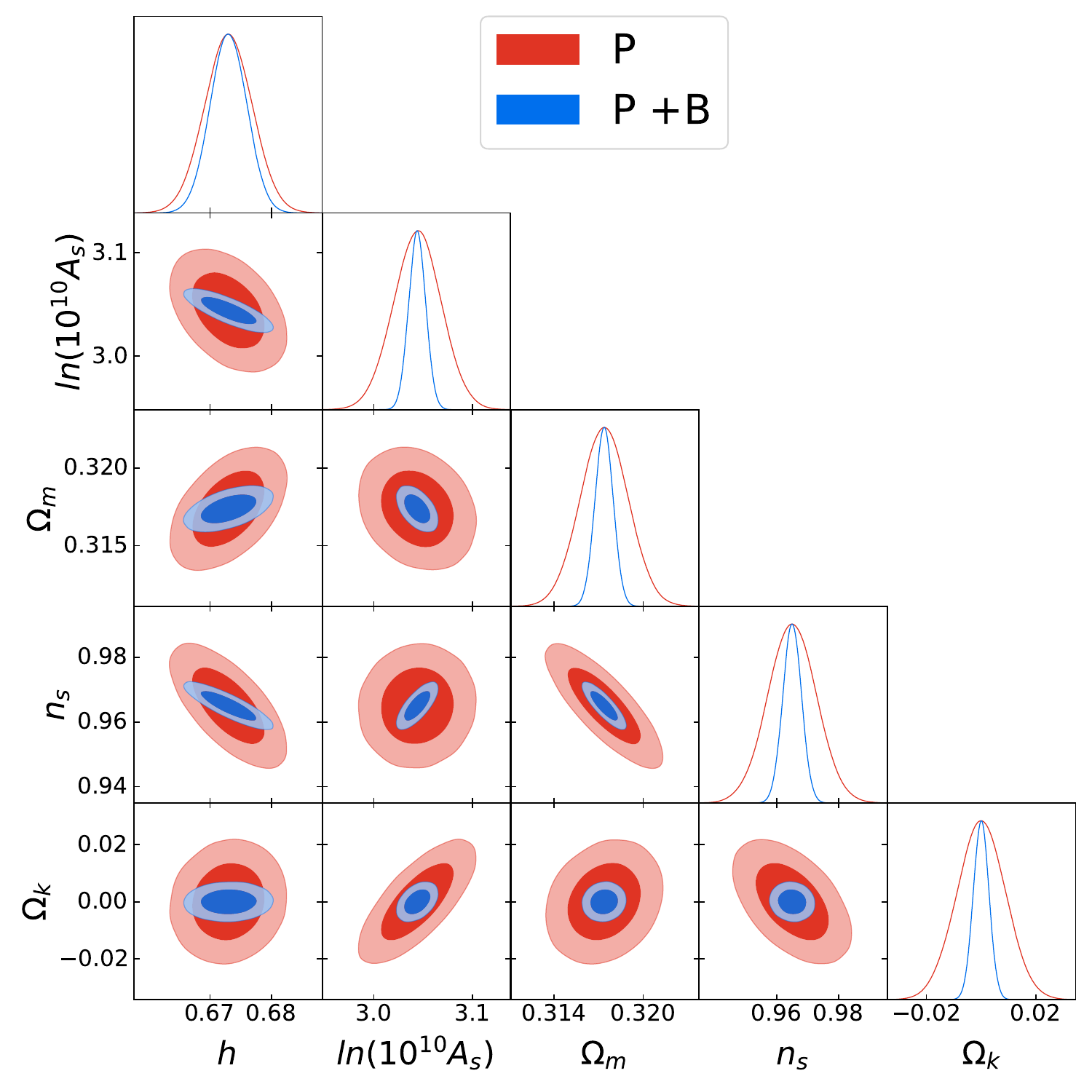} \hspace{-1 em}
 \includegraphics[width=0.5\textwidth]{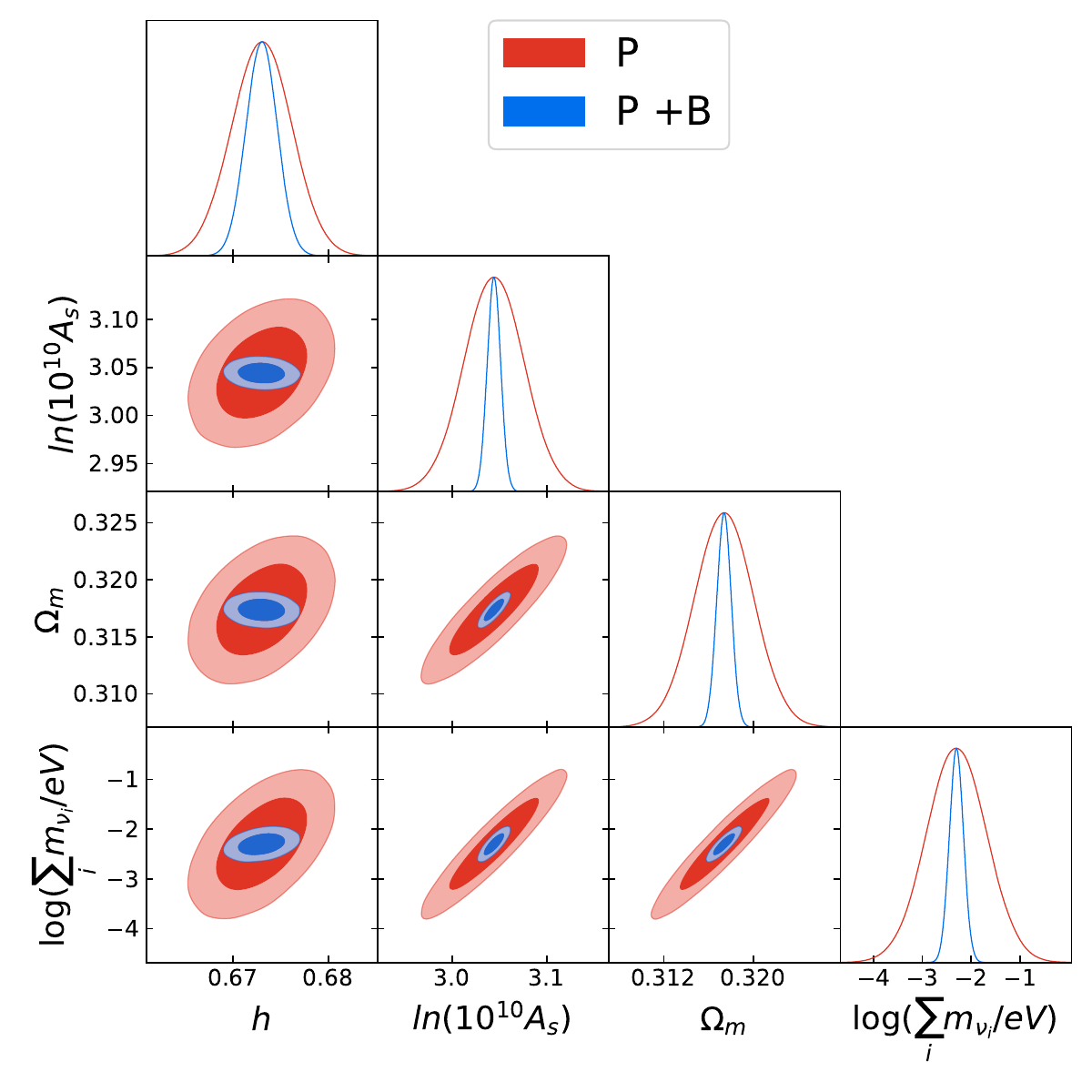}
 \scriptsize

 {{
 \begin{tabular}{|c|c|c|c|c|c|} \hline
 $\sigma(\cdot)$ & $h$ & $\ln(10^{10}A_s)$ & $\Omega_m$ & $n_s$& $\Omega_k$ \\ \hline
 $P$ & 0.0039 & 0.024 & 0.0016 & 0.0078 & 0.009 \\ \hline
 $P$+$B$ & 0.003 & 0.009 & 0.0006 & 0.0030 & 0.003 \\ \hline
\end{tabular}}
\begin{tabular}{|c|c|c|c|c|} \hline
 $\sigma(\cdot)$ & $h$ & $\ln(10^{10}A_s)$ & $\Omega_m$ &$\log m_{\nu}^{\text{tot.}} ({}^{\sigma^+}_{\sigma^-}) $ \\ \hline
 $P$ & 0.0032 & 0.031 & 0.0026 & 0.61(${}^{+0.08}_{-0.05}$) \\ \hline
 $P$+$B$ & 0.0016 & 0.007 & 0.0007 & 0.14 (${}^{+0.015}_{-0.013}$) \\ \hline
\end{tabular}}
 \\
 \vspace{1 em}
 {{
 \begin{tabular}{|c|c|c|c|} \hline
 $\sigma(\cdot)$ & $f_{\text{NL}}^{\text{loc.}}$&$f_{\text{NL}}^{\text{eq.}}$&$f_{\text{NL}}^{\text{orth.}}$ \\ \hline
 $P$+$B_{\text{Tree}}$ & 0.45 & 40.8 & 18.1 \\ \hline
 $P$+$B$ & 0.44 & 32.9 & 8.5 \\ \hline
 $P$+$B$+p.p. & 0.42& 29.5 & 8.0\\ \hline
 $P$+$B$+{p.p.+g.p. }& 0.42& 23.0 & 6.5 \\ \hline
 \end{tabular}}
 }
 \caption{\footnotesize 
 Triangle plots and errors from Fisher forecasts for MegaMapper (pessimistic) including the spectral tilt and spatial curvature (left) and massive neutrinos (right) and Non-Gaussianity (bottom). We use all power spectrum and bispectrum multipoles for the above results, and use the analytical covariance without cross-correlations. { In the table we also report the upper and lower bounds of the $68\%$ confidence interval for the sum of massive neutrinos, i.e $\mathbb{P}\left[\left(\sum_i m_{\nu_i}- \sum_i m_{\nu_i}^{\text{ref}}\right) \in (\sigma^-,\sigma^+)\right] =0.68$. }For non-Gaussianity, we also present results with only the inclusion of the tree-level bispectrum and with the inclusion of a perturbativity prior (p.p) and the ``galaxy-formation prior" (g.p.).}
 \label{fig:mmp}
 \end{figure}

 \subsection{Perturbativity prior effect on base cosmological parameters} \label{ppbase}
 
 We discuss here the effect of the perturbativity prior{, also in combination with the galaxy formation prior, }on base cosmological parameters. { Results for all surveys are shown in Fig.~\ref{fig:priorMastercosmo}. We see that the most notable effect is on $\ln(10^{10}A_s)$, $n_s$ and $\Omega_k$, and a smaller effect on the other parameters. Furthermore, we found almost no improvement on constraints for neutrino masses. }

\begin{figure}[htb]
 \centering
 \hspace{-0.3cm}
 \begin{minipage}{0.48\textwidth}
 \centering
 \caption*{BOSS}
 \includegraphics[width=\textwidth]{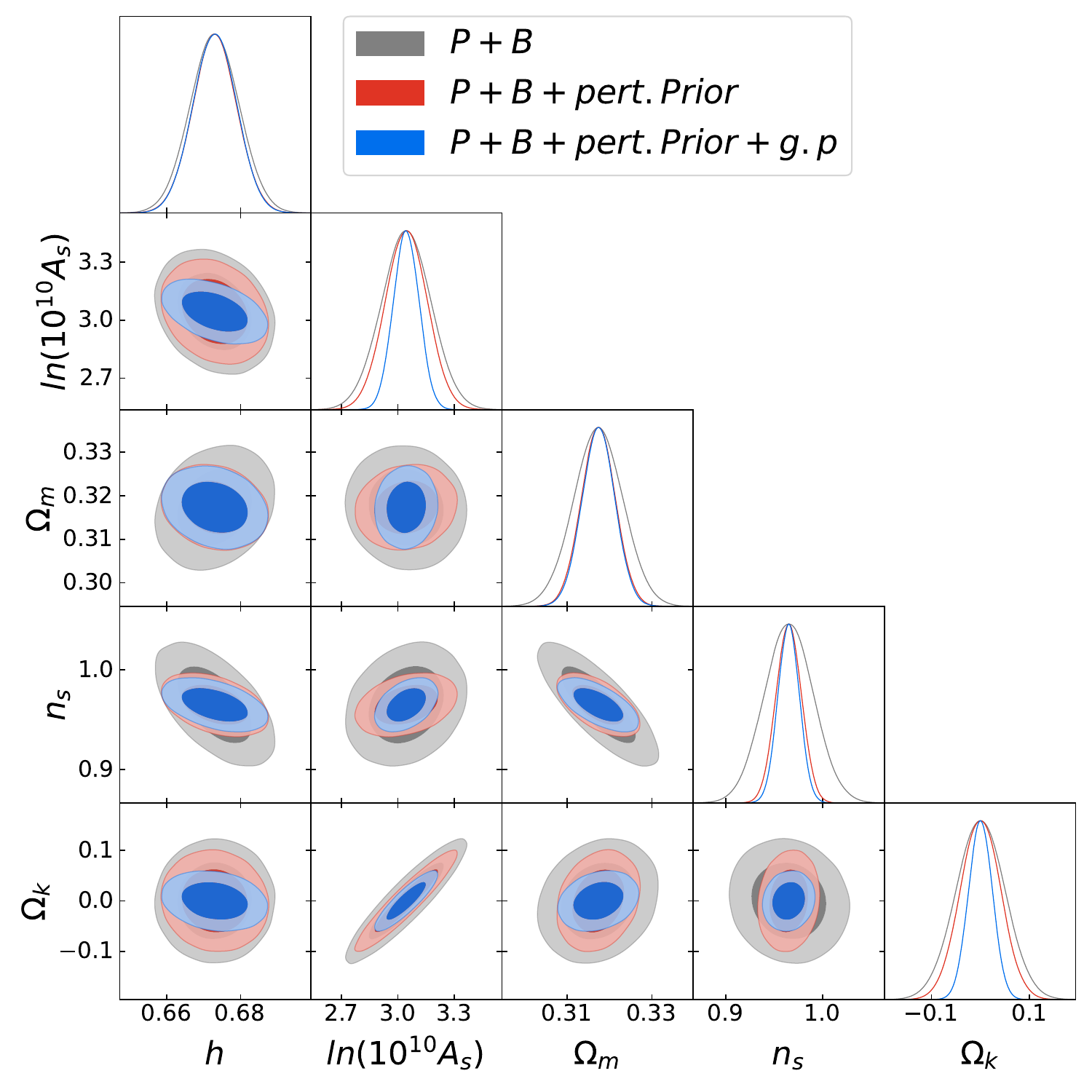}
 \end{minipage}
 \hfill
 \begin{minipage}{0.48\textwidth}
 \centering
 \caption*{DESI}
 \includegraphics[width=\textwidth]{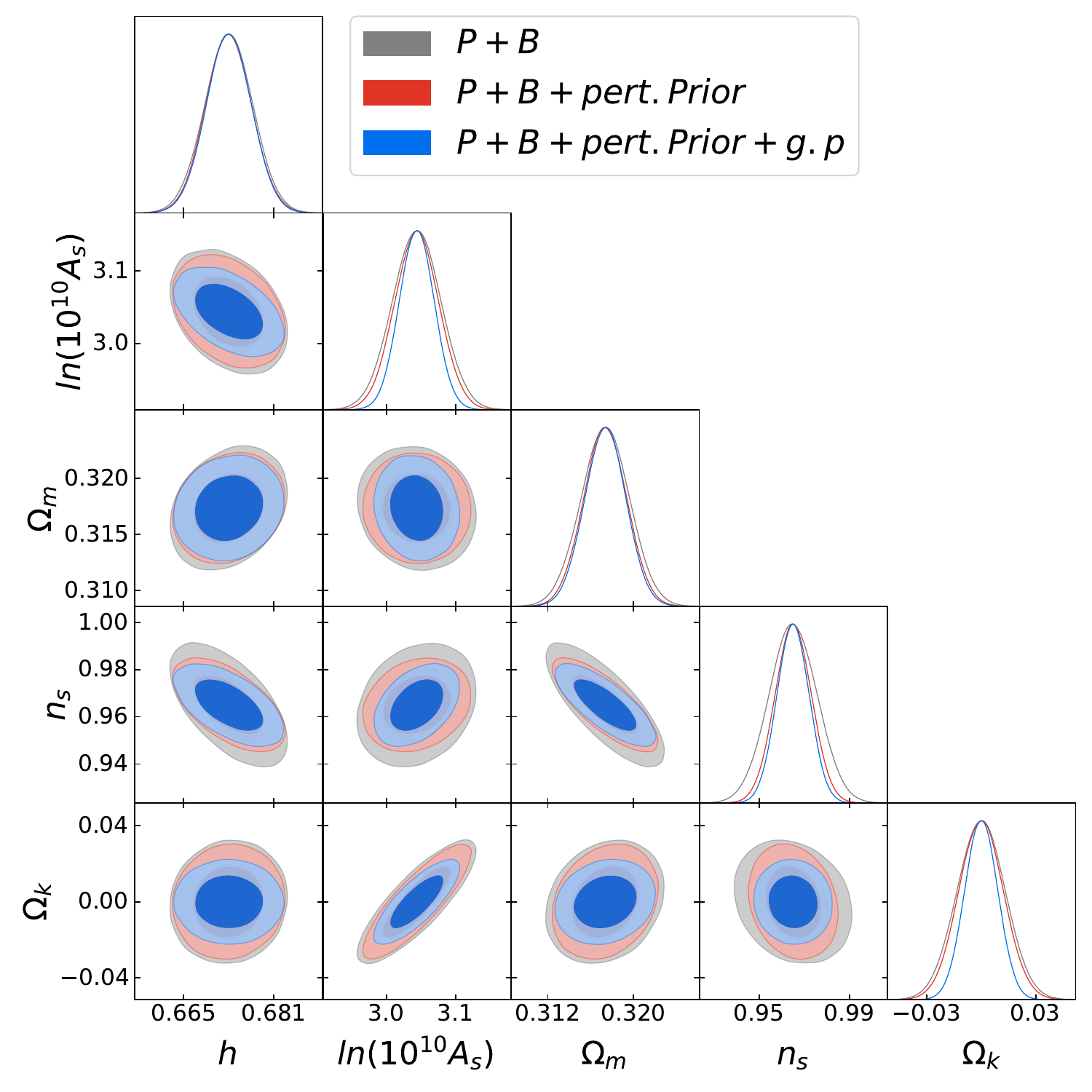}
 \end{minipage}
 \vspace{0.5cm}
 \begin{minipage}{0.47\textwidth}
 \centering
 
 \caption*{MegaMapper}
 \hspace{-0.625cm}
 \includegraphics[width=\textwidth]{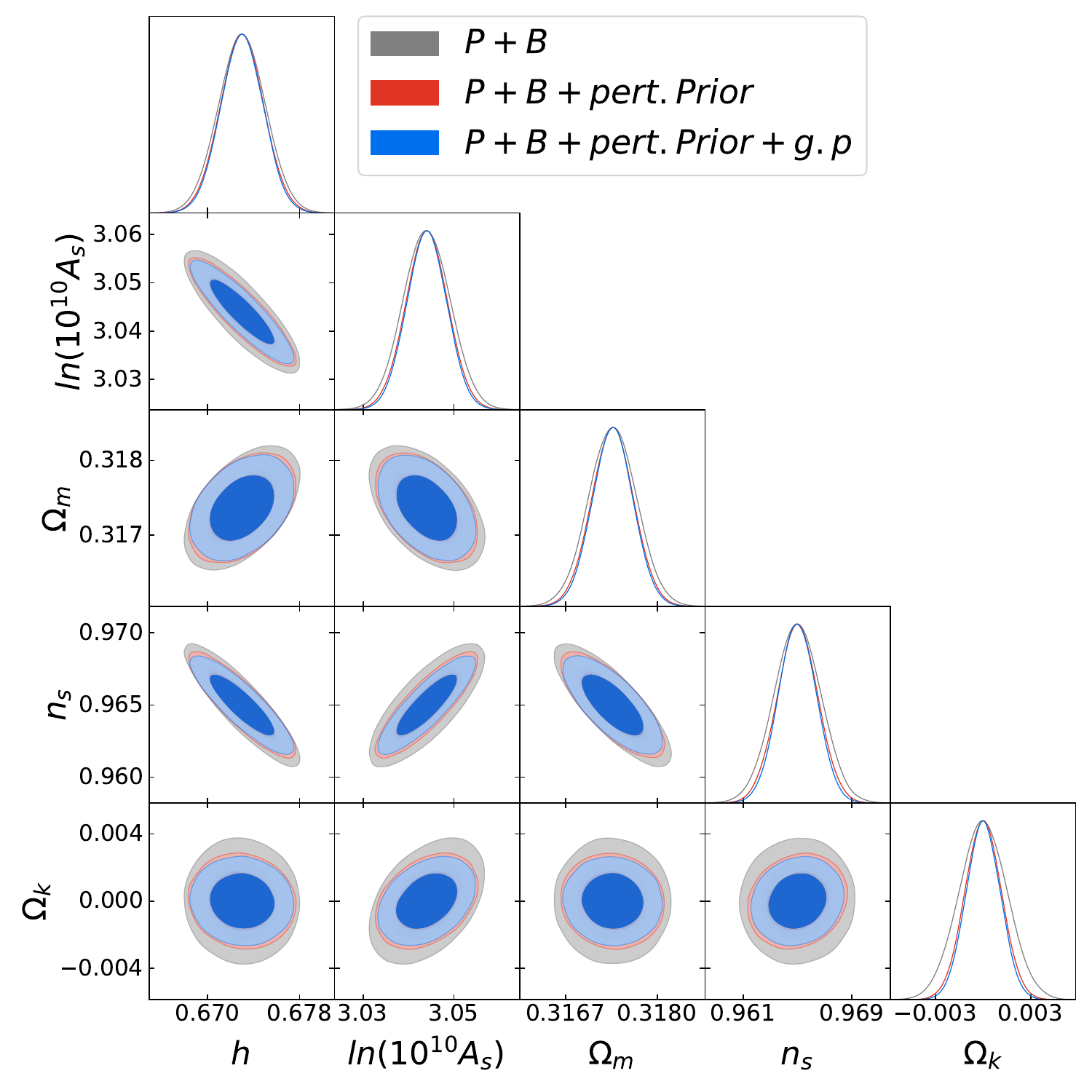}
 \end{minipage}
 \hfill
 \hspace{-2em}
 \begin{minipage}{0.53\textwidth}
 \centering
{
{\fontsize{7.7}{10}\selectfont
{
\begin{tabular}{|c|c|c|c|c|c|c|}
\hline
BOSS:$\sigma(\cdot)$ & $h$ &\hspace{-0.8em} $\ln(10^{10}A_s)$ \hspace{-0.8em} & $\Omega_m$ & $n_s$& $\Omega_k$ \\ 
\hline
$P$+$B$ & 0.007 & 0.130 & 0.006 & 0.025 & 0.050 \\
\hline
$P+B$+{p.p.} & 0.006 & 0.110 & 0.004 & 0.013 & 0.041 \\
\hline
$P$+$B$+{p.p.+g.p.} & 0.006 & 0.069 & 0.004 & 0.011 & 0.025 \\
\hline
\end{tabular}}
}
 \vspace{0.5cm}
 
 {\fontsize{7.7}{10}\selectfont
 {
\begin{tabular}{|c|c|c|c|c|c|c|}
\hline
DESI: $\sigma(\cdot)$ & $h$ &\hspace{-0.8em} $\ln(10^{10}A_s)$ \hspace{-0.8em} & $\Omega_m$ & $n_s$& $\Omega_k$ \\ 
\hline
$P$+$B$ & 0.004 & 0.035 & 0.002 & 0.011 & 0.013 \\
\hline
$P+B$+{p.p.} & 0.004 & 0.032 & 0.002 & 0.008 & 0.012 \\
\hline
$P$+$B$+{p.p.+g.p.} & 0.004 & 0.025 & 0.002 & 0.007 & 0.009 \\

\hline
\end{tabular}}
}
 \vspace{0.5cm}

{\fontsize{7.7}{10}\selectfont
{
\begin{tabular}{|c|c|c|c|c|c|c|}
\hline
MMo: $\sigma(\cdot)$ & $h$ &\hspace{-0.8em} $\ln(10^{10}A_s)$ \hspace{-0.8em} & $\Omega_m$ & $n_s$& $\Omega_k$ \\ 
\hline
$P$+$B$ & 0.002 & 0.0052 & 0.0003 & 0.002 & 0.0015 \\
\hline
$P+B$+{p.p.} & 0.002 & 0.0046 & 0.0003 & 0.002 & 0.0012 \\
\hline
$P$+$B$+{p.p.+g.p.} & 0.002 & 0.0044 & 0.0003 & 0.001 & 0.0011 \\
\hline
\end{tabular}}
}}
 \end{minipage}
 \hspace{0.15cm}
 
 \caption{Triangle plots and errors from Fisher forecasts for BOSS (top left), DESI (top right), and MegaMapper (bottom left), for base cosmological parameters including the spectral tilt
and spatial curvature. We always include the power spectrum at one loop order with the addition of either the loop bispectrum or the loop bispectrum with a perturbativity prior (p.p.) or also in combination with the ``galaxy-formation prior" (g.p.). We use all power spectrum and bispectrum multipoles in each case and use the analytical covariance without cross-correlations. }
 \label{fig:priorMastercosmo}
\end{figure}

	%
	%
	%
	%
	
	\bibliographystyle{JHEP}
	\small
	\bibliography{references}

\providecommand{\href}[2]{#2}\begingroup\raggedright\begin{thebibliography}{100}

\bibitem{Baumann:2010tm}
D.~Baumann, A.~Nicolis, L.~Senatore and M.~Zaldarriaga, \emph{{Cosmological
  Non-Linearities as an Effective Fluid}},
  \href{http://dx.doi.org/10.1088/1475-7516/2012/07/051}{\emph{JCAP} {\bf 1207}
  (2012) 051}, [\href{http://arxiv.org/abs/1004.2488}{{\tt 1004.2488}}].

\bibitem{Carrasco:2012cv}
J.~J.~M. Carrasco, M.~P. Hertzberg and L.~Senatore, \emph{{The Effective Field
  Theory of Cosmological Large Scale Structures}},
  \href{http://dx.doi.org/10.1007/JHEP09(2012)082}{\emph{JHEP} {\bf 09} (2012)
  082}, [\href{http://arxiv.org/abs/1206.2926}{{\tt 1206.2926}}].

\bibitem{DAmico:2019fhj}
G.~D'Amico, J.~Gleyzes, N.~Kokron, D.~Markovic, L.~Senatore, P.~Zhang et~al.,
  \emph{{The Cosmological Analysis of the SDSS/BOSS data from the Effective
  Field Theory of Large-Scale Structure}},
  \href{http://dx.doi.org/10.1088/1475-7516/2020/05/005}{\emph{JCAP} {\bf 05}
  (2020) 005}, [\href{http://arxiv.org/abs/1909.05271}{{\tt 1909.05271}}].

\bibitem{Ivanov:2019pdj}
M.~M. Ivanov, M.~Simonovi\'c and M.~Zaldarriaga, \emph{{Cosmological Parameters
  from the BOSS Galaxy Power Spectrum}},
  \href{http://dx.doi.org/10.1088/1475-7516/2020/05/042}{\emph{JCAP} {\bf 05}
  (2020) 042}, [\href{http://arxiv.org/abs/1909.05277}{{\tt 1909.05277}}].

\bibitem{Colas:2019ret}
T.~Colas, G.~D'amico, L.~Senatore, P.~Zhang and F.~Beutler, \emph{{Efficient
  Cosmological Analysis of the SDSS/BOSS data from the Effective Field Theory
  of Large-Scale Structure}},
  \href{http://dx.doi.org/10.1088/1475-7516/2020/06/001}{\emph{JCAP} {\bf 06}
  (2020) 001}, [\href{http://arxiv.org/abs/1909.07951}{{\tt 1909.07951}}].

\bibitem{Planck:2018vyg}
{\scshape Planck} collaboration, N.~Aghanim et~al., \emph{{Planck 2018 results.
  VI. Cosmological parameters}},
  \href{http://dx.doi.org/10.1051/0004-6361/201833910}{\emph{Astron.
  Astrophys.} {\bf 641} (2020) A6},
  [\href{http://arxiv.org/abs/1807.06209}{{\tt 1807.06209}}].

\bibitem{Zhang_2022}
P.~Zhang and Y.~Cai, \emph{{BOSS} full-shape analysis from the {EFTofLSS} with
  exact time dependence},
  \href{http://dx.doi.org/10.1088/1475-7516/2022/01/031}{\emph{Journal of
  Cosmology and Astroparticle Physics} {\bf 2022} (jan, 2022) 031}.

\bibitem{Carrilho_2023}
P.~Carrilho, C.~Moretti and A.~Pourtsidou, \emph{Cosmology with the {EFTofLSS}
  and {BOSS}: dark energy constraints and a note on priors},
  \href{http://dx.doi.org/10.1088/1475-7516/2023/01/028}{\emph{Journal of
  Cosmology and Astroparticle Physics} {\bf 2023} (jan, 2023) 028}.

\bibitem{Zhang:2021yna}
P.~Zhang, G.~D'Amico, L.~Senatore, C.~Zhao and Y.~Cai, \emph{{BOSS Correlation
  Function Analysis from the Effective Field Theory of Large-Scale Structure}},
   \href{http://arxiv.org/abs/2110.07539}{{\tt 2110.07539}}.

\bibitem{Chen:2021wdi}
S.-F. Chen, Z.~Vlah and M.~White, \emph{{A new analysis of the BOSS survey,
  including full-shape information and post-reconstruction BAO}},
  \href{http://arxiv.org/abs/2110.05530}{{\tt 2110.05530}}.

\bibitem{Simon:2022csv}
T.~Simon, P.~Zhang and V.~Poulin, \emph{{Cosmological inference from the
  EFTofLSS: the eBOSS QSO full-shape analysis}},
  \href{http://arxiv.org/abs/2210.14931}{{\tt 2210.14931}}.

\bibitem{Philcox:2021kcw}
O.~H.~E. Philcox and M.~M. Ivanov, \emph{{The BOSS DR12 Full-Shape Cosmology:
  $\Lambda$CDM Constraints from the Large-Scale Galaxy Power Spectrum and
  Bispectrum Monopole}},  \href{http://arxiv.org/abs/2112.04515}{{\tt
  2112.04515}}.

\bibitem{Tsedrik_2023}
M.~Tsedrik, C.~Moretti, P.~Carrilho, F.~Rizzo and A.~Pourtsidou,
  \emph{Interacting dark energy from the joint analysis of the power spectrum
  and bispectrum multipoles with the {EFTofLSS}},
  \href{http://dx.doi.org/10.1093/mnras/stad260}{\emph{Monthly Notices of the
  Royal Astronomical Society} {\bf 520} (jan, 2023) 2611--2632}.

\bibitem{damico2022boss}
G.~D'Amico, Y.~Donath, M.~Lewandowski, L.~Senatore and P.~Zhang, \emph{The boss
  bispectrum analysis at one loop from the effective field theory of
  large-scale structure},  2022.

\bibitem{Philcox_2022}
O.~H. Philcox, M.~M. Ivanov, G.~Cabass, M.~Simonovi{\'{c}}, M.~Zaldarriaga and
  T.~Nishimichi, \emph{Cosmology with the redshift-space galaxy bispectrum
  monopole at one-loop order},
  \href{http://dx.doi.org/10.1103/physrevd.106.043530}{\emph{Physical Review D}
  {\bf 106} (aug, 2022) }.

\bibitem{Ivanov_2023}
M.~M. Ivanov, O.~H. Philcox, G.~Cabass, T.~Nishimichi, M.~Simonovi{\'{c}} and
  M.~Zaldarriaga, \emph{Cosmology with the galaxy bispectrum multipoles:
  Optimal estimation and application to {BOSS} data},
  \href{http://dx.doi.org/10.1103/physrevd.107.083515}{\emph{Physical Review D}
  {\bf 107} (apr, 2023) }.

\bibitem{Philcox:2020vvt}
O.~H.~E. Philcox, M.~M. Ivanov, M.~Simonovi\'c and M.~Zaldarriaga,
  \emph{{Combining Full-Shape and BAO Analyses of Galaxy Power Spectra: A
  1.6\textbackslash{}\% CMB-independent constraint on H$_0$}},
  \href{http://dx.doi.org/10.1088/1475-7516/2020/05/032}{\emph{JCAP} {\bf 05}
  (2020) 032}, [\href{http://arxiv.org/abs/2002.04035}{{\tt 2002.04035}}].

\bibitem{DAmico:2020kxu}
G.~D'Amico, L.~Senatore and P.~Zhang, \emph{{Limits on $w$CDM from the EFTofLSS
  with the PyBird code}},
  \href{http://dx.doi.org/10.1088/1475-7516/2021/01/006}{\emph{JCAP} {\bf 01}
  (2021) 006}, [\href{http://arxiv.org/abs/2003.07956}{{\tt 2003.07956}}].

\bibitem{DAmico:2020tty}
G.~D'Amico, Y.~Donath, L.~Senatore and P.~Zhang, \emph{{Limits on Clustering
  and Smooth Quintessence from the EFTofLSS}},
  \href{http://arxiv.org/abs/2012.07554}{{\tt 2012.07554}}.

\bibitem{Riess:2019cxk}
A.~G. Riess, S.~Casertano, W.~Yuan, L.~M. Macri and D.~Scolnic, \emph{{Large
  Magellanic Cloud Cepheid Standards Provide a 1\% Foundation for the
  Determination of the Hubble Constant and Stronger Evidence for Physics beyond
  $\Lambda$CDM}},
  \href{http://dx.doi.org/10.3847/1538-4357/ab1422}{\emph{Astrophys. J.} {\bf
  876} (2019) 85}, [\href{http://arxiv.org/abs/1903.07603}{{\tt 1903.07603}}].

\bibitem{Freedman:2019jwv}
W.~L. Freedman et~al., \emph{{The Carnegie-Chicago Hubble Program. VIII. An
  Independent Determination of the Hubble Constant Based on the Tip of the Red
  Giant Branch}},  \href{http://arxiv.org/abs/1907.05922}{{\tt 1907.05922}}.

\bibitem{Verde:2019ivm}
L.~Verde, T.~Treu and A.~G. Riess, \emph{{Tensions between the Early and the
  Late Universe}},  in \emph{{Nature Astronomy 2019}}, 2019.
\newblock \href{http://arxiv.org/abs/1907.10625}{{\tt 1907.10625}}.
\newblock \href{http://dx.doi.org/10.1038/s41550-019-0902-0}{DOI}.

\bibitem{DAmico:2020ods}
G.~D'Amico, L.~Senatore, P.~Zhang and H.~Zheng, \emph{{The Hubble Tension in
  Light of the Full-Shape Analysis of Large-Scale Structure Data}},
  \href{http://dx.doi.org/10.1088/1475-7516/2021/05/072}{\emph{JCAP} {\bf 05}
  (2021) 072}, [\href{http://arxiv.org/abs/2006.12420}{{\tt 2006.12420}}].

\bibitem{Ivanov:2020ril}
M.~M. Ivanov, E.~McDonough, J.~C. Hill, M.~Simonovi\'c, M.~W. Toomey,
  S.~Alexander et~al., \emph{{Constraining Early Dark Energy with Large-Scale
  Structure}}, \href{http://dx.doi.org/10.1103/PhysRevD.102.103502}{\emph{Phys.
  Rev. D} {\bf 102} (2020) 103502},
  [\href{http://arxiv.org/abs/2006.11235}{{\tt 2006.11235}}].

\bibitem{Niedermann:2020qbw}
F.~Niedermann and M.~S. Sloth, \emph{{New Early Dark Energy is compatible with
  current LSS data}},
  \href{http://dx.doi.org/10.1103/PhysRevD.103.103537}{\emph{Phys. Rev. D} {\bf
  103} (2021) 103537}, [\href{http://arxiv.org/abs/2009.00006}{{\tt
  2009.00006}}].

\bibitem{Smith:2020rxx}
T.~L. Smith, V.~Poulin, J.~L. Bernal, K.~K. Boddy, M.~Kamionkowski and
  R.~Murgia, \emph{{Early dark energy is not excluded by current large-scale
  structure data}},
  \href{http://dx.doi.org/10.1103/PhysRevD.103.123542}{\emph{Phys. Rev. D} {\bf
  103} (2021) 123542}, [\href{http://arxiv.org/abs/2009.10740}{{\tt
  2009.10740}}].

\bibitem{Simon:2022adh}
T.~Simon, P.~Zhang, V.~Poulin and T.~L. Smith, \emph{{Updated constraints from
  the effective field theory analysis of BOSS power spectrum on Early Dark
  Energy}},  \href{http://arxiv.org/abs/2208.05930}{{\tt 2208.05930}}.

\bibitem{DAmico:2022gki}
G.~D'Amico, M.~Lewandowski, L.~Senatore and P.~Zhang, \emph{{Limits on
  primordial non-Gaussianities from BOSS galaxy-clustering data}},
  \href{http://arxiv.org/abs/2201.11518}{{\tt 2201.11518}}.

\bibitem{Cabass:2022wjy}
G.~Cabass, M.~M. Ivanov, O.~H.~E. Philcox, M.~Simonovi\'c and M.~Zaldarriaga,
  \emph{{Constraints on Single-Field Inflation from the BOSS Galaxy Survey}},
  \href{http://dx.doi.org/10.1103/PhysRevLett.129.021301}{\emph{Phys. Rev.
  Lett.} {\bf 129} (2022) 021301}, [\href{http://arxiv.org/abs/2201.07238}{{\tt
  2201.07238}}].

\bibitem{Moretti:2023drg}
C.~Moretti, M.~Tsedrik, P.~Carrilho and A.~Pourtsidou, \emph{{Modified gravity
  and massive neutrinos: constraints from the full shape analysis of BOSS
  galaxies and forecasts for Stage IV surveys}},
  \href{http://arxiv.org/abs/2306.09275}{{\tt 2306.09275}}.

\bibitem{Simon:2022hpr}
T.~Simon, \emph{{Constraining decaying dark matter with the effective field
  theory of large-scale structure}},  in \emph{{33rd Rencontres de Blois}:
  {Exploring the Dark Universe}}, 12, 2022.
\newblock \href{http://arxiv.org/abs/2212.03004}{{\tt 2212.03004}}.

\bibitem{Piga:2022mge}
L.~Piga, M.~Marinucci, G.~D'Amico, M.~Pietroni, F.~Vernizzi and B.~S. Wright,
  \emph{{Constraints on modified gravity from the BOSS galaxy survey}},
  \href{http://dx.doi.org/10.1088/1475-7516/2023/04/038}{\emph{JCAP} {\bf 04}
  (2023) 038}, [\href{http://arxiv.org/abs/2211.12523}{{\tt 2211.12523}}].

\bibitem{Chudaykin:2020ghx}
A.~Chudaykin, K.~Dolgikh and M.~M. Ivanov, \emph{{Constraints on the curvature
  of the Universe and dynamical dark energy from the Full-shape and BAO data}},
  \href{http://dx.doi.org/10.1103/PhysRevD.103.023507}{\emph{Phys. Rev. D} {\bf
  103} (2021) 023507}, [\href{http://arxiv.org/abs/2009.10106}{{\tt
  2009.10106}}].

\bibitem{Cabass:2022ymb}
G.~Cabass, M.~M. Ivanov, O.~H.~E. Philcox, M.~Simonovi\'c and M.~Zaldarriaga,
  \emph{{Constraints on multifield inflation from the BOSS galaxy survey}},
  \href{http://dx.doi.org/10.1103/PhysRevD.106.043506}{\emph{Phys. Rev. D} {\bf
  106} (2022) 043506}, [\href{http://arxiv.org/abs/2204.01781}{{\tt
  2204.01781}}].

\bibitem{Semenaite:2022unt}
A.~Semenaite, A.~G. S\'anchez, A.~Pezzotta, J.~Hou, A.~Eggemeier, M.~Crocce
  et~al., \emph{{Beyond \textendash{} \ensuremath{\Lambda}CDM constraints from
  the full shape clustering measurements from BOSS and eBOSS}},
  \href{http://dx.doi.org/10.1093/mnras/stad849}{\emph{Mon. Not. Roy. Astron.
  Soc.} {\bf 521} (2023) 5013--5025},
  [\href{http://arxiv.org/abs/2210.07304}{{\tt 2210.07304}}].

\bibitem{Rubira:2022xhb}
H.~Rubira, A.~Mazoun and M.~Garny, \emph{{Full-shape BOSS constraints on dark
  matter interacting with dark radiation and lifting the S8 tension}},
  \href{http://dx.doi.org/10.1088/1475-7516/2023/01/034}{\emph{JCAP} {\bf 01}
  (2023) 034}, [\href{http://arxiv.org/abs/2209.03974}{{\tt 2209.03974}}].

\bibitem{Carrilho:2022mon}
P.~Carrilho, C.~Moretti and A.~Pourtsidou, \emph{{Cosmology with the EFTofLSS
  and BOSS: dark energy constraints and a note on priors}},
  \href{http://dx.doi.org/10.1088/1475-7516/2023/01/028}{\emph{JCAP} {\bf 01}
  (2023) 028}, [\href{http://arxiv.org/abs/2207.14784}{{\tt 2207.14784}}].

\bibitem{Simon:2022ftd}
T.~Simon, G.~Franco~Abell\'an, P.~Du, V.~Poulin and Y.~Tsai,
  \emph{{Constraining decaying dark matter with BOSS data and the effective
  field theory of large-scale structures}},
  \href{http://dx.doi.org/10.1103/PhysRevD.106.023516}{\emph{Phys. Rev. D} {\bf
  106} (2022) 023516}, [\href{http://arxiv.org/abs/2203.07440}{{\tt
  2203.07440}}].

\bibitem{Porto:2013qua}
R.~A. Porto, L.~Senatore and M.~Zaldarriaga, \emph{{The Lagrangian-space
  Effective Field Theory of Large Scale Structures}},
  \href{http://dx.doi.org/10.1088/1475-7516/2014/05/022}{\emph{JCAP} {\bf 1405}
  (2014) 022}, [\href{http://arxiv.org/abs/1311.2168}{{\tt 1311.2168}}].

\bibitem{Carrasco:2013sva}
J.~J.~M. Carrasco, S.~Foreman, D.~Green and L.~Senatore, \emph{{The 2-loop
  matter power spectrum and the IR-safe integrand}},
  \href{http://dx.doi.org/10.1088/1475-7516/2014/07/056}{\emph{JCAP} {\bf 1407}
  (2014) 056}, [\href{http://arxiv.org/abs/1304.4946}{{\tt 1304.4946}}].

\bibitem{Carrasco:2013mua}
J.~J.~M. Carrasco, S.~Foreman, D.~Green and L.~Senatore, \emph{{The Effective
  Field Theory of Large Scale Structures at Two Loops}},
  \href{http://dx.doi.org/10.1088/1475-7516/2014/07/057}{\emph{JCAP} {\bf 1407}
  (2014) 057}, [\href{http://arxiv.org/abs/1310.0464}{{\tt 1310.0464}}].

\bibitem{Carroll:2013oxa}
S.~M. Carroll, S.~Leichenauer and J.~Pollack, \emph{{Consistent effective
  theory of long-wavelength cosmological perturbations}},
  \href{http://dx.doi.org/10.1103/PhysRevD.90.023518}{\emph{Phys. Rev.} {\bf
  D90} (2014) 023518}, [\href{http://arxiv.org/abs/1310.2920}{{\tt
  1310.2920}}].

\bibitem{Senatore:2014via}
L.~Senatore and M.~Zaldarriaga, \emph{{The IR-resummed Effective Field Theory
  of Large Scale Structures}},
  \href{http://dx.doi.org/10.1088/1475-7516/2015/02/013}{\emph{JCAP} {\bf 1502}
  (2015) 013}, [\href{http://arxiv.org/abs/1404.5954}{{\tt 1404.5954}}].

\bibitem{Baldauf:2015zga}
T.~Baldauf, E.~Schaan and M.~Zaldarriaga, \emph{{On the reach of perturbative
  methods for dark matter density fields}},
  \href{http://dx.doi.org/10.1088/1475-7516/2016/03/007}{\emph{JCAP} {\bf 1603}
  (2016) 007}, [\href{http://arxiv.org/abs/1507.02255}{{\tt 1507.02255}}].

\bibitem{Foreman:2015lca}
S.~{Foreman}, H.~{Perrier} and L.~{Senatore}, \emph{{Precision Comparison of
  the Power Spectrum in the EFTofLSS with Simulations}},
  \href{http://dx.doi.org/10.1088/1475-7516/2016/05/027}{\emph{JCAP} {\bf 1605}
  (2016) 027}, [\href{http://arxiv.org/abs/1507.05326}{{\tt 1507.05326}}].

\bibitem{Baldauf:2015aha}
T.~Baldauf, L.~Mercolli and M.~Zaldarriaga, \emph{{Effective field theory of
  large scale structure at two loops: The apparent scale dependence of the
  speed of sound}},
  \href{http://dx.doi.org/10.1103/PhysRevD.92.123007}{\emph{Phys. Rev.} {\bf
  D92} (2015) 123007}, [\href{http://arxiv.org/abs/1507.02256}{{\tt
  1507.02256}}].

\bibitem{Cataneo:2016suz}
M.~Cataneo, S.~Foreman and L.~Senatore, \emph{{Efficient exploration of
  cosmology dependence in the EFT of LSS}},
  \href{http://arxiv.org/abs/1606.03633}{{\tt 1606.03633}}.

\bibitem{Lewandowski:2017kes}
M.~Lewandowski and L.~Senatore, \emph{{IR-safe and UV-safe integrands in the
  EFTofLSS with exact time dependence}},
  \href{http://dx.doi.org/10.1088/1475-7516/2017/08/037}{\emph{JCAP} {\bf 1708}
  (2017) 037}, [\href{http://arxiv.org/abs/1701.07012}{{\tt 1701.07012}}].

\bibitem{Konstandin:2019bay}
T.~Konstandin, R.~A. Porto and H.~Rubira, \emph{{The Effective Field Theory of
  Large Scale Structure at Three Loops}},
  \href{http://arxiv.org/abs/1906.00997}{{\tt 1906.00997}}.

\bibitem{Pajer:2013jj}
E.~Pajer and M.~Zaldarriaga, \emph{{On the Renormalization of the Effective
  Field Theory of Large Scale Structures}},
  \href{http://dx.doi.org/10.1088/1475-7516/2013/08/037}{\emph{JCAP} {\bf 1308}
  (2013) 037}, [\href{http://arxiv.org/abs/1301.7182}{{\tt 1301.7182}}].

\bibitem{Abolhasani:2015mra}
A.~A. Abolhasani, M.~Mirbabayi and E.~Pajer, \emph{{Systematic Renormalization
  of the Effective Theory of Large Scale Structure}},
  \href{http://dx.doi.org/10.1088/1475-7516/2016/05/063}{\emph{JCAP} {\bf 1605}
  (2016) 063}, [\href{http://arxiv.org/abs/1509.07886}{{\tt 1509.07886}}].

\bibitem{Mercolli:2013bsa}
L.~Mercolli and E.~Pajer, \emph{{On the velocity in the Effective Field Theory
  of Large Scale Structures}},
  \href{http://dx.doi.org/10.1088/1475-7516/2014/03/006}{\emph{JCAP} {\bf 1403}
  (2014) 006}, [\href{http://arxiv.org/abs/1307.3220}{{\tt 1307.3220}}].

\bibitem{Senatore:2014eva}
L.~Senatore, \emph{{Bias in the Effective Field Theory of Large Scale
  Structures}},
  \href{http://dx.doi.org/10.1088/1475-7516/2015/11/007}{\emph{JCAP} {\bf 1511}
  (2015) 007}, [\href{http://arxiv.org/abs/1406.7843}{{\tt 1406.7843}}].

\bibitem{McQuinn:2015tva}
M.~McQuinn and M.~White, \emph{{Cosmological perturbation theory in 1+1
  dimensions}},
  \href{http://dx.doi.org/10.1088/1475-7516/2016/01/043}{\emph{JCAP} {\bf 1601}
  (2016) 043}, [\href{http://arxiv.org/abs/1502.07389}{{\tt 1502.07389}}].

\bibitem{Senatore:2014vja}
L.~Senatore and M.~Zaldarriaga, \emph{{Redshift Space Distortions in the
  Effective Field Theory of Large Scale Structures}},
  \href{http://arxiv.org/abs/1409.1225}{{\tt 1409.1225}}.

\bibitem{Baldauf:2015xfa}
T.~Baldauf, M.~Mirbabayi, M.~Simonovic and M.~Zaldarriaga, \emph{{Equivalence
  Principle and the Baryon Acoustic Peak}},
  \href{http://dx.doi.org/10.1103/PhysRevD.92.043514}{\emph{Phys. Rev.} {\bf
  D92} (2015) 043514}, [\href{http://arxiv.org/abs/1504.04366}{{\tt
  1504.04366}}].

\bibitem{Senatore:2017pbn}
L.~Senatore and G.~Trevisan, \emph{{On the IR-Resummation in the EFTofLSS}},
  \href{http://dx.doi.org/10.1088/1475-7516/2018/05/019}{\emph{JCAP} {\bf 1805}
  (2018) 019}, [\href{http://arxiv.org/abs/1710.02178}{{\tt 1710.02178}}].

\bibitem{Lewandowski:2018ywf}
M.~Lewandowski and L.~Senatore, \emph{{An analytic implementation of the
  IR-resummation for the BAO peak}},
  \href{http://arxiv.org/abs/1810.11855}{{\tt 1810.11855}}.

\bibitem{Blas:2016sfa}
D.~Blas, M.~Garny, M.~M. Ivanov and S.~Sibiryakov, \emph{{Time-Sliced
  Perturbation Theory II: Baryon Acoustic Oscillations and Infrared
  Resummation}},
  \href{http://dx.doi.org/10.1088/1475-7516/2016/07/028}{\emph{JCAP} {\bf 1607}
  (2016) 028}, [\href{http://arxiv.org/abs/1605.02149}{{\tt 1605.02149}}].

\bibitem{Lewandowski:2014rca}
M.~Lewandowski, A.~Perko and L.~Senatore, \emph{{Analytic Prediction of
  Baryonic Effects from the EFT of Large Scale Structures}},
  \href{http://dx.doi.org/10.1088/1475-7516/2015/05/019}{\emph{JCAP} {\bf 1505}
  (2015) 019}, [\href{http://arxiv.org/abs/1412.5049}{{\tt 1412.5049}}].

\bibitem{Braganca:2020nhv}
D.~P.~L. Bragan\c{c}a, M.~Lewandowski, D.~Sekera, L.~Senatore and R.~Sgier,
  \emph{{Baryonic effects in the Effective Field Theory of Large-Scale
  Structure and an analytic recipe for lensing in CMB-S4}},
  \href{http://arxiv.org/abs/2010.02929}{{\tt 2010.02929}}.

\bibitem{Angulo:2014tfa}
R.~E. Angulo, S.~Foreman, M.~Schmittfull and L.~Senatore, \emph{{The One-Loop
  Matter Bispectrum in the Effective Field Theory of Large Scale Structures}},
  \href{http://dx.doi.org/10.1088/1475-7516/2015/10/039}{\emph{JCAP} {\bf 1510}
  (2015) 039}, [\href{http://arxiv.org/abs/1406.4143}{{\tt 1406.4143}}].

\bibitem{Baldauf:2014qfa}
T.~Baldauf, L.~Mercolli, M.~Mirbabayi and E.~Pajer, \emph{{The Bispectrum in
  the Effective Field Theory of Large Scale Structure}},
  \href{http://dx.doi.org/10.1088/1475-7516/2015/05/007}{\emph{JCAP} {\bf 1505}
  (2015) 007}, [\href{http://arxiv.org/abs/1406.4135}{{\tt 1406.4135}}].

\bibitem{Bertolini:2016bmt}
D.~Bertolini, K.~Schutz, M.~P. Solon and K.~M. Zurek, \emph{{The Trispectrum in
  the Effective Field Theory of Large Scale Structure}},
  \href{http://arxiv.org/abs/1604.01770}{{\tt 1604.01770}}.

\bibitem{Baldauf:2015tla}
T.~Baldauf, E.~Schaan and M.~Zaldarriaga, \emph{{On the reach of perturbative
  descriptions for dark matter displacement fields}},
  \href{http://dx.doi.org/10.1088/1475-7516/2016/03/017}{\emph{JCAP} {\bf 1603}
  (2016) 017}, [\href{http://arxiv.org/abs/1505.07098}{{\tt 1505.07098}}].

\bibitem{Foreman:2015uva}
S.~Foreman and L.~Senatore, \emph{{The EFT of Large Scale Structures at All
  Redshifts: Analytical Predictions for Lensing}},
  \href{http://dx.doi.org/10.1088/1475-7516/2016/04/033}{\emph{JCAP} {\bf 1604}
  (2016) 033}, [\href{http://arxiv.org/abs/1503.01775}{{\tt 1503.01775}}].

\bibitem{Mirbabayi:2014zca}
M.~Mirbabayi, F.~Schmidt and M.~Zaldarriaga, \emph{{Biased Tracers and Time
  Evolution}},
  \href{http://dx.doi.org/10.1088/1475-7516/2015/07/030}{\emph{JCAP} {\bf 1507}
  (2015) 030}, [\href{http://arxiv.org/abs/1412.5169}{{\tt 1412.5169}}].

\bibitem{Angulo:2015eqa}
R.~Angulo, M.~Fasiello, L.~Senatore and Z.~Vlah, \emph{{On the Statistics of
  Biased Tracers in the Effective Field Theory of Large Scale Structures}},
  \href{http://dx.doi.org/10.1088/1475-7516/2015/09/029,
  10.1088/1475-7516/2015/9/029}{\emph{JCAP} {\bf 1509} (2015) 029},
  [\href{http://arxiv.org/abs/1503.08826}{{\tt 1503.08826}}].

\bibitem{Fujita:2016dne}
T.~Fujita, V.~Mauerhofer, L.~Senatore, Z.~Vlah and R.~Angulo, \emph{{Very
  Massive Tracers and Higher Derivative Biases}},
  \href{http://arxiv.org/abs/1609.00717}{{\tt 1609.00717}}.

\bibitem{Perko:2016puo}
A.~Perko, L.~Senatore, E.~Jennings and R.~H. Wechsler, \emph{{Biased Tracers in
  Redshift Space in the EFT of Large-Scale Structure}},
  \href{http://arxiv.org/abs/1610.09321}{{\tt 1610.09321}}.

\bibitem{Nadler:2017qto}
E.~O. Nadler, A.~Perko and L.~Senatore, \emph{{On the Bispectra of Very Massive
  Tracers in the Effective Field Theory of Large-Scale Structure}},
  \href{http://dx.doi.org/10.1088/1475-7516/2018/02/058}{\emph{JCAP} {\bf 1802}
  (2018) 058}, [\href{http://arxiv.org/abs/1710.10308}{{\tt 1710.10308}}].

\bibitem{Donath:2020abv}
Y.~Donath and L.~Senatore, \emph{{Biased Tracers in Redshift Space in the
  EFTofLSS with exact time dependence}},
  \href{http://dx.doi.org/10.1088/1475-7516/2020/10/039}{\emph{JCAP} {\bf 10}
  (2020) 039}, [\href{http://arxiv.org/abs/2005.04805}{{\tt 2005.04805}}].

\bibitem{McDonald:2009dh}
P.~McDonald and A.~Roy, \emph{{Clustering of dark matter tracers: generalizing
  bias for the coming era of precision LSS}},
  \href{http://dx.doi.org/10.1088/1475-7516/2009/08/020}{\emph{JCAP} {\bf 0908}
  (2009) 020}, [\href{http://arxiv.org/abs/0902.0991}{{\tt 0902.0991}}].

\bibitem{Lewandowski:2015ziq}
M.~Lewandowski, L.~Senatore, F.~Prada, C.~Zhao and C.-H. Chuang, \emph{{EFT of
  large scale structures in redshift space}},
  \href{http://dx.doi.org/10.1103/PhysRevD.97.063526}{\emph{Phys. Rev.} {\bf
  D97} (2018) 063526}, [\href{http://arxiv.org/abs/1512.06831}{{\tt
  1512.06831}}].

\bibitem{Senatore:2017hyk}
L.~Senatore and M.~Zaldarriaga, \emph{{The Effective Field Theory of
  Large-Scale Structure in the presence of Massive Neutrinos}},
  \href{http://arxiv.org/abs/1707.04698}{{\tt 1707.04698}}.

\bibitem{deBelsunce:2018xtd}
R.~de~Belsunce and L.~Senatore, \emph{{Tree-Level Bispectrum in the Effective
  Field Theory of Large-Scale Structure extended to Massive Neutrinos}},
  \href{http://arxiv.org/abs/1804.06849}{{\tt 1804.06849}}.

\bibitem{Lewandowski:2016yce}
M.~Lewandowski, A.~Maleknejad and L.~Senatore, \emph{{An effective description
  of dark matter and dark energy in the mildly non-linear regime}},
  \href{http://dx.doi.org/10.1088/1475-7516/2017/05/038}{\emph{JCAP} {\bf 1705}
  (2017) 038}, [\href{http://arxiv.org/abs/1611.07966}{{\tt 1611.07966}}].

\bibitem{Cusin:2017wjg}
G.~Cusin, M.~Lewandowski and F.~Vernizzi, \emph{{Dark Energy and Modified
  Gravity in the Effective Field Theory of Large-Scale Structure}},
  \href{http://dx.doi.org/10.1088/1475-7516/2018/04/005}{\emph{JCAP} {\bf 1804}
  (2018) 005}, [\href{http://arxiv.org/abs/1712.02783}{{\tt 1712.02783}}].

\bibitem{Bose:2018orj}
B.~Bose, K.~Koyama, M.~Lewandowski, F.~Vernizzi and H.~A. Winther,
  \emph{{Towards Precision Constraints on Gravity with the Effective Field
  Theory of Large-Scale Structure}},
  \href{http://dx.doi.org/10.1088/1475-7516/2018/04/063}{\emph{JCAP} {\bf 1804}
  (2018) 063}, [\href{http://arxiv.org/abs/1802.01566}{{\tt 1802.01566}}].

\bibitem{Assassi:2015jqa}
V.~Assassi, D.~Baumann, E.~Pajer, Y.~Welling and D.~van~der Woude,
  \emph{{Effective theory of large-scale structure with primordial
  non-Gaussianity}},
  \href{http://dx.doi.org/10.1088/1475-7516/2015/11/024}{\emph{JCAP} {\bf 1511}
  (2015) 024}, [\href{http://arxiv.org/abs/1505.06668}{{\tt 1505.06668}}].

\bibitem{Assassi:2015fma}
V.~Assassi, D.~Baumann and F.~Schmidt, \emph{{Galaxy Bias and Primordial
  Non-Gaussianity}},
  \href{http://dx.doi.org/10.1088/1475-7516/2015/12/043}{\emph{JCAP} {\bf 1512}
  (2015) 043}, [\href{http://arxiv.org/abs/1510.03723}{{\tt 1510.03723}}].

\bibitem{Bertolini:2015fya}
D.~Bertolini, K.~Schutz, M.~P. Solon, J.~R. Walsh and K.~M. Zurek,
  \emph{{Non-Gaussian Covariance of the Matter Power Spectrum in the Effective
  Field Theory of Large Scale Structure}},
  \href{http://arxiv.org/abs/1512.07630}{{\tt 1512.07630}}.

\bibitem{Bertolini:2016hxg}
D.~Bertolini and M.~P. Solon, \emph{{Principal Shapes and Squeezed Limits in
  the Effective Field Theory of Large Scale Structure}},
  \href{http://arxiv.org/abs/1608.01310}{{\tt 1608.01310}}.

\bibitem{Simonovic:2017mhp}
M.~Simonovic, T.~Baldauf, M.~Zaldarriaga, J.~J. Carrasco and J.~A. Kollmeier,
  \emph{{Cosmological perturbation theory using the FFTLog: formalism and
  connection to QFT loop integrals}},
  \href{http://dx.doi.org/10.1088/1475-7516/2018/04/030}{\emph{JCAP} {\bf 1804}
  (2018) 030}, [\href{http://arxiv.org/abs/1708.08130}{{\tt 1708.08130}}].

\bibitem{Nishimichi:2020tvu}
T.~Nishimichi, G.~D'Amico, M.~M. Ivanov, L.~Senatore, M.~Simonovi\'c, M.~Takada
  et~al., \emph{{Blinded challenge for precision cosmology with large-scale
  structure: results from effective field theory for the redshift-space galaxy
  power spectrum}},
  \href{http://dx.doi.org/10.1103/PhysRevD.102.123541}{\emph{Phys. Rev. D} {\bf
  102} (2020) 123541}, [\href{http://arxiv.org/abs/2003.08277}{{\tt
  2003.08277}}].

\bibitem{Chen:2020zjt}
S.-F. Chen, Z.~Vlah, E.~Castorina and M.~White, \emph{{Redshift-Space
  Distortions in Lagrangian Perturbation Theory}},
  \href{http://dx.doi.org/10.1088/1475-7516/2021/03/100}{\emph{JCAP} {\bf 03}
  (2021) 100}, [\href{http://arxiv.org/abs/2012.04636}{{\tt 2012.04636}}].

\bibitem{BOSS:2016wmc}
{\scshape BOSS} collaboration, S.~Alam et~al., \emph{{The clustering of
  galaxies in the completed SDSS-III Baryon Oscillation Spectroscopic Survey:
  cosmological analysis of the DR12 galaxy sample}},
  \href{http://dx.doi.org/10.1093/mnras/stx721}{\emph{Mon. Not. Roy. Astron.
  Soc.} {\bf 470} (2017) 2617--2652},
  [\href{http://arxiv.org/abs/1607.03155}{{\tt 1607.03155}}].

\bibitem{DESI:2016fyo}
{\scshape DESI} collaboration, A.~Aghamousa et~al., \emph{{The DESI Experiment
  Part I: Science,Targeting, and Survey Design}},
  \href{http://arxiv.org/abs/1611.00036}{{\tt 1611.00036}}.

\bibitem{Schlegel:2022vrv}
D.~J. Schlegel et~al., \emph{{The MegaMapper: A Stage-5 Spectroscopic
  Instrument Concept for the Study of Inflation and Dark Energy}},
  \href{http://arxiv.org/abs/2209.04322}{{\tt 2209.04322}}.

\bibitem{Chung:2023syw}
D.~J.~H. Chung, M.~M\"unchmeyer and S.~C. Tadepalli, \emph{{Search for
  Isocurvature with Large-scale Structure: A Forecast for Euclid and MegaMapper
  using EFTofLSS}},  \href{http://arxiv.org/abs/2306.09456}{{\tt 2306.09456}}.

\bibitem{Kajita:2016cak}
T.~Kajita, \emph{{Nobel Lecture: Discovery of atmospheric neutrino
  oscillations}},
  \href{http://dx.doi.org/10.1103/RevModPhys.88.030501}{\emph{Rev. Mod. Phys.}
  {\bf 88} (2016) 030501}.

\bibitem{Bousso:2013uia}
R.~Bousso, D.~Harlow and L.~Senatore, \emph{{Inflation after False Vacuum
  Decay}: {observational Prospects after Planck}},
  \href{http://dx.doi.org/10.1103/PhysRevD.91.083527}{\emph{Phys. Rev. D} {\bf
  91} (2015) 083527}, [\href{http://arxiv.org/abs/1309.4060}{{\tt 1309.4060}}].

\bibitem{Kleban:2012ph}
M.~Kleban and M.~Schillo, \emph{{Spatial Curvature Falsifies Eternal
  Inflation}},
  \href{http://dx.doi.org/10.1088/1475-7516/2012/06/029}{\emph{JCAP} {\bf 06}
  (2012) 029}, [\href{http://arxiv.org/abs/1202.5037}{{\tt 1202.5037}}].

\bibitem{Babich:2004gb}
D.~Babich, P.~Creminelli and M.~Zaldarriaga, \emph{{The Shape of
  non-Gaussianities}},
  \href{http://dx.doi.org/10.1088/1475-7516/2004/08/009}{\emph{JCAP} {\bf 08}
  (2004) 009}, [\href{http://arxiv.org/abs/astro-ph/0405356}{{\tt
  astro-ph/0405356}}].

\bibitem{Creminelli:2005hu}
P.~Creminelli, A.~Nicolis, L.~Senatore, M.~Tegmark and M.~Zaldarriaga,
  \emph{{Limits on non-gaussianities from wmap data}},
  \href{http://dx.doi.org/10.1088/1475-7516/2006/05/004}{\emph{JCAP} {\bf 0605}
  (2006) 004}, [\href{http://arxiv.org/abs/astro-ph/0509029}{{\tt
  astro-ph/0509029}}].

\bibitem{Senatore:2009gt}
L.~Senatore, K.~M. Smith and M.~Zaldarriaga, \emph{{Non-Gaussianities in Single
  Field Inflation and their Optimal Limits from the WMAP 5-year Data}},
  \href{http://dx.doi.org/10.1088/1475-7516/2010/01/028}{\emph{JCAP} {\bf 1001}
  (2010) 028}, [\href{http://arxiv.org/abs/0905.3746}{{\tt 0905.3746}}].

\bibitem{Cheung:2007st}
C.~Cheung, P.~Creminelli, A.~L. Fitzpatrick, J.~Kaplan and L.~Senatore,
  \emph{{The Effective Field Theory of Inflation}},
  \href{http://dx.doi.org/10.1088/1126-6708/2008/03/014}{\emph{JHEP} {\bf 03}
  (2008) 014}, [\href{http://arxiv.org/abs/0709.0293}{{\tt 0709.0293}}].

\bibitem{Bernardeau:2002jy}
F.~Bernardeau and J.-P. Uzan, \emph{{NonGaussianity in multifield inflation}},
  \href{http://dx.doi.org/10.1103/PhysRevD.66.103506}{\emph{Phys. Rev. D} {\bf
  66} (2002) 103506}, [\href{http://arxiv.org/abs/hep-ph/0207295}{{\tt
  hep-ph/0207295}}].

\bibitem{Lyth:2002my}
D.~H. Lyth, C.~Ungarelli and D.~Wands, \emph{{The Primordial density
  perturbation in the curvaton scenario}},
  \href{http://dx.doi.org/10.1103/PhysRevD.67.023503}{\emph{Phys. Rev. D} {\bf
  67} (2003) 023503}, [\href{http://arxiv.org/abs/astro-ph/0208055}{{\tt
  astro-ph/0208055}}].

\bibitem{Zaldarriaga:2003my}
M.~Zaldarriaga, \emph{{Non-Gaussianities in models with a varying inflaton
  decay rate}}, \href{http://dx.doi.org/10.1103/PhysRevD.69.043508}{\emph{Phys.
  Rev. D} {\bf 69} (2004) 043508},
  [\href{http://arxiv.org/abs/astro-ph/0306006}{{\tt astro-ph/0306006}}].

\bibitem{Senatore:2010wk}
L.~Senatore and M.~Zaldarriaga, \emph{{The Effective Field Theory of Multifield
  Inflation}}, \href{http://dx.doi.org/10.1007/JHEP04(2012)024}{\emph{JHEP}
  {\bf 04} (2012) 024}, [\href{http://arxiv.org/abs/1009.2093}{{\tt
  1009.2093}}].

\bibitem{Cabass:2022epm}
G.~Cabass, M.~M. Ivanov, O.~H.~E. Philcox, M.~Simonovic and M.~Zaldarriaga,
  \emph{{Constraining single-field inflation with MegaMapper}},
  \href{http://dx.doi.org/10.1016/j.physletb.2023.137912}{\emph{Phys. Lett. B}
  {\bf 841} (2023) 137912}, [\href{http://arxiv.org/abs/2211.14899}{{\tt
  2211.14899}}].

\bibitem{Flauger:2016idt}
R.~Flauger, M.~Mirbabayi, L.~Senatore and E.~Silverstein, \emph{{Productive
  Interactions: heavy particles and non-Gaussianity}},
  \href{http://dx.doi.org/10.1088/1475-7516/2017/10/058}{\emph{JCAP} {\bf 10}
  (2017) 058}, [\href{http://arxiv.org/abs/1606.00513}{{\tt 1606.00513}}].

\bibitem{Anastasiou:2022udy}
C.~Anastasiou, D.~P.~L. Bragan\c{c}a, L.~Senatore and H.~Zheng,
  \emph{{Efficiently evaluating loop integrals in the EFTofLSS using QFT
  integrals with massive propagators}},
  \href{http://arxiv.org/abs/2212.07421}{{\tt 2212.07421}}.

\bibitem{Zhang:2024thl}
H.~Zhang, M.~Bonici, G.~D'Amico, S.~Paradiso and W.~J. Percival,
  \emph{{HOD-informed prior for EFT-based full-shape analyses of LSS}},
  \href{http://dx.doi.org/10.1088/1475-7516/2025/04/041}{\emph{JCAP} {\bf 04}
  (2025) 041}, [\href{http://arxiv.org/abs/2409.12937}{{\tt 2409.12937}}].

\bibitem{Ivanov:2025qie}
M.~M. Ivanov, \emph{{Simulation-Based Priors without Simulations: an Analytic
  Perspective on EFT Parameters of Galaxies}},
  \href{http://arxiv.org/abs/2503.07270}{{\tt 2503.07270}}.

\bibitem{Planck:2019kim}
{\scshape Planck} collaboration, Y.~Akrami et~al., \emph{{Planck 2018 results.
  IX. Constraints on primordial non-Gaussianity}},
  \href{http://dx.doi.org/10.1051/0004-6361/201935891}{\emph{Astron.
  Astrophys.} {\bf 641} (2020) A9},
  [\href{http://arxiv.org/abs/1905.05697}{{\tt 1905.05697}}].

\bibitem{Tegmark:1997rp}
M.~Tegmark, \emph{{Measuring cosmological parameters with galaxy surveys}},
  \href{http://dx.doi.org/10.1103/PhysRevLett.79.3806}{\emph{Phys. Rev. Lett.}
  {\bf 79} (1997) 3806--3809},
  [\href{http://arxiv.org/abs/astro-ph/9706198}{{\tt astro-ph/9706198}}].

\bibitem{Agarwal:2020lov}
N.~Agarwal, V.~Desjacques, D.~Jeong and F.~Schmidt, \emph{{Information content
  in the redshift-space galaxy power spectrum and bispectrum}},
  \href{http://dx.doi.org/10.1088/1475-7516/2021/03/021}{\emph{JCAP} {\bf 03}
  (2021) 021}, [\href{http://arxiv.org/abs/2007.04340}{{\tt 2007.04340}}].

\bibitem{Yankelevich:2018uaz}
V.~Yankelevich and C.~Porciani, \emph{{Cosmological information in the
  redshift-space bispectrum}},
  \href{http://dx.doi.org/10.1093/mnras/sty3143}{\emph{Mon. Not. Roy. Astron.
  Soc.} {\bf 483} (2019) 2078--2099},
  [\href{http://arxiv.org/abs/1807.07076}{{\tt 1807.07076}}].

\bibitem{Baumann:2017gkg}
D.~Baumann, D.~Green and B.~Wallisch, \emph{{Searching for light relics with
  large-scale structure}},
  \href{http://dx.doi.org/10.1088/1475-7516/2018/08/029}{\emph{JCAP} {\bf 08}
  (2018) 029}, [\href{http://arxiv.org/abs/1712.08067}{{\tt 1712.08067}}].

\bibitem{DAmico:2022osl}
G.~D'Amico, Y.~Donath, M.~Lewandowski, L.~Senatore and P.~Zhang, \emph{{The
  BOSS bispectrum analysis at one loop from the Effective Field Theory of
  Large-Scale Structure}},  \href{http://arxiv.org/abs/2206.08327}{{\tt
  2206.08327}}.

\bibitem{DAmico:2022ukl}
G.~D'Amico, Y.~Donath, M.~Lewandowski, L.~Senatore and P.~Zhang, \emph{{The
  one-loop bispectrum of galaxies in redshift space from the Effective Field
  Theory of Large-Scale Structure}},
  \href{http://arxiv.org/abs/2211.17130}{{\tt 2211.17130}}.

\bibitem{Feldman:1993ky}
H.~A. Feldman, N.~Kaiser and J.~A. Peacock, \emph{{Power spectrum analysis of
  three-dimensional redshift surveys}},
  \href{http://dx.doi.org/10.1086/174036}{\emph{Astrophys. J.} {\bf 426} (1994)
  23--37}, [\href{http://arxiv.org/abs/astro-ph/9304022}{{\tt
  astro-ph/9304022}}].

\bibitem{Scoccimarro:1997st}
R.~Scoccimarro, S.~Colombi, J.~N. Fry, J.~A. Frieman, E.~Hivon and A.~Melott,
  \emph{{Nonlinear evolution of the bispectrum of cosmological perturbations}},
  \href{http://dx.doi.org/10.1086/305399}{\emph{Astrophys. J.} {\bf 496} (1998)
  586}, [\href{http://arxiv.org/abs/astro-ph/9704075}{{\tt astro-ph/9704075}}].

\bibitem{Chan:2016ehg}
K.~C. Chan and L.~Blot, \emph{{Assessment of the Information Content of the
  Power Spectrum and Bispectrum}},
  \href{http://dx.doi.org/10.1103/PhysRevD.96.023528}{\emph{Phys. Rev. D} {\bf
  96} (2017) 023528}, [\href{http://arxiv.org/abs/1610.06585}{{\tt
  1610.06585}}].

\bibitem{Sefusatti:2006pa}
E.~Sefusatti, M.~Crocce, S.~Pueblas and R.~Scoccimarro, \emph{{Cosmology and
  the Bispectrum}},
  \href{http://dx.doi.org/10.1103/PhysRevD.74.023522}{\emph{Phys. Rev. D} {\bf
  74} (2006) 023522}, [\href{http://arxiv.org/abs/astro-ph/0604505}{{\tt
  astro-ph/0604505}}].

\bibitem{Desjacques:2016bnm}
V.~Desjacques, D.~Jeong and F.~Schmidt, \emph{{Large-Scale Galaxy Bias}},
  \href{http://dx.doi.org/10.1016/j.physrep.2017.12.002}{\emph{Phys. Rept.}
  {\bf 733} (2018) 1--193}, [\href{http://arxiv.org/abs/1611.09787}{{\tt
  1611.09787}}].

\bibitem{Ferraro:2019uce}
S.~Ferraro et~al., \emph{{Inflation and Dark Energy from Spectroscopy at $z >
  2$}}, {\emph{Bull. Am. Astron. Soc.} {\bf 51} (2019) 72},
  [\href{http://arxiv.org/abs/1903.09208}{{\tt 1903.09208}}].

\bibitem{Kokron:2021faa}
N.~Kokron, J.~DeRose, S.-F. Chen, M.~White and R.~H. Wechsler, \emph{{Priors on
  red galaxy stochasticity from hybrid effective field theory}},
  \href{http://dx.doi.org/10.1093/mnras/stac1420}{\emph{Mon. Not. Roy. Astron.
  Soc.} {\bf 514} (2022) 2198--2213},
  [\href{http://arxiv.org/abs/2112.00012}{{\tt 2112.00012}}].

\bibitem{Alcock:1979mp}
C.~Alcock and B.~Paczynski, \emph{{An evolution free test for non-zero
  cosmological constant}},
  \href{http://dx.doi.org/10.1038/281358a0}{\emph{Nature} {\bf 281} (1979)
  358--359}.

\bibitem{Baldauf:2010vn}
T.~Baldauf, U.~Seljak and L.~Senatore, \emph{{Primordial non-Gaussianity in the
  Bispectrum of the Halo Density Field}},
  \href{http://dx.doi.org/10.1088/1475-7516/2011/04/006}{\emph{JCAP} {\bf 04}
  (2011) 006}, [\href{http://arxiv.org/abs/1011.1513}{{\tt 1011.1513}}].

\bibitem{Donath:inprog}
Y.~Donath and L.~Senatore, \emph{{Exact Trispectrum covariances and Fisher
  matrices}},  \href{http://arxiv.org/abs/in progress}{{\tt in progress}}.

\bibitem{Steele:2021lnz}
T.~Steele and T.~Baldauf, \emph{{Precise Calibration of the One-Loop
  Trispectrum in the Effective Field Theory of Large Scale Structure}},
  \href{http://dx.doi.org/10.1103/PhysRevD.103.103518}{\emph{Phys. Rev. D} {\bf
  103} (2021) 103518}, [\href{http://arxiv.org/abs/2101.10289}{{\tt
  2101.10289}}].

\bibitem{Reid:2015gra}
B.~Reid et~al., \emph{{SDSS-III Baryon Oscillation Spectroscopic Survey Data
  Release 12: galaxy target selection and large scale structure catalogues}},
  \href{http://dx.doi.org/10.1093/mnras/stv2382}{\emph{Mon. Not. Roy. Astron.
  Soc.} {\bf 455} (2016) 1553--1573},
  [\href{http://arxiv.org/abs/1509.06529}{{\tt 1509.06529}}].

\bibitem{BOSS:2016psr}
{\scshape BOSS} collaboration, F.~Beutler et~al., \emph{{The clustering of
  galaxies in the completed SDSS-III Baryon Oscillation Spectroscopic Survey:
  Anisotropic galaxy clustering in Fourier-space}},
  \href{http://dx.doi.org/10.1093/mnras/stw3298}{\emph{Mon. Not. Roy. Astron.
  Soc.} {\bf 466} (2017) 2242--2260},
  [\href{http://arxiv.org/abs/1607.03150}{{\tt 1607.03150}}].

\bibitem{Font-Ribera:2013rwa}
A.~Font-Ribera, P.~McDonald, N.~Mostek, B.~A. Reid, H.-J. Seo and A.~Slosar,
  \emph{{DESI and other dark energy experiments in the era of neutrino mass
  measurements}},
  \href{http://dx.doi.org/10.1088/1475-7516/2014/05/023}{\emph{JCAP} {\bf 05}
  (2014) 023}, [\href{http://arxiv.org/abs/1308.4164}{{\tt 1308.4164}}].

\end{thebibliography}\endgroup
\end{document}